\documentclass[a4paper]{article}

\usepackage[english]{babel}
\usepackage{booktabs}

\usepackage[
nochapters, 
beramono, 
eulermath,
pdfspacing, 
dottedtoc 
]{classicthesis} 

\usepackage[T1]{fontenc} 

\usepackage[utf8]{inputenc} 

\usepackage{graphicx} 
\graphicspath{{Figures/}} 

\usepackage{enumitem} 

\usepackage{lipsum} 

\usepackage{amsmath,amssymb,amsthm} 

\usepackage{varioref} 

\theoremstyle{definition} 

\theoremstyle{plain} 

\theoremstyle{remark} 

\hypersetup{
colorlinks=true, breaklinks=true, bookmarks=true,bookmarksnumbered,
urlcolor=webbrown, linkcolor=RoyalBlue, citecolor=webgreen, 
pdftitle={}, 
pdfauthor={\textcopyright}, 
pdfsubject={}, 
pdfkeywords={}, 
pdfcreator={pdfLaTeX}, 
} 

\usepackage{setspace} 
\usepackage{changepage} 
\usepackage[export]{adjustbox}
\usepackage{caption}
\usepackage{subcaption}
\usepackage{float}
\usepackage{amsmath}
\usepackage{comment}

\usepackage{titlesec}

\title{Dynamical Modularity in Automata Models of  Biochemical Networks}
\author{Thomas Parmer\textsuperscript{1}, Luis M. Rocha\textsuperscript{2,3}}

\date{} 

\begin{document}




\titleformat*{\section}{\normalfont\Large}
\titleformat*{\subsection}{\large\bfseries}

\maketitle

\begin{abstract}

Given the large size and complexity of most biochemical regulation and signaling networks, 
there is a non-trivial relationship between the micro-level logic of component interactions and the observed macro-dynamics.  Here we address this issue by formalizing
the concept of pathway modules developed by \textit{Marques-Pita and Rocha} \cite{marques2013canalization}, which are sequences of state updates that are guaranteed to occur (barring outside interference) in the causal dynamics of automata networks after the perturbation of a subset of driver nodes.  
We present a novel algorithm to automatically extract pathway modules from networks and characterize the interactions that may take place between the modules. This methodology uses only the causal logic of individual node variables (micro-dynamics) without the need to compute the dynamical landscape of the networks (macro-dynamics). 
Specifically, we identify complex modules, which 
maximize pathway length and require synergy between their components.
This allows us to propose a new take on dynamical modularity that partitions complex networks into causal pathways of variables that are guaranteed to transition to specific dynamical states given a perturbation to a set of driver nodes. Thus, the same node variable can take part in distinct modules depending on the state it takes.
Our measure of dynamical modularity of a network is then inversely proportional to the overlap among complex modules and maximal when complex modules are completely decouplable from one another in the network dynamics.
We estimate dynamical modularity for several genetic regulatory networks, including the full \textit{Drosophila melanogaster} segment-polarity network.
We discuss how identifying complex modules and the dynamical modularity portrait of networks explains the macro-dynamics of biological networks, such as uncovering the (more or less) decouplable building blocks of emergent computation (or collective behavior) in biochemical regulation and signaling.

\end{abstract}


{\let\thefootnote\relax\footnotetext{1 \textit{Luddy School of Informatics, Computing, and Engineering, Indiana University}}}

{\let\thefootnote\relax\footnotetext{2 \textit{Systems Science and Industrial Engineering Department, Binghamton University (State University of New York), Binghamton NY 13902, USA}}}

{\let\thefootnote\relax\footnotetext{3 \textit{Instituto Gulbenkian de Ciência, Oeiras 2780-156, Portugal}}}



\section{Introduction}


Biological intracellular networks are often composed of modules that are formed from molecular components and perform a specific function within the cell \cite{hartwell1999molecular,yeger2004network,peter2009modularity,davidson2010emerging,jaimovich2010modularity,vidal2011interactome}. These separable components are thought to provide the biological organism with robustness to environmental perturbations and genotypic mutations while still allowing enough flexibility for evolution \cite{stelling2004robustness,tanay2005conservation,pigliucci2008evolvability,hernandez2022effects}.  Though this kind of modularity is a general and intuitive concept, there are many formal definitions and it is not clear how to best model this phenomenon in living systems.

A popular method to model biological systems is with discrete dynamic network models because they are simple to understand and do not rely on the estimation of kinetic parameters (unlike continuous models such as ordinary differential equations) \cite{albert2009discrete,zanudo2018discrete}.
The network nodes (automata) typically represent genes, transcripts, proteins, molecular species, external inputs, or other qualitative states affecting the biological cell, and they are connected to one another by an edge if they interact in some way.  Automata can take any number of discrete states, but the simplest example is Boolean automata which are in one of two states at any given time \cite{kauffman1969metabolic, thomas1973boolean, albert2014boolean}.
These models are often applied to intracellular signaling networks because they can recreate phenotypical states using discrete variables, while ignoring kinetic details.  

Even given a simple model like a Boolean network (BN), it is unclear how to optimally decompose the system into separable modules.  Furthermore, networks observed in the real-world may contain a large number of nodes and interactions and be arbitrarily complex.  Although the interactions between components in such a BN are well defined, it is unclear how the individual node interactions give rise to global properties of the network such as its modularity or attractor landscape. It is, therefore, useful to understand how the micro-dynamics of component interactions give rise to the macro-dynamic behavior observed in networks.

Modularity methods have become a standard approach to decompose graphs and solve problems such as community detection. These approaches typically focus on topological modules, where a module is defined by a higher density of links between nodes in the same module and a lower density of links to nodes in different modules \cite{clauset2004finding,blondel2008fast}.  
Such structural modularity measures ignore the dynamical information of the network and may, therefore, be limited when it comes to discovering functional modules \cite{alexander2009understanding,verd2019modularity}.
It has also been shown that relying on structural information alone cannot account well for dynamics in control problems on BNs \cite{gates2016control}.
Additionally, the same circuit topologies in biological systems can switch between distinct steady states and dynamical behaviors (multi-stability and multifunctionality), which indicates that there is not a clear relationship between structure and function \cite{jimenez2017spectrum}.

For these reasons, dynamical modularity measures have also been proposed.
\textit{Irons et al.} proposed a method to decompose BNs into dynamical modules based on attractors of the system without the details of the underlying network, although with a suitable mathematical model the method can also determine how robust the subsystems are and how they are regulated \cite{irons2007identifying}. \textit{Kolchinsky et al.} proposed a method based on the idea that modules constrain the spread of perturbations and defined perturbation modularity as the autocovariance of perturbed trajectories \cite{kolchinsky2011prediction}.
Other methods combine structural and dynamical approaches by decomposing a network into structural components and associating these components with dynamical properties \cite{paul2018decomposition,kadelka2022decomposition}.
\textit{Za{\~n}udo and Albert} decompose an expanded version of a BN (taking into account its dynamics) into stable motifs, which are strongly-connected components representing partial fixed points of the system. They show that controlling these motifs can drive a network to a desired attractor \cite{zanudo2015cell}.

We use the threshold network representation of automata network dynamics proposed by \textit{Marques-Pita and Rocha}, the \textit{dynamics canalization map} (DCM), which represents both Boolean states of every node.  They simplified the description of a network's dynamics by explicitly removing redundancy in the logical expressions governing each node's update (described in the following subsections).  This redescription allows to describe dynamical modularity, critical nodes that guarantee convergence to steady states (with incomplete knowledge of initial conditions), and measures of macro-level canalization \cite{marques2013canalization}.
Using the DCM, they identified the dynamical modules on the \textit{Drosophila melanogaster} single-cell segment polarity network (SPN) \cite{albert2003topology} that are controlled by the network's inputs.

Here we formally define modules on the DCM.  We expand the analysis of \cite{marques2013canalization} by finding such modules not just for the inputs but for all possible nodes and low-order interactions on the \textit{drosophila} segment polarity network (SPN).  We develop theoretical results to explain the interaction of different dynamical modules and define \textit{synergy} between modules. The concept of synergy is used to define a subset of modules, called complex modules, that capture all unique dynamical information in the BN.

We then use complex modules to define a measure of dynamical modularity for BNs.  This measure is driven purely by the network's dynamics rather than its structure.  Unlike other measures, our dynamical modularity measure differentiates between different states of a variable. It accounts for the fact that a single variable may be part of multiple functional modules depending on its behavior and thus may be useful in models of multifunctionality in biological systems. Additionally, our measure makes explicit the influence of each variable in each state and the control that a given variable has on the rest of the network.  We also use the mean dynamical modularity of a network to quantify its decomposability and use this metric to compare different biological networks.

This paper is organized as follows: In the remainder of the Introduction, we give some background on schematic redescription, canalyzing maps, and the DCM introduced in \cite{marques2013canalization}.  In section II, we give theoretical results concerning the formal definition of pathway modules (based on the perturbation of seed nodes), complex modules (which have synergy between seed nodes and are of maximal size), and dynamical modularity (based on an optimal partition of the DCM).
In section III, we apply our methodology to the \textit{drosophila} single-cell and parasegment SPN to find complex modules on these networks, quantify their dynamical modularity, and compare them to other biological networks of comparable size.  In section IV, we offer our conclusions and note some limitations to our methodology.

\subsection{Boolean Networks}

A Boolean automaton $x$ can be in one of two states, x $\in \{0,1\}$ (traditionally ON or OFF, representing activation of the molecule of interest above or below a certain threshold) \cite{kauffman1969metabolic, thomas1973boolean, albert2014boolean}.  The state of an automaton is updated in discrete time steps governed by a logical function of its inputs (e.g., $A = B \lor C$).  More formally, x$^{t+1} = f($x$_{1}^t$,...,x$_{k}^t)$ for automaton x with $k$ inputs at time step $t$, where the mapping $f:\{0,1\}^k \to \{0,1\}$ can be read from the automaton's look-up table $F$ that denotes the output x$^{t+1}$ for each of its $2^k$ input combinations (refer to Figure \ref{fig:example_OR}).

A \textit{BN} is a graph $B=(X,E)$ where $X$ is a set of $n$ Boolean automata nodes and $E$ is a set of directed edges $e_{i,j} \in E$ that indicate that node $x_i$ is an input to node $x_j$ \cite{kauffman1969metabolic,gershenson2004introduction}.  The \textit{configuration} of a BN is 
the value of all automata states $x \in X$.
Automata nodes within the network may be updated synchronously in discrete time steps or asynchronously, where each node has a different update schedule (deterministic asynchronous BNs) or is selected at random with some probability \cite{harvey1997time, gershenson2004introduction}.  Deterministic BNs are guaranteed to eventually resolve in some attractor, either a fixed point or a limit cycle of repeating states.  In biological networks, attractors are typically associated with some phenotype (such as a wildtype expression pattern) \cite{albert2014boolean}.

\subsection{Node canalization}

\begin{figure}
\centering
\includegraphics[width=1.2\columnwidth]{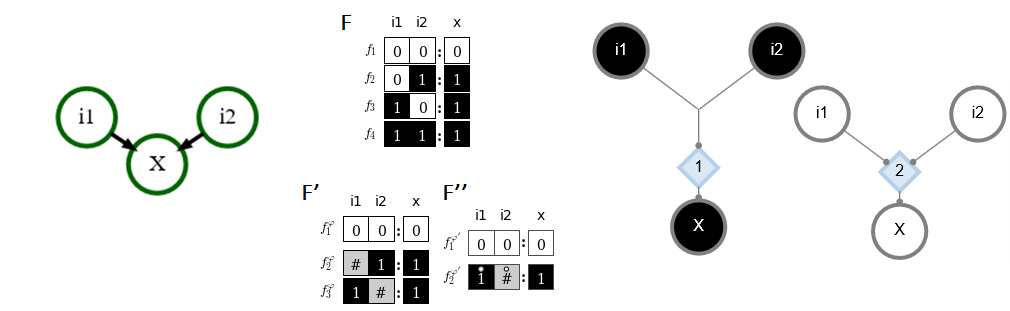}
\caption{\textbf{Example automaton with a logical 'OR' transition function}.  Left: The node $x$ has two inputs.  Center: The look-up table $F$ describing the transition of $x$ and the one and two-symbol schemata redescription ($F'$ and $F''$; \# is a wildcard enput, $^o$ is a group-invariant enput).  Right: The CM for $x$ with the associated logical function representing both state updates.  Black s-units represent nodes in their ON state, white s-units represent nodes in their OFF state, and blue diamonds represent t-units labeled with their threshold value ($\tau$).  Each edge (fibre) represents the logical contribution of a source unit to an end unit and thus dictates how X will turn ON or OFF.  The CM shows that one ON input is sufficient to turn X ON but that two OFF inputs are needed to turn X OFF.}
\label{fig:example_OR}
\end{figure}

\textit{Canalization} refers to the ability of a subset of inputs to control an automaton's transition \cite{waddington1942canalization,kauffman1984emergent, siegal2002waddington, kauffman2004genetic, marques2013canalization, gates2016control, manicka2017role}.  The Boolean transition function of the automaton is thus satisfied by this subset of inputs and the logical states of its other inputs can be ignored.  This shows inherent redundancy in the transition function that can be removed via schematic redescription, a process introduced in \cite{marques2013canalization}.  The Quine-McCluskey Boolean minimization algorithm is used to reduce an automaton's look-up table $F$ to a set of prime implicants \cite{Quine1955truth} that are then combined to form a set of wildcard schemata $F^{'}$.  
Within the wildcard schemata only essential inputs (enputs) are defined as ON or OFF. Every other input is replaced with the wildcard symbol (\#) which indicates that those inputs are redundant (given the states of the enputs) and do not affect the automaton's state transition \cite{marques2013canalization}.  Additionally, permutations of inputs that leave the transition unchanged are marked with a position-free symbol ($^o$) and combined to form group-invariant enputs.  This captures the input-symmetry within the wildcard schemata and further reduces them to a set of two-symbol schemata without permutation redundancy $F^{''}$ (see Figure \ref{fig:example_OR}).  It is not surprising to find input redundancy and input-symmetry at the automaton level (known as micro-canalization) in biological networks because they are known to be robust, which buffers perturbations to individual inputs and can help maintain wildtype phenotypes  \cite{conrad1990geometry, siegal2002waddington, pigliucci2008evolvability}.

After all redundancy is removed, the necessary and sufficient logic of automata nodes can also be represented by a canalyzing map (CM) \cite{marques2013canalization},
a type of threshold network \cite{mcculloch1943logical}.
The CM of a variable $x$ represents all logical transitions (two-symbol schemata) that can result in an update of the state of $x$.  It is composed of \textit{s-units} which represent nodes ($x$ and its inputs) in a particular discrete state (e.g., ON or OFF), \textit{t-units} that implement the transition function of $x$ via thresholds, and fibres that connect s-units and t-units together.  A fibre may connect one s-unit to one t-unit (indicating that the s-unit sends one \textit{signal} to the t-unit), or multiple fibres may be merged together so that several s-units send one signal to a t-unit (representing permutation redundancy).  T-units have a threshold value $\tau$ and only fire if they receive at least $\tau$ simultaneous incoming signals from s-units; s-units fire if they receive any incoming signal from a t-unit.  All logical interactions between $x$ and its inputs are represented such that each t-unit corresponds to one schema in $F^{''}$.
Thus the CM is equivalent to the schemata necessary to ensure the logical transition from $x^t$ to $x^{t+1}$ (see Figure \ref{fig:example_OR}).

\subsection{Dynamics canalization maps}

The canalyzing map of each automaton can be be integrated into a larger threshold network called the DCM \cite{marques2013canalization}.  This parsimonious network, with input and symmetry redundancy removed, represents the control logic and dynamics of the entire BN. Each possible state of each variable is represented in the DCM; thus there are $2n$ s-units for a BN of size $n$ and t-units that represent each schema, as defined for each individual CM.

Because the DCM captures the state updates of all automata in the network, it is possible to determine a logical sequence of signals based on the deterministic firing of s-units and t-units.  Given an initial set of s-units (and barring any outside interference), a logical sequence of state updates is guaranteed to occur based on the logic embedded in the schema $F''$ and represented by the t-units in the DCM.  This sequence of state updates is referred to as a \textit{pathway module}
because it is a sequence of s-units and t-units firing in the DCM and the state updates occur independently of all other node states that are not involved in the firing of t-units \cite{marques2013canalization}.

Importantly, the initial conditions of a pathway module may only define the state of a (small) subset of nodes in the network, while the complement of this set is assumed to be in an unknown state (represented by `\#').  
These known and unknown node states are referred to as a \textit{partial configuration} $\hat{x}$ of the network.
Because dynamics are deterministic, this partial configuration causes a sequence of state updates (which we refer to as dynamical unfolding) that results in an outcome 
configuration $P$, 
which may be either another partial configuration or an attractor of the network.  If $P$ is an attractor then we know that $\hat{x}$ controls the full network dynamics because the attractor is guaranteed to occur based on the subset of known s-units in the initial condition if the initial node states are sustained \cite{marques2013canalization}.
Thus, we can compute global (macro-scale) dynamical information from only partial knowledge of the network configuration by integrating knowledge of the local (micro-scale) dynamics. 

We note that the calculation of dynamical unfolding is similar to the calculation of partial steady states in the logical interaction hypergraph \cite{klamt2006methodology}, cascading effects of node removal and the calculation of elementary signaling modes on the expanded network \cite{wang2011elementary}, and calculation of the logical domain of influence on the expanded network \cite{yang2018target}.  However, none of these methods explicitly remove input symmetry from the transition functions of nodes or consider different types of node perturbation.

\begin{figure}[h!]
\includegraphics[width=1.2\columnwidth]{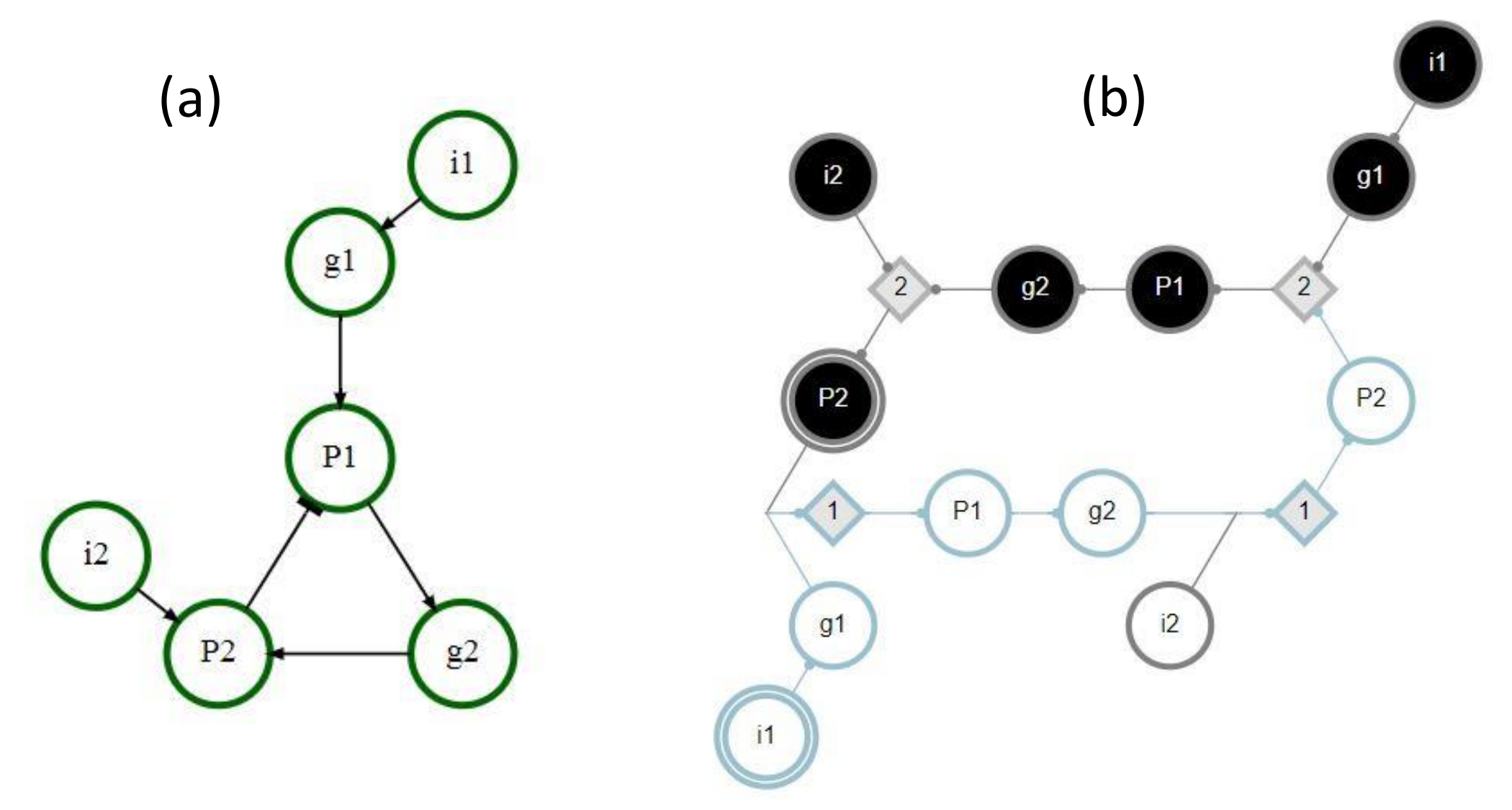}
\caption{\textbf{Example GRN}.  The interaction graph is shown on the left, the DCM on the right.  This network has two inputs (\textit{i1},\textit{i2}), two genes (\textit{g1},\textit{g2}), and two proteins (P1,P2).  There is a negative feedback loop present as P1 promotes the creation of P2, an inhibitory protein.  P1 requires both \textit{g1} and the absence of P2, while P2 requires both \textit{g2} and \textit{i2}.  The exact logic of the dynamics is captured in the DCM, where black s-units represent nodes in their active (ON) state, white s-units represent nodes in their inactive (OFF) state, and grey diamonds represent t-units.  The seeds of pathway modules $\textbf{M}_{P2-1}$ and $\textbf{M}_{i1-0}$ are indicated by double edges; the pathway module $\textbf{M}_{i1-0}$ is shown in blue.  
Note that t-units with $\tau=1$ and no permutation redundancy are left out of the DCM for simplicity.
}
\label{fig:example_network}
\end{figure}

In the next section we give a formal treatment of pathway modules and provide definitions of novel concepts such as pathway module interaction, complex modules, and measures of dynamical modularity.

\section{Formal Results}

Pathway modules can be discovered using a breadth-first search algorithm on the DCM starting from the initial seeds.  However, unlike a search on a traditional graph, the different types of units on the DCM must be considered.  For a t-unit to be included, its threshold $\tau$ must be met.  Furthermore, the threshold is met only if its inputs fire simultaneously; therefore, the firing of a t-unit with multiple inputs requires confidence that all of its inputs can fire at the same time.  If the initial seeds are pinned (not allowed to change), then they fire at every time step and all downstream s-units fire repeatedly as well. However, if the initial seeds are only perturbed once (pulse perturbation), then their future state cannot be guaranteed after time $t=0$.  The type of initial perturbation, therefore, affects the pathway module that unfolds on the DCM. 

\begin{figure}
\begin{adjustwidth}{-1cm}{}
\centering
\includegraphics[width=1.1\columnwidth]{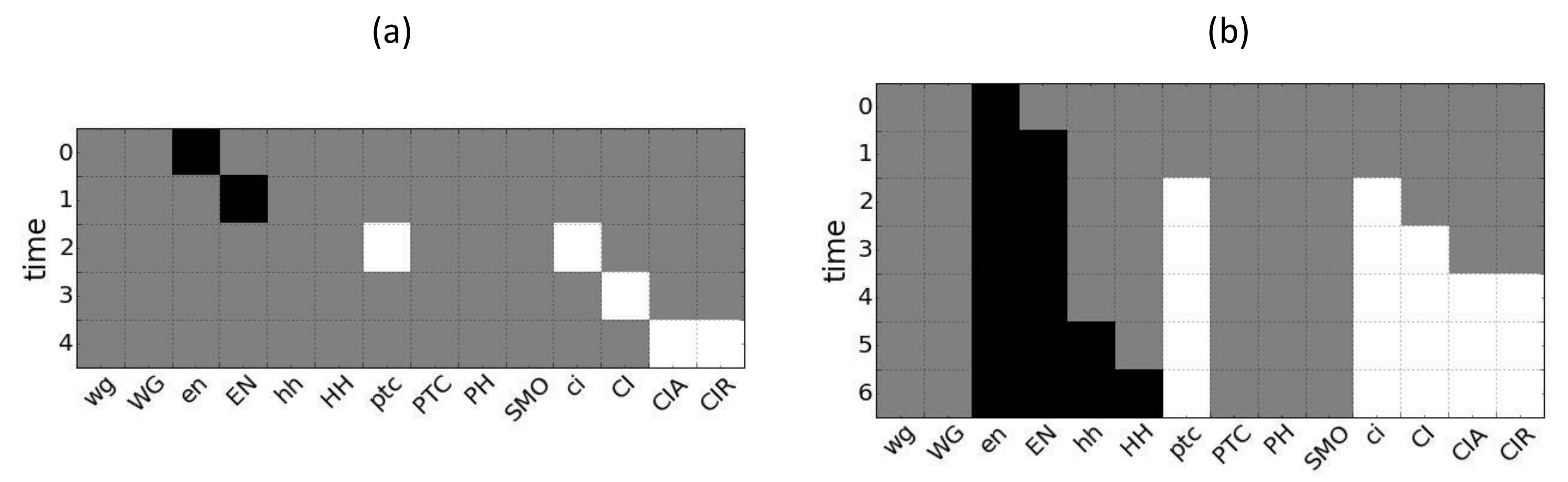}
\caption{\textbf{Example of different perturbations in the \textit{drosophila} SPN.}  
Black represents nodes in their active (ON) state, white represents nodes in their inactive (OFF) state, and grey represents nodes whose state is unknown.
(a) The dynamical unfolding of the pathway module $\textbf{M}_{en-1}$ in the \textit{drosophila} single-cell SPN under pulse perturbation. 
Node states are not maintained over time, and as a result, cannot activate \textit{hh} or HH which require the activation of EN.
(b) $\textbf{M}_{en-1}$ unfolding under pinning perturbation.  The s-units \textit{hh}-1 and HH-1 fire due to the ongoing activation of EN.}
\label{fig:control_types}
\end{adjustwidth}
\end{figure}

As an example, consider the case of the \textit{engrailed} (\textit{en}) protein from the \textit{drosophila} SPN (see Fig. \ref{fig:control_types}).
If \textit{en} is perturbed to ON for only a single time step, the subsequent pathway module 
takes four time steps to unfold
(panel a).  However, if \textit{en} is constitutively active, then the resulting pathway module 
takes six time steps to unfold
and reaches two extra node states (\textit{hh-1} and HH-1), which can fire because the logical condition for \textit{hh-1} (\textit{hh}$_{t+1}$=EN$_{t}$ $\wedge$ $\lnot$CIR$_{t}$) is satisfied by the ongoing firing of EN-1 (panel b).
Perturbations may be more complicated than these two extremes of pulse or pinning perturbation. For example, an s-unit may be fired in $k$ time step intervals or may be controlled for a specific length of time and then released.  

It is also important to factor logical contradiction into the search algorithm when determining pathway modules.  If an s-unit representing the state of variable $x$ fires at time $t$, then an s-unit representing a different state of $x$ cannot also fire at time $t$. For pinning perturbation, this guarantees that only one state of any variable is ever reached, while for pulse perturbation s-units representing different states of $x$ may fire.  The consequence of this is that modules based on pinning perturbation may uncover fixed points in the network but 
not limit cycles.

In the example GRN in Fig. \ref{fig:example_network}, the module $\textbf{M}_{i1-0}$ includes the same s-units (g1-0, P1-0, g2-0, and P2-0) and t-units (not shown in the figure), regardless of perturbation type.  The module $\textbf{M}_{P2-1}$, however, includes the s-units P1-0, g2-0, and P2-0 under pulse perturbation but only P1-0 and g2-0 under pinning perturbation (because P2-0 is a logical contradiction to P2-1).

Multiple pathway modules may simultaneously unfold within the complex dynamics of a network. Thus, it is natural to explore how different pathway modules may interact, assuming that their initial conditions are simultaneously activated.  This leads to concepts such as independence, logical obstruction, and synergy.
Synergy is used to define complex modules, which are maximal pathway modules whose seeds have synergy between them; complex modules are further used to define the dynamical modularity of the network.
Formal definitions of these concepts are given in the following subsections.


\subsection{Defining Pathway Modules}

The definition of the DCM is provided in the Introduction.
Formally, let $X$ be the set of all nodes in an automata network $B$ and $\mathcal{S}$ be the set of all s-units in the DCM (which represents all node states) of $B$.  For Boolean networks, $| \mathcal{S} | = 2 * | X |$.  Let $\Theta$ be the set of all threshold units in the DCM (its size $| \Theta |$ depends on the number 
of redescribed schemata for each node in $B$). 
At any given time $t$, a set of s-units, $S^t \subset \mathcal{S}$, and a set of t-units, $\theta^t \subset \Theta$, fire in the DCM.
A t-unit fires at time $t$ if its logical conditions are met (determined by its threshold value $\tau$), causing its neighboring s-unit to fire at time $t+1$.  The firing of a t-unit means that there is sufficient information to satisfy a schema in the look-up table of the variable associated with the downstream s-unit.  The output state of this schema is the state associated with the s-unit (i.e., 0 or 1).
An s-unit fires if its predecessor (a t-unit) fires; when it fires, it sends signals to all of its neighboring t-units.
An s-unit may also fire at time $t$ due to an external signal, i.e., a perturbation that is applied to that node variable of $B$. 
Such a signal can immediately trigger 
downstream t-units to fire if their logical conditions are met.
In other words, while signals from t-units to s-units propagate with a one time step delay ($dt=1$), signals from s-units to t-units propagate immediately ($dt=0$)  \footnote{Alternatively, the s-unit to t-unit transition can have delay $dt=1$ and the t-unit to s-unit transition can have delay $dt=0$. The choice of which one has $dt=1$ is arbitrary because t-units can only connect to s-units in the DCM and vice versa (i.e., either preserves the dynamics of $B$). Importantly, it takes one time step for an s-unit to fire after its inputs have fired.}. This asymmetry guarantees that the original dynamics of state transitions in $B$ is preserved in the DCM. 
Here, we denote the state of a Boolean variable $x$ by $x$-1 if $x$ is ON and $x$-0 if $x$ is OFF.

There is an important constraint on the set $S^t$ of s-units that fire at time $t$. If an s-unit that represents the state of $x$ fires at time $t$, the s-unit that represents the negation of the state of $x$ cannot fire.
Thus, $x$-1 $\in S^t \implies x$-0 $\notin S^t$ and vice versa.
$S_i^t$ represents a partial configuration whereby the logical state of the corresponding node variables $x \in X$ is known,
but any other variable may be in either state (ON or OFF). The number of (full) configurations that a partial configuration $S^t$ describes is $2^{|X| - | S^t|}$.
As a corollary, $\max(|S_i^t|)=|X|$ denotes the situation when the logical state of all variables is known, and we know the precise configuration the network is in at time $t$.

A \textit{seed set} is a set of s-units $S^0 \subset S$ that, without loss of generality, are assumed to fire  at time $t=0$. This functions as an external perturbation applied to a subset of variables $X^0 \subset X$ of network $B$.  Note that these s-units can represent variables in either the ON or OFF states.

Canalized dynamics is represented by a function $\mu (S^t) \rightarrow S^{t+1}$, whereby the set of s-units $S^t$ that fire at time $t$ transition to another set of s-units that fire at time $t+1$.
Naturally, this function can be applied sequentially for $t=0,1,..., \infty$, which we refer to as dynamical unfolding (see below). 
Note also that $S^{t+1} \cap S^t \not\equiv  \emptyset$ if any s-units in $S^t$ fire again due to external perturbation, regulation by other members of $S^t$, or self-reinforcement.
Importantly, unlike the computation of dynamics in the original Boolean network $B$, dynamical unfolding in the DCM via $\mu$ is typically computed with no information about the logical state of many variables.
This is possible because the schemata redescription of logical functions introduces the wildcard and position-free symbols.

$S^0$ can be treated differently in the dynamical unfolding process. When studying \textit{pulse perturbations}, s-units $s \in S^0$ are assumed to fire only at $t=0$, unless the logic of the canalized dynamics ($\mu$) leads them to fire again. In contrast, \textit{pinning perturbations} lock the s-units $s \in S^0$ into firing at every time step $t$ 
\footnote{In control theory, pinning may refer to controlling a variable to any pattern of set states (e.g., an oscillation), but here we use the stricter sense of pinning to a single state.  This is equivalent to the node state overrides assumed by feedback vertex set theory and applied to biological networks \cite{fiedler2013dynamics,mochizuki2013dynamics,zanudo2017structure}.}. 
In this case, we always have $S^t \subset S^{t+1}$ in the canalized dynamics.
Unless otherwise noted, we use pinning perturbation, which represents situations where the perturbation signal is assumed to be in steady-state and last much longer than the transient dynamics that it affects. This is a realistic scenario, especially for gene regulatory network models.

A \textit{module} in the DCM is defined as $M^t \equiv (S^t, \theta^t)$.  It is a tuple consisting of a set of s-units $S^t \subset \mathcal{S}$ and t-units $\theta^t \subset \Theta$ that fire at time $t$ due to a perturbation applied to seed set $S^0$ at $t=0$.
For convenience we define set operations on modules by applying the operation to both elements of the tuple.  Thus, $M_i^t \subset  M_j^u$ if and only if $S_i^{t} \subset S_j^{u}$ and $\theta_i^{t} \subset \theta_j^{u}$, for modules $M_i$ and $M_j$ at time steps $t$ and $u$.  Other operators, such as union and intersection, are defined similarly.

Modules \textit{dynamically unfold} in time via the function $\mu (S^t) \rightarrow S^{t+1}$ until a time $t=T$ when the module either repeats ($M^T \equiv M^{t<T}$) or is empty ($M^T \equiv (\emptyset, \emptyset)$) because no s- or t-units can logically fire with the known information. 
This leads to a \textit{pathway module}, a time-ordered sequence of all modules that unfold  due to a perturbation given by seed set $S^0$ : $\textbf{M}(S^0) = [M^0, M^1,..., M^T]$ for all time steps $t \leq T$.

In turn, the \textit{pathway module set} includes all s-units that fire in at least one module $M^t \in \textbf{M}(S^0) $, and is defined as $S = S^{0 \rightarrow T} = \bigcup_{t\leq T} S^t$. 
Under pulse perturbation (or other perturbation types), not all elements of $S$ are guaranteed to fire at time $T$.
However, in the case of pinning perturbation, 
$S = S^T = S^\infty$, since $S^t \subset S^{t+1}$
\footnote{We assume that external signals take precedence over natural dynamics of the network. As such, a perturbation that causes an s-unit to fire will prevent any logically contradictory s-units from firing during the same time step.  For pinning perturbation, a variable may only fire in one state during the dynamical unfolding of a pathway module; that is, $x$-1 $\in S \implies x$-0 $\notin S$ and vice versa.  Thus, pinning perturbation cannot lead to limit cycles or other oscillations.}.
The \textit{length} of a pathway module $\textbf{M}(S^0)$ is $T$, whereas its \textit{size} is simply the cardinality of its corresponding pathway module set $| S |$.
Finally, $P_s$ denotes the set of all pathway modules $\textbf{M}(S^0)$ that unfold in a DCM from seed sets of size $s = |S^0|$.

\subsection{Interaction of pathway modules}

Given that there is an exponential number of pathway modules, for computational purposes we must narrow down the possible dynamical interactions of the system. Therefore, we define how pathway modules may interact to hone in on those that best define the computational dynamics of the network.
Specifically, we consider that
two pathway modules $\textbf{M}_i$ and $\textbf{M}_j$ can be \textit{combined} by considering the dynamical unfolding of the union of their seed sets, $\textbf{M}_{i,j} = \mu (S_i^0 \cup S_j^0)$
\footnote{Seed sets can, of course, be combined via set-theoretical operations at distinct time steps, but here we only consider unions of seed sets at time $t=0$.}.

Pathway module $\textbf{M}_i$ is \textit{subsumed} by $\textbf{M}_j$
if $S_i \subset S_j$.  
$\textbf{M}_i$ is \textit{partially subsumed} by $\textbf{M}_j$ if $S_i \cap S_j \neq \emptyset$ but $S_i \not\subset S_j$
(i.e., there is some overlap between the dynamical unfolding of the two modules). 
$\textbf{M}_i$ is \textit{temporally subsumed} by $\textbf{M}_j$
if there exists a $k \geq 0$ such that $M_i^t \subset M_j^{t+k}$ for all $t \leq T$ (the sequence is preserved temporally, where constant $k \geq 0$ indicates that the dynamical unfolding of the subsequence $\textbf{M}_i$ within $\textbf{M}_j$ begins $k$ time steps after $t=0$).

Pathway module $\textbf{M}_i$ is a \textit{submodule} of $\textbf{M}_j$ if $S_i^0 \subset S_j^0$ and $S_i^0 \neq \emptyset$.
Note that when $\textbf{M}_i$ is a submodule of $\textbf{M}_j$, it does not imply that $\textbf{M}_i$ is subsumed by $\textbf{M}_j$
due to the possibility of logical obstruction (see below).

Pathway module $\textbf{M}_i$ is considered \textit{maximal} if 
it cannot be temporally subsumed by any other pathway modules with the same seed set size. For pinning perturbation, $S_{i} \not\subset{S_{j}}$ for all pathway modules $\textbf{M}_j$ with seed set size $|S_j^0| \leq |S_i^0|$.  In other words, $\mu(S_{i}^0)$ results in a unique set of s-units that cannot be obtained from dynamically unfolding any other pathway module with the same seed set size. $\Lambda_s \subset P_s$ is the set of all maximal pathway modules with seed set size $s$.

\textit{Logical obstruction} exists between pathway modules $\textbf{M}_i$ and $\textbf{M}_j$ 
if their combination $\textbf{M}_{i,j}$ results in an s-unit not firing during time step $t=k$ when that s-unit does fire in either pathway module $M_i$ or $M_j$ at time $t=k$.  For pinning perturbation, this situation occurs when
$\mu(S_i^0) \cup{\mu(S_j^0)} - \mu(S_i^0 \cup S_j^0) \neq \emptyset$. 
In cases where there is also no synergy between $\textbf{M}_i$ and $\textbf{M}_j$ (see below), $\mu(S_i^0 \cup S_j^0) \subset  \mu(S_i^0) \cup{\mu(S_j^0)}$.

Logical obstruction can only occur when the union of the seed sets results in a logical contradiction whereby s-units for the state of at least one node variable and its negation are both included (in the seed sets or their dynamical unfolding logic), for example, $x$-1 $\in S_i^0 \land x$-0 $\in S_j^0$.
This means that combining the pathway modules results in one or more s-units not firing that would have fired had the pathway modules been separate.

\textit{Synergy} exists between pathway modules $\textbf{M}_i$ and $\textbf{M}_j$ if their combination $\textbf{M}_{i,j}$ results in an s-unit (or t-unit) firing during time step $t=k$ when that s-unit (or t-unit) does not fire in either pathway module $\textbf{M}_i$ or $\textbf{M}_j$ at time $t=k$.  For pinning perturbation we have:
$\mu(S_i^0 \cup S_j^0) - \mu(S_i^0) \cup \mu(S_j^0) \neq \emptyset$. 
If there is no logical obstruction (see above) between $\textbf{M}_i$ and $\textbf{M}_j$, then:
$\mu(S_i^0 \cup S_j^0) \supset  \mu(S_i^0) \cup{\mu(S_j^0)}$.
This means that additional s-units fire as a result of combining the pathway modules.

Pathway modules $\textbf{M}_i$ and $\textbf{M}_j$ are \textit{decoupled} if there is no partial subsumption (s-unit overlap), logical obstruction, or synergy between them.  This means that $\mu(S_i^0 \cup S_j^0) \equiv \mu(S_i^0) \cup{\mu(S_j^0)}$ and $|\mu(S_i^0 \cup S_j^0) | = |\mu(S_i^0)| + |\mu(S_j^0)| $ are necessary and sufficient conditions for pinning perturbation.

Pathway module $\textbf{M}_i$ is a \textit{complex module} if 
$\textbf{M}_i \in \Lambda_{s}$ and 
any submodule $\textbf{M}_j$ of $\textbf{M}_i$ is synergistic with the submodule $\textbf{M}_k$ whose seed set is $S_k^0 = S_i^0 - S_j^0$.
For $s=1$, this means that all modules $\textbf{M}_i \in \Lambda_1$ are complex.
We define $I_s \subset P_s$ as the set of all complex modules with seed size $s$.
	
To be a complex module then, a pathway module must be maximal and every submodule (seed combination) within it must add synergy.  For a given $s$, each complex module $\textbf{M}_i \in I_s$ represents the introduction of a unique firing sequence that is not part of any other pathway module with seed set size $x \leq s$ (otherwise it would not be maximal) or any combination of pathway modules with seed set sizes $x \leq s$ (because every submodule adds synergy).  As a consequence, complex modules are the building blocks of a network's dynamics.  
Importantly, complex modules allow the vast 
number of possible dynamical interactions to be reduced to a smaller set of
unique signaling sequences that may represent (full or partial) biologically functional pathways.

Given a DCM, its complex modules are defined using only parameter $s$, the size of the seed set. Moreover, the complex modules are hierarchical in the sense that modules with larger seed set sizes are composed of (partially or fully subsumed) modules with smaller seed set sizes.  Higher-level complex modules thus capture the emergent behavior (e.g., synergy) that results from combining lower-level modules. For $s=1$, the set of complex modules $I_1 = \Lambda_1$ is the minimal set of pathway modules that covers the DCM.  However, synergy allows for the DCM to be covered using fewer pathway modules (with larger seed sets).

\begin{figure}
\includegraphics[width=20cm,height=20cm,keepaspectratio]
{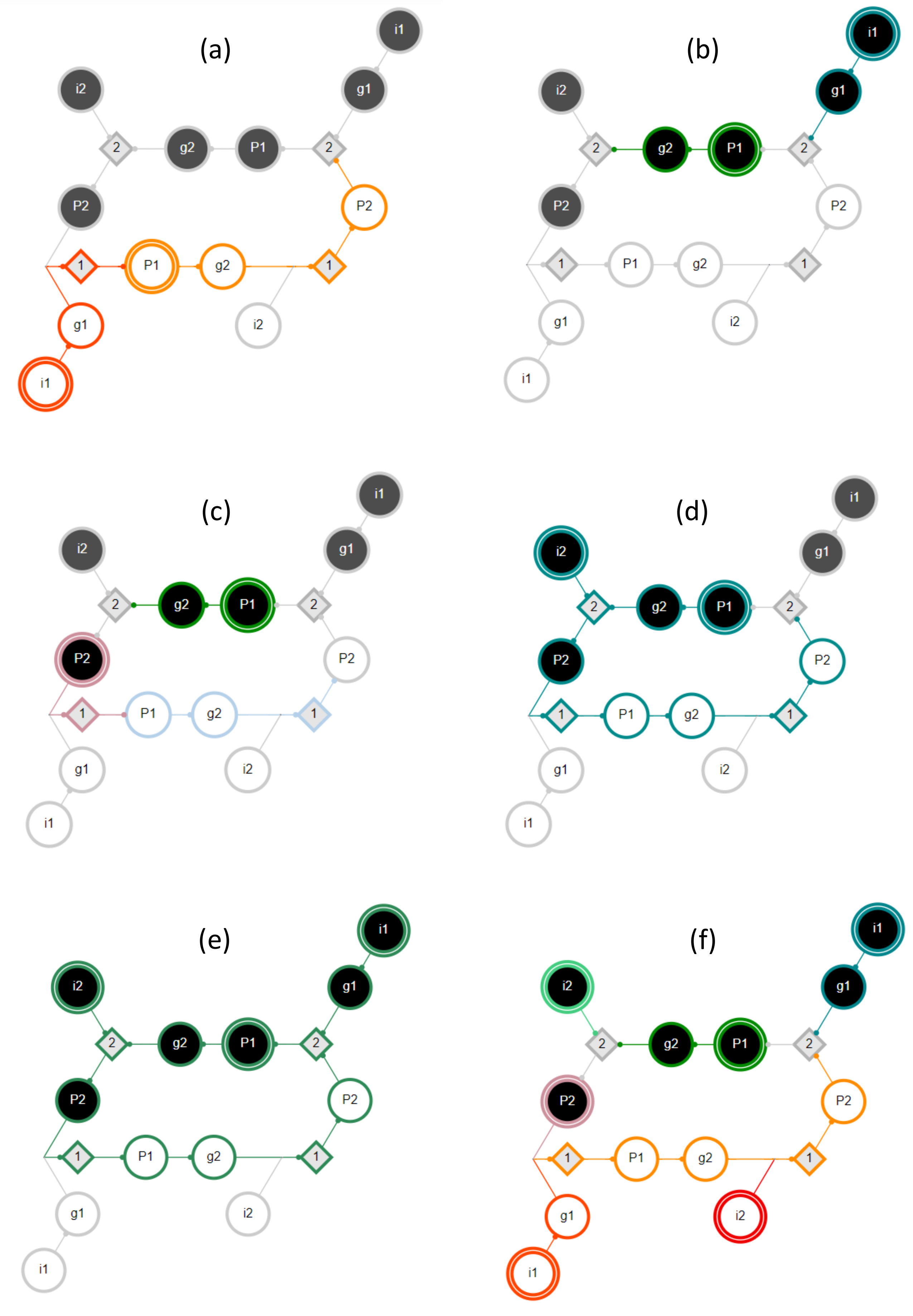}
\caption{\textbf{Example GRN interactions.}  The DCM shown here is the same as in Figure \ref{fig:example_network}.  
 Individual pathway modules are differentiated by color; the seed s-unit has a double edge.
(a) $\textbf{M}_{P1-0}$ is subsumed by module $\textbf{M}_{i1-0}$.
(b) $\textbf{M}_{P1-1}$ and $\textbf{M}_{i1-1}$ are decoupled.
(c) There is logical obstruction between $\textbf{M}_{P1-1}$ and $\textbf{M}_{P2-1}$.
(d) There is synergy between $\textbf{M}_{P1-1}$ and $\textbf{M}_{i2-1}$.  
(e) A limit cycle of the network due to $\textbf{M}_{P1-1}$, $\textbf{M}_{i1-1}$, and $\textbf{M}_{i2-1}$.  The inputs are held constant but the node P1 is not (pulse perturbation).
(f) All six complex modules with seed set size $s=1$.  The nodes shown in yellow are subsumed by multiple other modules.
}
\label{fig:example_interactions}
\end{figure}

Possible interactions between pathway modules are demonstrated in Fig. \ref{fig:example_interactions}.  In this example network, module $\textbf{M}_{P1-0}$ is subsumed by module $\textbf{M}_{i1-0}$ with either pinning or pulse perturbation (panel a). Thus, $\textbf{M}_{i1-0} \in \Lambda_1$ is a maximal module, whereas $\textbf{M}_{P1-0}$ is not.
Modules $\textbf{M}_{P1-1}$ and $\textbf{M}_{i1-1}$ are decoupled (dynamically independent) as they share no overlap or synergy, regardless of perturbation type (panel b).
Module $\textbf{M}_{P1-1}$ logically obstructs module $\textbf{M}_{P2-1}$ with pinning perturbation as the pinning of P1 prevents the firing of P1-0 and the subsequent unfolding of $\textbf{M}_{P2-1}$ (panel c).
By contrast, there is synergy between the modules $\textbf{M}_{P1-1}$ and $\textbf{M}_{i2-1}$ that results in the firing of P2-1 (panel d).  
If the input i2 is constant, but the node P1 is not (pulse perturbation), several additional downstream s-units will be reached.
Panel e shows a limit cycle of the network reached by the combination of modules $\textbf{M}_{P1-1}$, $\textbf{M}_{i1-1}$, and $\textbf{M}_{i2-1}$.  In this case, the inputs i1 and i2 are held constant but the node P1 is not (pulse perturbation).
Finally, all six complex modules of seed set size $s=1$ are shown, assuming pinning perturbation (panel f).

\subsection{Dynamical Modularity}

With the characterization of pathway module interaction and the introduction of complex modules, we may now characterize the macro-scale properties of the network, such as its organization and modularity.
The concept of dynamical modularity is analogous to structural modularity in graphs, whereby it is considered to occur when interactions within modules are strong and interactions between modules are weak. Dynamical modularity is, however, a more general concept derived from Simon's framing of complex systems in terms of near-decomposability \cite{simon1962architecture}, which includes both structural and dynamical interactions \cite{kolchinsky2015modularity,kolchinsky2011prediction,gates2016control}. We use it to find subsets of variables and their dynamical states that are more or less  easy to decouple in a given multivariate dynamical system.
Thus, we can compare complex (dynamical) networks (or systems) to determine which ones are easier to decouple into independently-functioning parts.

The \textit{cover} of a DCM is a set of pathway modules $\Pi = \{\textbf{M}_i, \textbf{M}_j, ...\}$ such that the union of their corresponding pathway modules sets is equivalent to the set of all s-units in the DCM; that is, $S_{\Pi} = \bigcup_i{S_i} = \mathcal{S}$ for set $S_i$ 
with corresponding pathway module $\textbf{M}_i \in \Pi$.  Because pathway module sets may overlap, this results in a fuzzy partition of s-units.  $|\Pi|$ is the size of the cover in terms of the number of pathway modules that comprise it.
In practice, we are interested in finding a cover $\Pi_s$ that is composed of modules with no more than $s$ seeds; additionally, we focus on small $s$ because modules with smaller seed sets are generally easier to control.

The \textit{independence} of a pathway module $\textbf{M}_i$ from a set of pathway modules $\Sigma$ is the fraction of unique s-units that fire within the pathway module:
\begin{equation} \label{eq:1}
ind(\textbf{M}_i,\Sigma) = |S_i - S_{\Sigma}|/|S_i| 
\end{equation}
where $S_{\Sigma} = \bigcup_j{S_j}$ for set $S_j$ 
with corresponding pathway module $\textbf{M}_j \in \Sigma$.  If $\{\textbf{M}_i\} \cup \Sigma = \Pi$, then $S_i \cup S_{\Sigma} = S_{\Pi}$ and $ind(\textbf{M}_i,\Sigma)$ gives us the independence of pathway module $\textbf{M}_i$ within a cover $\Pi$.

The \textit{dynamical modularity} of a cover $\Pi$ is the distribution of independence scores for all its pathway  modules: $D(\Pi) \equiv \{ind(\textbf{M}_i, \Pi-\textbf{M}_i)\},\; \forall_{\textbf{M}_i \in \Pi}$.
One can derive several meaningful statistics from this distribution, but here we use its mean value as a characterization of dynamical modularity: 
\begin{equation} \label{eq:2}
\overline{D}(\Pi) = (\sum_{\textbf{M}_i \in \Pi}{ind(\textbf{M}_i, \Pi-\textbf{M}_i)}) / |\Pi| 
\end{equation}

As there are many possible covers of the DCM, it is useful to find one that is optimal for measuring modularity.
One can define optimality of a cover in different ways, for example, the cover with the minimal size, $\Pi_{min}$. 
However, as the goal is to find modules that are decoupable from one another, here we define optimality by maximizing the mean dynamical modularity $\overline{D}(\Pi)$.
We define $\Pi^*_s$ as the \textit{optimal cover} of the DCM, where $\Pi^*_s$ 
has the maximum mean dynamical modularity among covers that satisfy the property that $\textbf{M}_i \in I$ for all pathway modules $\textbf{M}_i \in \Pi_s$ (i.e., the cover is composed only of complex modules with seed set size $|S^0| \leq s$).  The corresponding mean dynamical modularity of this cover indicates how decouplable a network is into its constituent dynamical building blocks.

We note that the optimal cover may be difficult to calculate exactly due to the exponential number of possible module combinations. When networks and seed set sizes are too large to find an exact solution, we use a \textit{greedy selection} to estimate $\Pi_s^*$.  The algorithm starts with an empty set $\Sigma=\{\}$ and then iteratively adds the complex module $\textbf{M}_i \in I_s$ that has the highest independence score from $\Sigma$, $max(ind(\textbf{M}_i, \Sigma))$, until a cover is reached, such that $S_\Sigma = \mathcal{S}$.  We use this cover $\Sigma$ to estimate the optimal cover and its mean dynamical modularity $\overline{D}(\Sigma)$ to estimate the optimal mean dynamical modularity of the network $\overline{D}(\Pi_s^*)$.
This quasi-optimal solution provides a lower-bound on $\overline{D}(\Pi_s^*)$.
In practice, multiple covers may have the same mean dynamical modularity, meaning that there may be multiple optimal solutions.

The modules that compose a cover $\Pi_s$ have a distribution of seed set sizes that are bounded by $s$. By definition, the mean dynamical modularity of $\Pi^*_s$ should increase or remain the same as $s$ increases, because each increase in maximum seed set size allows for more modules to choose from when finding an optimal cover.  The \textit{characteristic seed number} $s*$ occurs when increasing $s$ no longer increases the mean dynamical modularity of $\Pi^*_s$.  This represents the minimal seed set size that allows for a network to optimally be dynamically decoupled.  

In Fig. \ref{fig:example_interactions} panel f, the six complex modules shown comprise $\Lambda_1$ and the associated optimal cover $\Pi^*_1$; these are $\textbf{M}_{i1-1}$, $\textbf{M}_{P1-1}$, $\textbf{M}_{i2-1}$, $\textbf{M}_{P2-1}$, $\textbf{M}_{i1-0}$, and $\textbf{M}_{i2-0}$.
The mean dynamical modularity of this cover is $\overline{D}(\Pi_1^*)=0.71$. 
Additionally, there are two complex modules (assuming pinning perturbation) with $s=2$ ($\textbf{M}_{P1-1,i2-1}$ and $\textbf{M}_{i1-1,i2-0}$).  With pinning perturbation, this network can be covered by three modules: $\Pi_2=$ \{$\textbf{M}_{i1-1,i2-0}$, $\textbf{M}_{P1-1,i2-1}$, $\textbf{M}_{i1-0}$\}, with mean dynamical modularity $d=0.6$.
Note that this minimal cover is not optimal because $\overline{D}(\Pi_1^*)$ has a higher mean dynamical modularity.
Rather, the optimal cover at $s=2$ is $\Pi^*_2=$ \{$\textbf{M}_{i1-1}, \textbf{M}_{i2-0}$, $\textbf{M}_{P1-1,i2-1}$, $\textbf{M}_{i1-0}$\}; $\overline{D}(\Pi_2^*) = 0.83$.
The dynamical modularity does not increase with greater $s$ so the characteristic seed number of this network is $s*=2$.

In summary, dynamical modularity allows for the comparison of dynamical organization between networks.  Dynamically simple networks will be easier to decouple and will tend to have higher mean dynamical modularity and lower characteristic seed number.  More complex networks will have more overlap dynamically and will tend to have lower mean dynamical modularity and higher characteristic seed number.

We note that, because our dynamical cover partitions \textit{s-units} in the DCM, the same node will belong to different modules depending on what state it is in. This is a more granular but lengthier description of the dynamics of the system than considering the nodes themselves.

Computationally, our optimal measure of dynamical modularity is NP-hard as it entails the set cover optimization problem \cite{karp1972reducibility,korte2012combinatorial} (by comparing all possible sets of complex modules that cover the set $\mathcal{S}$).
However, the number of modules to consider is greatly reduced by considering only complex modules rather than all pathway modules. For all real-world networks that we observed, $I_s << P_s$ because many pathway modules are either subsumed into longer modules or they have non-synergistic seeds.  Thus, by only considering complex modules, we can greatly reduce the computational complexity of our dynamical modularity optimization problem.


\section{Experimental Results}

We demonstrate our approach by using the above formalism to analyze the dynamical modularity of experimentally-validated models of biochemical regulation from systems biology.
We focus on the single-cell and full parasegment \textit{drosophila} SPN models \cite{albert2003topology} and compare the analysis to that of additional models such as the yeast cell-cycle \cite{li2004yeast}, \textit{Arabidopsis thaliana} floral development \cite{chaos2006genes}, and T-LGL leukemia \cite{zhang2008network} networks.

Two versions of the SPN exist to model how a pathway of segment polarity genes (and their protein products) regulate segmentation of the body of the fruit fly in its development.
The \textbf{single-cell} model is a regulatory network of $n=17$ interacting genes and proteins represented as Boolean variables \cite{willadsen2007robustness,marques2013canalization}.
It has 3 input nodes that are not regulated by other nodes and are usually assumed to be pinned ON or OFF: $SLP$ (Sloppy-Paired proteins), $nWG$ (neighboring Wingless protein), and $nHH$ (neighboring Hedgehog protein).
The remaining 14 (internal) nodes are regulated by other nodes in the network.

A larger \textbf{parasegment} network of $n=60$ gene and protein variables is built by connecting the single-cell model in a linear lattice of 4 cells with periodic boundary conditions.
This network models intra- and inter-cellular regulatory pathways where each of the 4 cells has the same 14 internal nodes as the single-cell model, plus one $SLP$ input per cell (which represents the output of upstream developmental pathways and is usually assumed to be pinned ON or OFF).
Furthermore, each of the 4 cells in the parasegment 
regulates its two neighboring cells via its internal $WG$ and $HH$ nodes, which act as external, inter-cellular inputs to the two neighboring cells. Thus, in the parasegment SPN there are no input $nWG$ or $nHH$ nodes for each cell, which results in a network of $n=60$ variables. 
This model has been extensively studied \cite{von2000segment, albert2003topology, ingolia2004topology, chaves2005robustness, chaves2006methods,chaves2008studying,marques2013canalization,gates2016control} and its attractors are fully known for the single-cell case, which makes it a useful example to test our modularity methodology.

We compute all complex modules present in the dynamics of the single-cell and full parasegment SPN models for seed set size $s \leq 6$.
This analysis allows us to calculate the dynamical modularity of the SPN and compare it to the other systems biology models we analyzed.
In the present work we only consider the dynamics that ensues from pinning perturbation of the seed set. This form of perturbation is feasible for experimental work with the target biological models (e.g., gene silencing in \textit{drosophila} \cite{Correia2022.03.02.482557}), but our approach also allows for other forms of perturbation (e.g., pulse perturbations) that can be explored in the future.

\subsection{Complex modules in the \textit{Drosophila} Single-Cell SPN}

Analysis of the \textit{drosophila} single-cell SPN highlights many useful features of our dynamical modularity methodology.  It allows us to reduce the enormous complexity of this dynamical system's state space to only a few building blocks, characterize which dynamics are involved in transient configurations and attractors, understand the dynamical overlap between similar modules, and study the specific effects of individual nodes and sets of nodes (e.g., comparing network inputs to internal nodes). 

To understand the effect of perturbing a specific node in the single-cell SPN with a brute-force approach would require analysis of $2^{17}$ configurations. Additionally, there are $2^s$~$17 \choose s$ different ways to perturb $s$ seeds, where each node $x$ in the seed set of size $s$ may be in the state $x$-0 or $x$-1.  However, our dynamical modularity methodology offers a much more direct approach without enumerating all network configurations and by reducing the total possible number of seed sets to only those that are complex modules.

For seed sets of size $s=1$, there are 34 possible pathway modules to consider (because there are 34 possible seeds; each node $x$ of the $n=17$ nodes may be in the state $x$-0 or $x$-1), but only 14 complex modules.  This indicates that the other 20 modules do not reveal any dynamical information that is not contained in the 14 complex modules.  These complex modules cover the DCM, meaning that they include all 34 node states.  For seed sets of size $s=2$, there are 9 complex modules out of 544 pathway modules; for size $s=3$, there are 19 complex modules out of 5440 pathway modules; for size $s=4$, there are 14 complex modules out of 38080 pathway modules.  
For seed set sizes $s>4$ there are no additional complex modules. Any higher-order interaction between variables (i.e., modules with larger seed sets) can be understood as a combination of lower-order interactions that may include overlap or contradictions between the respective modules but no synergy.  In other words, no novel synergy emerges in the network by pinning more than 4 seeds. Thus, all of the macro-scale computation that the dynamical system is doing involves building blocks that are controllable by a small number of nodes.

This results in a total of 56 complex modules.  However, this number can be further reduced using the \textit{maximal seed heuristic} where modules are discarded if they contain seeds that do not initiate maximal pathway modules.  That is, we consider \textit{core} complex modules to be those where each seed's pathway module is itself maximal, $\textbf{M}_x \in \Lambda_1 \,, \forall x \in S^0$.
This reduction removes modules that would have been subsumed if not for contradiction.
For example, both $\textbf{M}_{ptc-1,nHH-0}$ and $\textbf{M}_{PTC-1,nHH-0}$ are complex modules in the single-cell SPN; however, the only reason that $\textbf{M}_{PTC-1,nHH-0}$ is not subsumed by $\textbf{M}_{ptc-1,nHH-0}$ is because the former includes the state $ptc$-0, which contradicts the seed set of the latter.
Moreover, $\textbf{M}_{PTC-1,nHH-0}$ is not a core complex module because its submodule $\textbf{M}_{PTC-1}$ is subsumed by $\textbf{M}_{ptc-1}$.  
Using the maximal seed heuristic, there are 28 core complex modules for the single-cell SPN with maximal seed set size $s=3$.  These 28 are sufficient to describe all attractors in the network (with a slight caveat mentioned below) and are listed in the supplementary materials.

\begin{figure}
\begin{adjustwidth}{-3cm}{}
\centering
\includegraphics[width=1.5\columnwidth]{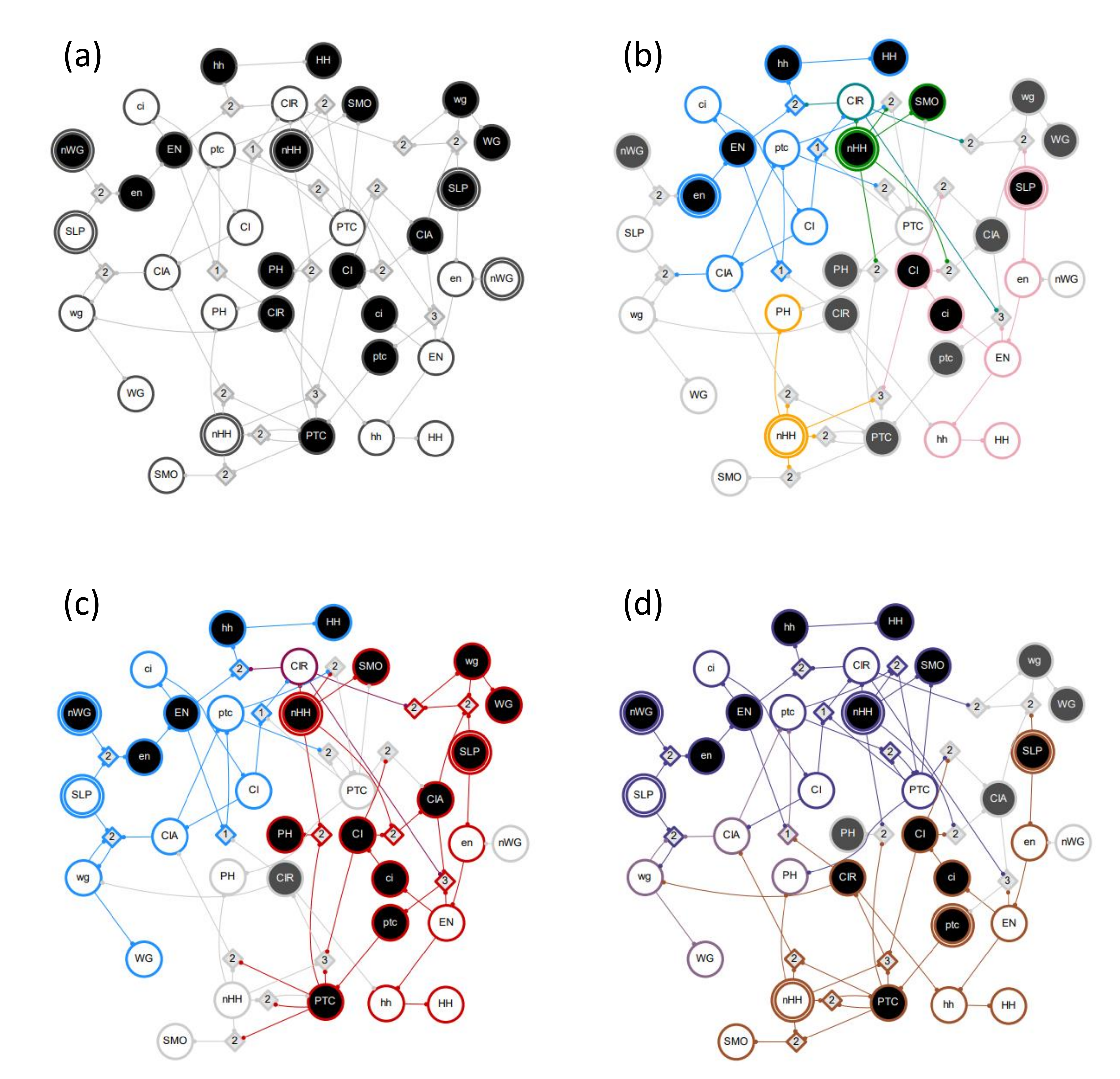}
\caption{\textbf{Complex modules in the \textit{drosophila} single-cell SPN.}
(a) The DCM of the \textit{drosophila} single-cell SPN (input nodes have a double edge).  Black represents variables in their active (ON) state, white represents variables in their inactive (OFF) state, and grey diamonds represent t-units.  The DCM consists of 34 s-units.
Note that several t-units have been removed or modified in the figure for clarity of the transition function.
(b) Select complex modules (indicated by color) are shown for seed set size $|S_i^0|=1$.
These modules dynamically unfold from their respective seed sets 
(highlighted with a double edge) and include the s and t-units that are guaranteed to fire given pinning perturbation.  
Modules $\textbf{M}_{en-1}$ (blue),  $\textbf{M}_{nHH-1}$ (green),  $\textbf{M}_{SLP-1}$ (pink), and  $\textbf{M}_{nHH-0}$ (gold) are shown.  The s-unit CIR-0 is in both $\textbf{M}_{en-1}$ and $\textbf{M}_{nHH-1}$.  Together these four modules cover 59\% of the s-units in the DCM.
(c) The same as b, but for $|S_i^0|=2$.
Modules $\textbf{M}_{nWG-1, SLP-0}$ (blue) and $\textbf{M}_{nHH-1, SLP-1}$ (red) are shown.  The s-unit CIR-0 is in both modules.  Together both modules cover 82\% of the s-units in the DCM.
(d) The same as b, but for $|S_i^0|=3$. Modules $\textbf{M}_{nHH-1, nWG-1, SLP-0}$ (violet) and $\textbf{M}_{nHH-0, ptc-1, SLP-1}$ (brown) are shown.  The s-units CIA-0, PH-0, wg-0 and WG-0 are in both modules.  Together both modules cover 85\% of the s-units in the DCM.
Additional modules are shown in supplementary materials.}
\label{fig:drosophila_modules_paper}
\end{adjustwidth}
\end{figure}

Only one complex module directly results in an attractor, meaning that the states of all $n=17$ variables are resolved: $|S_{(nWG-1, SLP-0, nHH-1)}|=17$, as shown in Fig. \ref{fig:drosophila_modules_paper}d.  The other 9 attractors of the single-cell SPN can be recovered by combining two or three complex modules (see the supplementary materials for a full description of each attractor).  Certain modules appear in multiple attractors.  For example, $|S_{(SLP-1, nHH-1)}|=16$ (shown in Fig. \ref{fig:drosophila_modules_paper}c) resolves every node except the input $nWG$.  The module $|S_{(nWG-0, nHH-1)}|=14$ is identical in its unfolding except that it does not resolve $wg$ or $WG$; therefore, this module must be combined with $\textbf{M}_{wg-0}$ or $\textbf{M}_{wg-1}$ and $\textbf{M}_{SLP-0}$ or $\textbf{M}_{SLP-1}$ to reach an attractor.  The module $|S_{(nWG-1, SLP-0)}|=13$ (shown in Fig. \ref{fig:drosophila_modules_paper}c) resolves most nodes and is in the basin of two attractors when $nHH$ is OFF (when $nHH$ is ON, it is subsumed by $\textbf{M}_{(nWG-1, SLP-0, nHH-1)}$ mentioned above).  Finally, $|S_{(ptc-1, SLP-1, nHH-0)}|=16$ (shown in Fig. \ref{fig:drosophila_modules_paper}d) and the similar module $|S_{(ptc-1, nWG-0, nHH-0)}|=16$ reach one of three attractors, depending on the state of the input nodes $SLP$ and $nWG$ \footnote{
Here we see the caveat mentioned above: these three attractors are not quite reached if $ptc$ is left pinned.  The modules $\textbf{M}_{(ptc-1, SLP-1, nHH-0)}$ and $\textbf{M}_{(ptc-1, nWG-0, nHH-0)}$ pin $ptc$ in the ON state; however, if $ptc$ is unpinned, these modules will turn $ptc$ OFF (as seen in the respective attractors).  This behavior is captured by the complex modules $\textbf{M}_{(PTC-1, SLP-1, nHH-0)}$ and $\textbf{M}_{(PTC-1, nWG-0, nHH-0)}$, which do result in $ptc$-$0$; however, these are not core complex modules because $\textbf{M}_{ptc-1}$ subsumes $\textbf{M}_{PTC-1}$.
}.

Even modules that do not result in a steady-state of the network give us important information about partial control of the network state.  The size of the module indicates how many other nodes can be controlled.  For example, if a researcher can only perturb one node in the single-cell SPN, then turning $en$ ON has the most effect because $|S_{en-1}|=9$ is the longest module at $s=1$ (note that this one seed alone controls half of the other network nodes).  If perturbing two nodes, then $|S_{(SLP-1, nHH-1)}|=16$ gives the most control.  In addition to quantifying how influential a node $x$ is, our methodology also identifies which other node states are reached if $x$ is perturbed. This is useful information when a researcher is interested in reaching certain states in the network.  For example, for the T-LGL leukemia network, modules of interest may include those that result in $Apoptosis$-$1$ because perturbation of the respective seed set guarantees that the cell death state is reached.

Our methodology also indicates how synergy occurs between seeds and the time step at which it occurs.  For example, $|S_{SLP-1}|=7$ and $|S_{PTC-1}|=4$; however, combining the modules results in novel synergy, $|S_{(SLP-1, PTC-1)}|=15$ (and thus resolves every node in the network except the two additional inputs $nWG$ and $nHH$),
with synergy between the two modules occurring at time steps $t=5$ and $t=6$.
Knowing specific states of specific nodes allows for additional inferences when making further perturbations.  For example, $\textbf{M}_{(SLP-1, nHH-1)}$ results in $ptc-1$, so if the state of $nHH$ is flipped, then $\textbf{M}_{(ptc-1, SLP-1, nHH-0)}$ will unfold, resulting in a different attractor (see supplementary materials for a description of the modules).

Complex modules also allow us to see how dynamics overlap between different modules.  In some cases, this offers alternative control strategies.  For example, the unfolding of $\textbf{M}_{en-1}$ is identical to the unfolding of $\textbf{M}_{(SLP-0, nWG-1)}$, except that in the former the states of $wg$ and $WG$ remain unknown.  Perhaps less obvious is that $PTC$-$0$ behaves similarly to $nHH$-$1$: they both fire $SMO$-$1$ and $CIR$-$0$ and contribute to firing $CIA$-$1$. This shared behavior means that $S_{(SLP-1, nHH-1)}$ and $S_{(SLP-1, PTC-0)}$ are identical except for the seed nodes and the state of the $PH$ node.

In other cases, the lack of overlap tells us that two modules are independent.  For example, $\textbf{M}_{en-1}$ and $\textbf{M}_{SLP-1}$ have no overlap and all of their downstream products are in opposing states to one another. This indicates that only one of these modules will ever be active at a time in the dynamical system.  Because of this independence, these modules and modules of greater seed set size that subsume them appear in the optimal covers of the DCM for seed set sizes $s=[1,4]$.

Finally, these complex modules help to elucidate the role of specific nodes, such as the inputs, in the single-cell SPN dynamics.  It is already known that the input nodes control much of the network dynamics \cite{marques2013canalization,gates2016control}; however, our analysis shows when the inputs themselves may result in an attractor and when an additional node such as $wg$ or $PTC$ is needed \footnote{Our analysis matches the results in \cite{parmer2022influence}. We similarly overestimate the size of the minimal control sets for three attractors because additional dynamics of non-seed nodes are not accounted for. 
}.  Additionally, we see the effects that internal nodes, such as $PTC$ or $CIR$, have by themselves.  This is especially useful in analyzing the parasegment SPN ($n=60$) because all nodes in this network are internal except for the four SLP inputs.

For help in analyzing the parasegment SPN, we group modules that have nearly identical (possibly differing in the inclusion of $wg$ or $WG$) downstream products together: let $\Sigma_1 = \{\textbf{M}_{nWG-0},$ $  \textbf{M}_{SLP-1}$\}, $\Sigma_2 = \{\textbf{M}_{\textit{en-1}}, \textbf{M}_{nWG-1, SLP-0}\}$, $\Sigma_3 = \{\textbf{M}_{nWG-0, nHH-1}, \textbf{M}_{SLP-1, nHH-1}\}$, $\Sigma_4 = \{\textbf{M}_{\textit{en-1}, nHH-1}, \textbf{M}_{nWG-1, SLP-0, nHH-1}\}$, and $\Sigma_5 = \{\textbf{M}_{nWG-0, PTC-0}$, $\textbf{M}_{SLP-1, PTC-0}\}$.
As discussed below, these groups are seen often in the parasegment's dynamics.

Even though the complex modules discussed here are based on pinning perturbation, many of the results are generalizable to any type of perturbation because the input node states are self-sustaining and, therefore, all of their downstream products are guaranteed to fire at some future point.  Other modules may be self-sustaining as well. For example, $|S_{(PTC-1, nHH-0)}|=6$ maintains the state of its inputs (and, therefore, its downstream products) due to a positive feedback loop between $PTC$-$1$ and $nHH$-$0$.

\subsection{Complex modules in the \textit{Drosophila} Parasegment SPN}

We use our dynamical modularity methodology to aid the analysis of the \textit{drosophila} parasegment SPN.  As in the single-cell case, this allows for a great reduction in the complexity of the dynamical system model and enables us to characterize dynamical trajectories within the system. 
Using complex modules, we are able to find a minimal seed set that reaches the wildtype attractor of drosophila that is lower than previous estimates seen in the literature. Furthermore, we are able to characterize the transient dynamics toward this attractor in terms of the intracellular modules we found in the single-cell SPN. 

It is computationally restrictive to discover all complex modules for the \textit{drosophila} parasegment model, but as in the single-cell case, the number of complex modules found for low seed set sizes is substantially less than the number of possible pinned perturbations.  We again use the maximal seed heuristic to speed up computation time.  For $s=1$, we find 40 complex modules out of 120 possible pathway modules; for $s=2$, we find 40 core complex modules out of 7080 pathway modules; for $s=3$, we find 152 core complex modules out of 273760; for $s=6$, we find 1475 out of more than 3.2 billion.  Despite this great reduction, there are still enough core complex modules to make analysis difficult; therefore, we focus only on the modules with the greatest size for each $s$ value.  We are aided also by the symmetry in the model (the index of each of the four cells does not matter because each one is identical to the next) and the fact that the intracellular dynamics of each cell in the parasegment is the exact same as in the single-cell case, except that they are influenced by inputs from neighboring cells ($WG$ and $HH$) as well as the external input $SLP$.

\begin{figure}
\centering
\includegraphics[width=1.1\columnwidth]{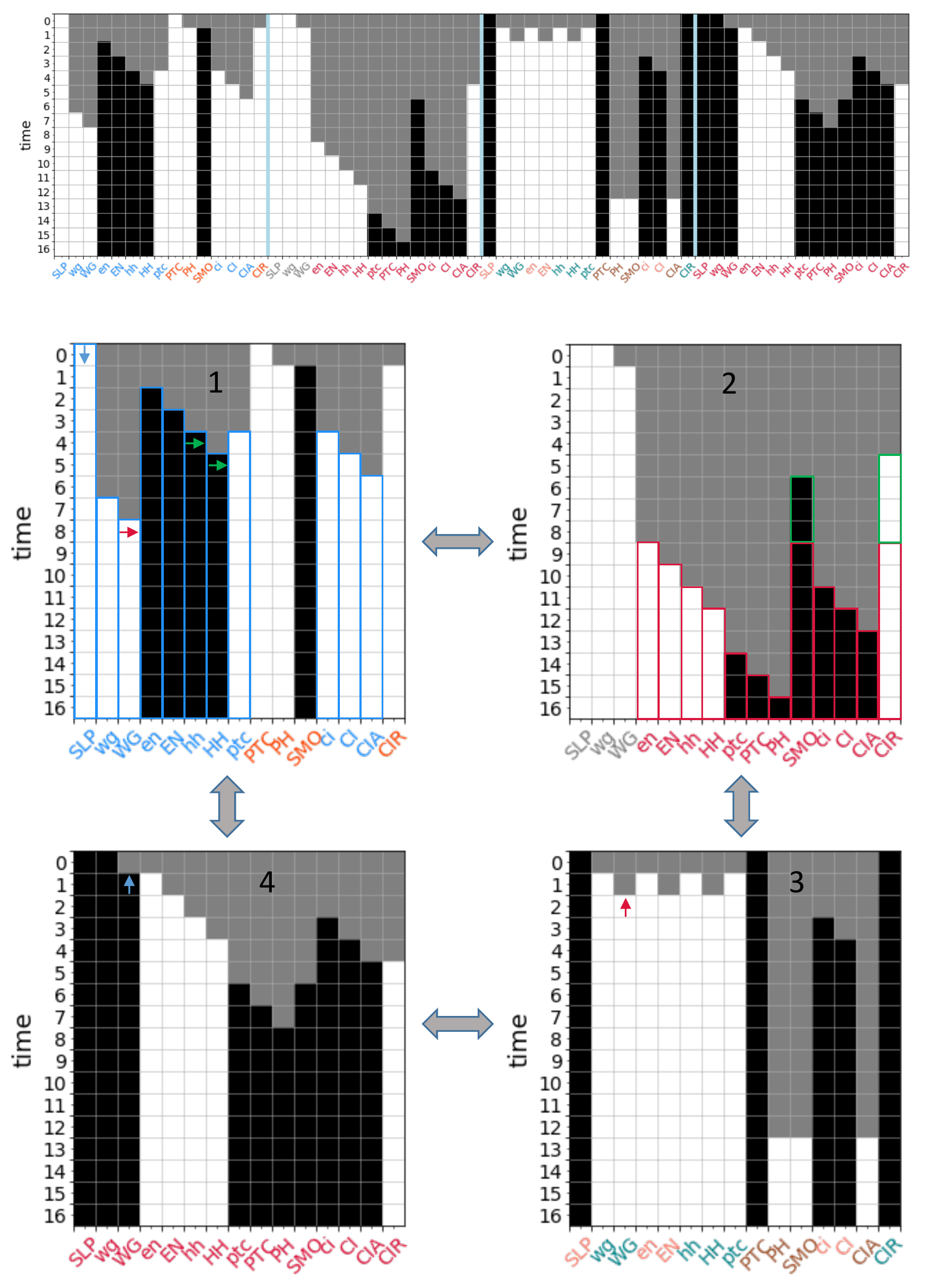}
\caption{\textbf{Dynamical unfolding of the wildtype attractor in the \textit{drosophila} parasegment.}  The minimal steady-state perturbation that we found sufficient to fully resolve the wildtype attractor is shown.  Cell boundaries are separated by a blue line; white indicates that a node is OFF in that time step, black indicates that a node is ON, and grey indicates that the node state is unknown.  The color of the node label indicates an associated complex module.  
Select modules are also highlighted in the dynamical unfolding process by colored borders. Arrows indicate the initiation of a new module; arrows pointing to the side or upward indicate that the respective s-unit is initiating a module in a neighboring cell.
For example, $\textbf{M}_{SLP-0}$ is present in cell 1 based on initial conditions (blue arrow).  The influence of WG-1 in cell 4 (blue arrow) initiates $\textbf{M}_{SLP-0, nWG-1}$ in cell 1 (blue borders). This module turns ON \textit{hh} and $HH$, which initiates $\textbf{M}_{nHH-1}$ in cell 2 (green borders) and then turns OFF $WG$ in cell 1 which, together with $WG$-0 in cell 3 (red arrows), initiates $\textbf{M}_{nWG-0}$ in cell 2; the synergy between $\textbf{M}_{nHH-1}$ and $\textbf{M}_{nWG-0}$ results in $\textbf{M}_{nHH-1, nWG-0}$ in cell 2 (which subsumes both individual modules, shown by red borders).
This unfolding demonstrates how developmental dynamics interact across cells in the parasegment. 
See supplementary materials for a full description of the unfolding of the wildtype attractor and the intracellular complex modules involved.}
\label{fig:wildtype_unfolding}
\end{figure}

As in the single-cell case, the size of complex modules gives us information about partial control in the network.  For $s=1$, the largest module is similarly $|S_{en-1}|=13$ in any of the four cells. For $s=2$, we see behavior not seen in the single-cell case: the largest modules are $|S_{(en-1_i,en-1_{i \pm 1})}|=28$ and $|S_{(CIR-1_i,CIR-1_{i \pm 2})}|=28$, where $i$ denotes the index of the cell.  For $s=3$, the largest modules are $|S_{(SLP-1_i,SLP-1_{i \pm 2},wg-1_{i \pm 1})}|=52$.  We already see that three seeds alone can resolve $87\%$ of the states of the $n=60$ nodes in the network.  For $s=4$, the largest complex modules resolve 56 node states; for $s=5$, the largest complex modules resolve 59 node states; for $s=6$, the largest complex modules resolve all 60 node states, which indicates a network attractor (see supplementary materials).  By sampling larger pathway modules, we find that 9 nodes are sufficient to reach the wildtype attractor of the parasegment SPN, which is lower than previous estimates seen in the literature \cite{marques2013canalization,zanudo2017structure,parmer2022influence} (see Fig. \ref{fig:wildtype_unfolding}).

The discovery of the complex modules allows us to describe the transient dynamics that occur in the parasegment model toward the wildtype attractor (see Fig. \ref{fig:wildtype_unfolding} and supplementary materials) and other attractors in the network.  We find that inter-cellular interactions can be described by the intracellular complex modules found for the single-cell SPN, where the unfolding of one intracellular module $\textbf{M}_i$ may initiate another module $\textbf{M}_j$ by causing the seed of that module to fire ($S_j^0 \in S_i$).
We observe that these transient dynamics move back and forth across cell boundaries as the unfolding of one intracellular module affects its neighboring cells (see the example of the wildtype attractor in Fig. \ref{fig:wildtype_unfolding} and further examples in supplementary materials).
For example, a $\Sigma_2$ module will initiate $\textbf{M}_{nHH-1}$ in both adjacent cells (which will interact synergistically with any $\Sigma_1$ or $\Sigma_2$ modules already active in those cells) and $\Sigma_2$ modules in non-adjacent cells will activate $\Sigma_3$ modules in their neighboring cells (via neighboring $WG$-0 and $HH$-1).
Interestingly, some intracellular modules that were dynamically independent in the single-cell SPN work together across cell boundaries in the parasegment.  For example, $\textbf{M}_{(SLP-1,nHH-1)}$ (a $\Sigma_3$ module) will activate $WG$-1, which can interact with $SLP$-0 in an adjacent cell to initiate $\textbf{M}_{nWG-1, SLP-0}$ (a $\Sigma_2$ module) in that cell, even though $\Sigma_2$ and $\Sigma_3$ modules are dynamically independent (with the exception of $CIR$-$0$) within the same cell.

By analyzing the largest complex modules, we find that internal nodes play a large role in control of the dynamics, particularly $en$, $wg$, $PTC$, and $CIR$.  For example, the expression of $\textbf{M}_{CIR-1}$ in non-adjacent cells activates $\textbf{M}_{nWG-0}$ and $\textbf{M}_{nHH-0}$ in the other two cells.  As in the single-cell case, we also find alternative control strategies.  For example, either $\textbf{M}_{PTC-0, SLP-1}$ or $\textbf{M}_{nHH-1, SLP-1}$ activates $\textbf{M}_{nWG-1}$ in the two neighboring cells. 

We observe that the unfolding of pathway modules indicates a temporal order of events, such that one intracellular module may need to completely unfold to initiate another module in the same cell or an adjacent cell.
Of course, during the natural dynamics of the network, multiple modules will be initiated simultaneously based on initial conditions and this natural activation may speed up convergence to a final state.  However, the unfolding of a pathway module $\textbf{M}_i$ guarantees that any downstream node states $x \in S_i$ will occur (and gives an upper bound on the number of iterations it will take to do so via the module length), as long as the seed nodes are pinned for a long enough period.
As an example, pinning the 9 seeds indicated in Fig. \ref{fig:wildtype_unfolding} (4 of which are the SLP inputs) is guaranteed to drive the network to the wildtype attractor in no more than 16 iterations.  Given that this is a steady state, 16 iterations is also an upper bound on how long the seed nodes must be controlled, as the network will remain in the wildtype configuration once it reaches it.

As seen above,
using the easily discoverable intracellular complex modules allows for a concise description and qualitative understanding of what is happening in the parasegment network, even though the network is too large for an exhaustive description.  Because the \textit{drosophila} parasegment is divided into four identical and modular cells, we can understand the dynamics of the larger network by understanding the complex modules within the individual cells and how those modules interact across cell boundaries.  This gives us a clearer picture of the multivariate dynamics in the network than we could get via simple attractor analysis, while at the same time (due to its Boolean nature) being easier to analyze than a more detailed model (e.g., parameter estimation of differential equations).  To better understand how signals propagate between dynamically dependent modules, we simplify the macro-level dynamics of the system and, thus,  how the network computes.

\subsection{Dynamical Modularity of the \textit{Drosophila} SPN}

We use the complex modules found in the \textit{drosophila} SPN to estimate dynamical modularity of the single-cell and parasegment models by finding an optimal cover of the DCM, as described in section 2.3.  
For seed set size $s=1$, the optimal cover is defined by the set of complex modules $\Lambda_1$; however, the general problem of finding an optimal cover composed of modules with at most $s$ seeds is NP-hard based on the set-cover optimization problem \cite{karp1972reducibility,korte2012combinatorial}.  We can, however, restrict the number of modules in the cover to a maximum defined by parameter $q$.  For $\textit{I}$ complex modules there are, therefore, $\textit{I} \choose q$ options to consider for the optimal cover.
In general, dynamical modularity scores increase with $q$ because there are a greater number of module combinations to choose from.  However, for larger networks like the parasegment SPN, it is computationally prohibitive to estimate covers with even a moderately high $q$.  In this case, we estimate the cover using a greedy algorithm (see section 2.3), where at each step the module with the highest independence score is added to the cover until $q$ selections have been made.
In both cases, we only consider core complex modules using the maximal seed heuristic for inclusion in the cover.

\begin{figure}[h!]
\begin{adjustwidth}{-4cm}{-4cm}
\centering
\includegraphics[width=1.25\columnwidth]{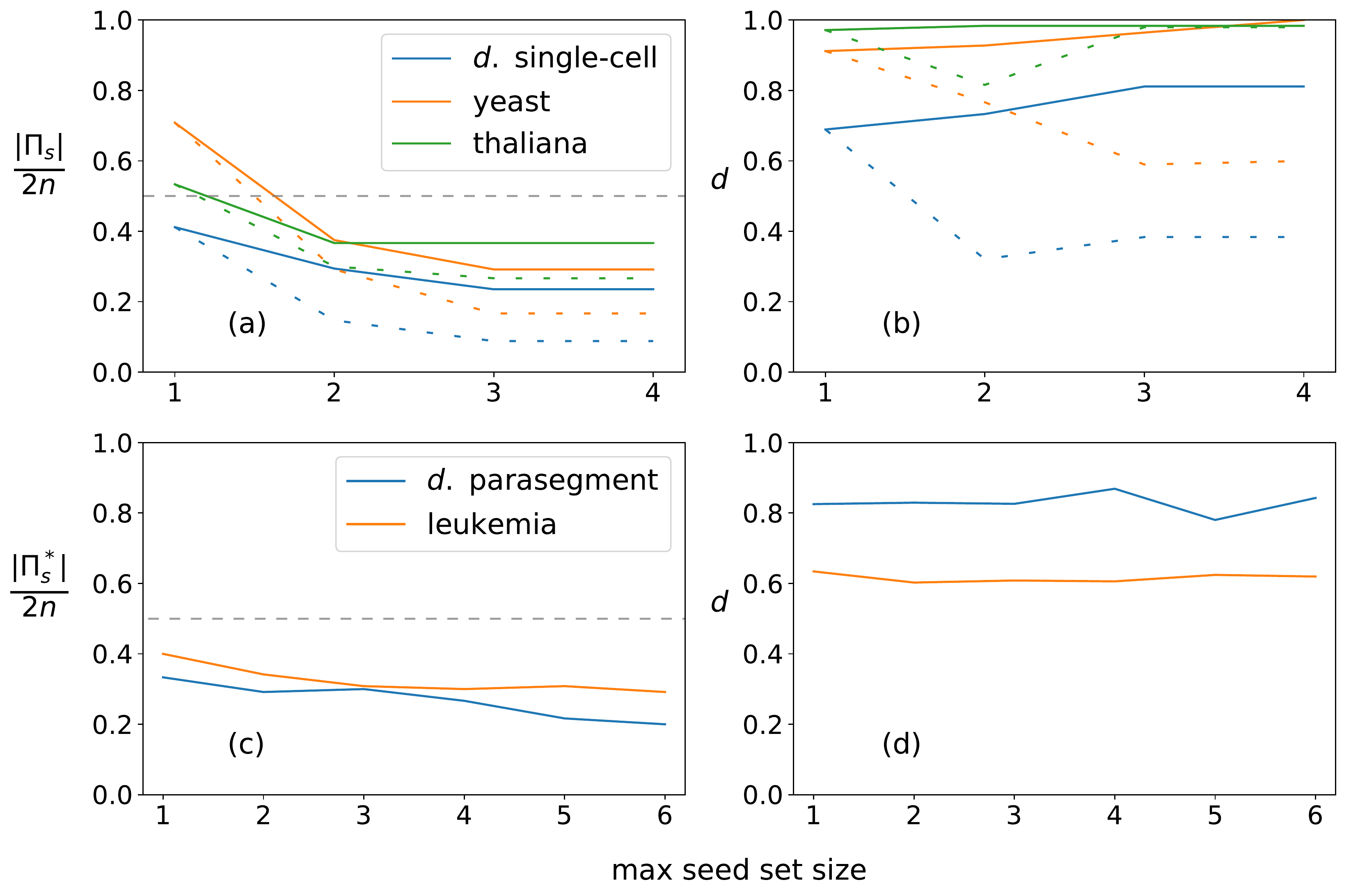}
\end{adjustwidth}
\caption{\textbf{Comparison of dynamical modularity across GRNs.}  (a) Results for the \textit{drosophila} single-cell SPN, the yeast cell cycle network, and \textit{thaliana arabadopsis}. The relative cover size $|\Pi_s|/2n$ 
is shown for each network's optimal cover $\Pi^*$ (based on maximizing $\overline{D}(\Pi_s)$, shown as full lines) and minimally-sized cover $\Pi_{min}$ (dashed lines) per maximum seed set size $s$.  Note that $|\Pi_s|/2n \leq 1$ because there are $2n$ s-units that together create a trivial cover (every s-unit acts as a module's seed). The intermediate line $|\Pi_s| = n$ is indicated by grey dashes. We see that the optimal cover has more than the minimal number of modules for $s>1$.
(b) Same as panel a, but the dynamical modularity score $d=\overline{D}(\Pi_s)$ is shown for each network's optimal cover $\Pi^*$ (full lines) and minimally-sized cover $\Pi_{min}$ (dashed lines).
(c) Results for the \textit{drosophila} parasegment SPN and the T-LGL leukemia network. We estimate the optimal cover $\Pi_s^*$ by greedy selection of the maximum module independence, $max(ind(M_i, \Pi_s-M_i))$, as described in Section 2.3.
The relative cover size $|\Pi_s^*|/2n$ is shown for each network.  Grey dashes indicate the line $|\Pi_s| = n$.
(d) Same as panel c, but the dynamical modularity score $d=\overline{D}(\Pi_s^*)$ is shown for each network. Due to the suboptimal nature of the greedy selection, $d$ is sometimes lower for larger maximum seed set sizes.
}
\label{fig:cover_comparison_small}
\end{figure}

For the single-cell SPN, the optimal cover at $s=1$ is composed of 14 complex modules and has mean dynamical modularity $\overline{D}(\Pi_1^*)=0.69$.  For $s=2$, we calculate the optimal cover $\Pi^*_2$ for $q \leq 11$; for $s=3$, we calculate the optimal cover $\Pi^*_3$ for $q \leq 8$; for $s \geq 4$ there are no more core complex modules so the solution does not change.
We find that the optimal covers have more than the minimal number of modules necessary to cover the DCM for $s>1$ (see Fig. \ref{fig:cover_comparison_small}, a-b).  For $s=2$, the DCM can be covered with 5 modules, but the optimal cover is $|\Pi^*_2|=10$, $\overline{D}(\Pi_2^*)=0.73$.  For $s=3$, the minimal cover is only three modules, but the optimal cover is $|\Pi^*_3|=8$, $\overline{D}(\Pi_3^*)=0.81$
($s*=3$ is also
the characteristic seed number of the single-cell SPN).  
We also find that the DCM cannot be covered by only pinning the s-units representing the input nodes ($SLP$, $nWG$, $nHH$) because none of these modules include the s-units $SMO$-0 or $CIR$-1.

Comparing \textit{drosophila} to other small GRNs, the yeast cell-cycle network \cite{li2004yeast} ($|\Pi^*_3|=7$, $\overline{D}(\Pi_3^*)=1.0$) and the \textit{thaliana} cell-fate specification network \cite{chaos2006genes} ($|\Pi^*_3|=11$, $\overline{D}(\Pi_3^*)=0.98$) have higher modularity scores and characteristic seed numbers $s*=4$ and $s*=2$ respectively.  
Interestingly, in both these networks, we find covers composed of complex modules that are (nearly) completely independent.

For the parasegment SPN, the optimal cover for seed set size $s=1$ is $|\Pi^*_1|=40$, $\overline{D}(\Pi_1^*)=0.83$.  Using our greedy algorithm, we find similar or increased scores for higher $s$ values: $|\Pi^*_2|=35$, $\overline{D}(\Pi_2^*)=0.83$, $|\Pi^*_3|=36$, $\overline{D}(\Pi_3^*)=0.83$, and $|\Pi^*_4|=32$, $\overline{D}(\Pi_4^*)=0.87$.  Due to the sub-optimality of the algorithm, the estimated score is somewhat lower at higher $s$ values, with $|\Pi^*_6|=24$, $\overline{D}(\Pi_6^*)=0.84$.
Our results suggest that the \textit{drosophila} parasegment SPN is more modular than both the single-cell SPN and the similarly-sized leukemia T-LGL network \cite{zhang2008network} (estimated $|\Pi^*_4|=36$, $\overline{D}(\Pi_4^*)=0.61$, see Figure \ref{fig:cover_comparison_small}, c-d).

We can also identify the modules that are included in the optimal cover.  For the single-cell SPN, at $s=2$, this includes $|S_{(nHH-1, CIR-1)}|=10$ and $|S_{(nHH-0, ptc-1)}|=6$, and single seed modules such as $|S_{en-1}|=9$.  For $s=3$, this includes $|S_{(SLP-1, nHH-0, ptc-1)}|=16$ (shown in Fig. \ref{fig:drosophila_modules_paper}d), $|S_{(nHH-1, en-1)}|=13$, and additional single seed modules.  In both cases, larger complex modules that are mostly independent cover much of the DCM and single seed modules fill in the missing node states.

For the parasegment SPN, the majority of the modules in our estimated covers are also of seed set size $s=1$ with only a few larger complex modules. Additionally, we see in Fig. \ref{fig:cover_comparison_small}d that the estimated mean dynamical modularity is roughly constant for both the parasegment SPN and the T-LGL leukemia network as $s$ is increased.
This suggests that even though the optimal cover is impossible to calculate at higher $s$ values, and larger complex modules that include useful synergies may be missed, the mean dynamical modularity at $s=1$ is a useful lower bound 
that may estimate well the true value for higher seed set sizes and may be used for comparison to other networks.

\section{Conclusion}

We have formally defined the pathway modules first discovered in \cite{marques2013canalization} and offered an algorithm to discover them via a modified breadth-first search on a network's DCM.  We have also formally analyzed interactions between pathway modules and defined complex modules, which are maximal pathway modules with synergy (dynamic dependence) between seed s-units.  This enables us to define the optimal cover of a DCM (a partition of s-units based on complex modules) and the associated mean dynamical modularity.

This modularity methodology is advantageous in that it greatly reduces the combinatoric number of pathway modules in Boolean GRNs for a given seed set size, $2^s {n \choose s}$, and only considers those that are complex modules.  It also allows a bottom-up approach to defining modularity based on a single parameter, the maximum seed set size $s$.  We were thus able to analyze all complex modules that occur in the \textit{drosophila} single-cell SPN, and sample many that occur in the parasegment.  This analysis of dynamical modularity allows an in-depth understanding of how dynamic signals propagate between cells in the parasegment and how dynamics unfold toward biologically-relevant attractors.  It also provides a means to compare different GRNs based on the size and dynamical modularity of the optimal cover of their DCM for a given seed set size.  This method can be applied to any automata network DCM, including those outside of the biological domain.

Importantly, this analysis describes the emergent computation that takes place in the example models as single seeds synergistically interact, causing their combined dynamical influence to logically propagate downstream, resulting in either an attractor or a partial configuration that represents the (much reduced) possible network configurations remaining.
These partial configurations are useful for control problems because the description of pathway modules elucidates exactly which downstream nodes are guaranteed to be affected and which state they will be changed to.  This can be especially helpful for target control \cite{gao2014target,yang2018target}, where reaching a certain variable state is desirable (such as an apoptotic state in a cancer growth model) or not desirable (such as a proliferative state).
Pathway modules, and complex modules in particular, can also be used to find seeds that have a similar effect on downstream targets.
Alternative control strategies such as these have potential uses in biological networks, such as designing optimal therapeutic targets.

Pathway modules describe the network dynamics in a mechanistic, causal way.  All s-units in a pathway module set are guaranteed to be reached given perturbation of the seed set, as long as there are no other interfering signals.  
This suggests that modules should be somewhat robust to the updating scheme chosen, although additional work is needed to explore module robustness in depth.
Pathway module sets, furthermore, provide the logical domain of influence of the seed set \cite{yang2018target}. This is a way to estimate which other variables are stabilized (guaranteed to be in a certain state) based on perturbation of a given seed set, regardless of the state of the other variables in the network.

One drawback to this dynamical modularity methodology is that it is still NP-hard in general, and analysis of  moderately-sized networks (such as the \textit{drosophila} parasegment or the T-LGL leukemia network) becomes difficult.  However, it is possible to sample even very large networks with low seed set size and greedy selection criteria to estimate optimal covers.
We leave additional analysis of efficient heuristics for 
finding optimal covers for future work as it is outside the scope of this paper.
 
Despite the computational difficulty of the problem, our methodology is a step toward understanding how network components interact based on the micro-dynamics of node state updates, and how these dynamical building blocks give rise to the macro-dynamics of network behavior.  





\section*{Acknowledgements}
Luis M. Rocha was partially funded by: National Institutes of Health, National Library of Medicine grant 1R01LM012832; Fundação para Ciencia e Tecnologia grant: 2022.09122.PTDC; National Science Foundation Research Traineeship “Interdisciplinary Training in Complex Networks and Systems” Grant 1735095.
The funders had no role in study design, data collection and analysis, decision to publish, or  any opinions, findings, conclusions or recommendations expressed in the manuscript.

The authors would also like to thank Deborah Rocha for scientific copy editing.


\section*{Data availability}
Network data can be retrieved at \url{https://github.com/tjparmer/dynamical_modularity} and \url{https://github.com/rionbr/CANA/tree/master/cana/datasets} \cite{correia2018cana}.

\section*{Code availability}
The code developed for this paper is made available at \url{https://github.com/tjparmer/dynamical_modularity}.

\pagebreak

\bibliographystyle{unsrt}

\bibliography{dynamics.bib} 

\begin{thebibliography}{10}

\bibitem{marques2013canalization}
Manuel Marques-Pita and Luis~M Rocha.
\newblock Canalization and control in automata networks: body segmentation in
  drosophila melanogaster.
\newblock {\em PloS one}, 8(3):e55946, 2013.

\bibitem{hartwell1999molecular}
Leland~H Hartwell, John~J Hopfield, Stanislas Leibler, and Andrew~W Murray.
\newblock From molecular to modular cell biology.
\newblock {\em Nature}, 402(6761):C47--C52, 1999.

\bibitem{yeger2004network}
Esti Yeger-Lotem, Shmuel Sattath, Nadav Kashtan, Shalev Itzkovitz, Ron Milo,
  Ron~Y Pinter, Uri Alon, and Hanah Margalit.
\newblock Network motifs in integrated cellular networks of
  transcription--regulation and protein--protein interaction.
\newblock {\em Proceedings of the National Academy of Sciences},
  101(16):5934--5939, 2004.

\bibitem{peter2009modularity}
Isabelle~S Peter and Eric~H Davidson.
\newblock Modularity and design principles in the sea urchin embryo gene
  regulatory network.
\newblock {\em FEBS letters}, 583(24):3948--3958, 2009.

\bibitem{davidson2010emerging}
Eric~H Davidson.
\newblock Emerging properties of animal gene regulatory networks.
\newblock {\em Nature}, 468(7326):911--920, 2010.

\bibitem{jaimovich2010modularity}
Ariel Jaimovich, Ruty Rinott, Maya Schuldiner, Hanah Margalit, and Nir
  Friedman.
\newblock Modularity and directionality in genetic interaction maps.
\newblock {\em Bioinformatics}, 26(12):i228--i236, 2010.

\bibitem{vidal2011interactome}
Marc Vidal, Michael~E Cusick, and Albert-L{\'a}szl{\'o} Barab{\'a}si.
\newblock Interactome networks and human disease.
\newblock {\em Cell}, 144(6):986--998, 2011.

\bibitem{stelling2004robustness}
J{\"o}rg Stelling, Uwe Sauer, Zoltan Szallasi, Francis~J Doyle~III, and John
  Doyle.
\newblock Robustness of cellular functions.
\newblock {\em Cell}, 118(6):675--685, 2004.

\bibitem{tanay2005conservation}
Amos Tanay, Aviv Regev, and Ron Shamir.
\newblock Conservation and evolvability in regulatory networks: the evolution
  of ribosomal regulation in yeast.
\newblock {\em Proceedings of the National Academy of Sciences},
  102(20):7203--7208, 2005.

\bibitem{pigliucci2008evolvability}
Massimo Pigliucci.
\newblock Is evolvability evolvable?
\newblock {\em Nature Reviews Genetics}, 9(1):75, 2008.

\bibitem{hernandez2022effects}
U~Hern{\'a}ndez, L~Posadas-Vidales, and Carlos Espinosa-Soto.
\newblock On the effects of the modularity of gene regulatory networks on
  phenotypic variability and its association with robustness.
\newblock {\em Biosystems}, 212:104586, 2022.

\bibitem{albert2009discrete}
R{\'e}ka Albert and Rui-Sheng Wang.
\newblock Discrete dynamic modeling of cellular signaling networks.
\newblock {\em Methods in enzymology}, 467:281--306, 2009.

\bibitem{zanudo2018discrete}
Jorge~GT Za{\~n}udo, Steven~N Steinway, and R{\'e}ka Albert.
\newblock Discrete dynamic network modeling of oncogenic signaling: Mechanistic
  insights for personalized treatment of cancer.
\newblock {\em Current Opinion in Systems Biology}, 9:1--10, 2018.

\bibitem{kauffman1969metabolic}
Stuart~A Kauffman.
\newblock Metabolic stability and epigenesis in randomly constructed genetic
  nets.
\newblock {\em Journal of theoretical biology}, 22(3):437--467, 1969.

\bibitem{thomas1973boolean}
Ren{\'e} Thomas.
\newblock Boolean formalization of genetic control circuits.
\newblock {\em Journal of theoretical biology}, 42(3):563--585, 1973.

\bibitem{albert2014boolean}
Reka Albert and Juilee Thakar.
\newblock Boolean modeling: a logic-based dynamic approach for understanding
  signaling and regulatory networks and for making useful predictions.
\newblock {\em Wiley Interdisciplinary Reviews: Systems Biology and Medicine},
  6(5):353--369, 2014.

\bibitem{clauset2004finding}
Aaron Clauset, Mark~EJ Newman, and Cristopher Moore.
\newblock Finding community structure in very large networks.
\newblock {\em Physical review E}, 70(6):066111, 2004.

\bibitem{blondel2008fast}
Vincent~D Blondel, Jean-Loup Guillaume, Renaud Lambiotte, and Etienne Lefebvre.
\newblock Fast unfolding of communities in large networks.
\newblock {\em Journal of statistical mechanics: theory and experiment},
  2008(10):P10008, 2008.

\bibitem{alexander2009understanding}
Roger~P Alexander, Philip~M Kim, Thierry Emonet, and Mark~B Gerstein.
\newblock Understanding modularity in molecular networks requires dynamics.
\newblock {\em Science signaling}, 2(81):pe44--pe44, 2009.

\bibitem{verd2019modularity}
Berta Verd, Nicholas~AM Monk, and Johannes Jaeger.
\newblock Modularity, criticality, and evolvability of a developmental gene
  regulatory network.
\newblock {\em Elife}, 8:e42832, 2019.

\bibitem{gates2016control}
Alexander~J Gates and Luis~M Rocha.
\newblock Control of complex networks requires both structure and dynamics.
\newblock {\em Scientific reports}, 6:24456, 2016.

\bibitem{jimenez2017spectrum}
Alba Jim{\'e}nez, James Cotterell, Andreea Munteanu, and James Sharpe.
\newblock A spectrum of modularity in multi-functional gene circuits.
\newblock {\em Molecular systems biology}, 13(4):925, 2017.

\bibitem{irons2007identifying}
David~J Irons and Nicholas~AM Monk.
\newblock Identifying dynamical modules from genetic regulatory systems:
  applications to the segment polarity network.
\newblock {\em BMC bioinformatics}, 8(1):413, 2007.

\bibitem{kolchinsky2011prediction}
A~Kolchinsky and LM~Rocha.
\newblock Prediction and modularity in dynamical systems.
\newblock In {\em Advances in artificial life. Proceedings of the Eleventh
  European conference on the synthesis and simulation of living systems (ECAL
  2011)}, pages 423--430. MIT Press, 2011.

\bibitem{paul2018decomposition}
Soumya Paul, Cui Su, Jun Pang, and Andrzej Mizera.
\newblock A decomposition-based approach towards the control of boolean
  networks.
\newblock In {\em Proceedings of the 2018 ACM International Conference on
  Bioinformatics, Computational Biology, and Health Informatics}, pages 11--20,
  2018.

\bibitem{kadelka2022decomposition}
Claus Kadelka, Reinhard Laubenbacher, David Murrugarra, Alan Veliz-Cuba, and
  Matthew Wheeler.
\newblock Decomposition of boolean networks: An approach to modularity of
  biological systems.
\newblock {\em arXiv preprint arXiv:2206.04217}, 2022.

\bibitem{zanudo2015cell}
Jorge~GT Zanudo and R{\'e}ka Albert.
\newblock Cell fate reprogramming by control of intracellular network dynamics.
\newblock {\em PLoS computational biology}, 11(4):e1004193, 2015.

\bibitem{albert2003topology}
R{\'e}ka Albert and Hans~G Othmer.
\newblock The topology of the regulatory interactions predicts the expression
  pattern of the segment polarity genes in drosophila melanogaster.
\newblock {\em Journal of theoretical biology}, 223(1):1--18, 2003.

\bibitem{gershenson2004introduction}
Carlos Gershenson.
\newblock Introduction to random boolean networks.
\newblock {\em arXiv preprint nlin/0408006}, 2004.

\bibitem{harvey1997time}
Inman Harvey and Terry Bossomaier.
\newblock Time out of joint: Attractors in asynchronous random boolean
  networks.
\newblock In {\em Proceedings of the Fourth European Conference on Artificial
  Life}, pages 67--75. MIT Press, Cambridge, 1997.

\bibitem{waddington1942canalization}
Conrad~H Waddington.
\newblock Canalization of development and the inheritance of acquired
  characters.
\newblock {\em Nature}, 150(3811):563--565, 1942.

\bibitem{kauffman1984emergent}
Stuart~A Kauffman.
\newblock Emergent properties in random complex automata.
\newblock {\em Physica D: Nonlinear Phenomena}, 10(1-2):145--156, 1984.

\bibitem{siegal2002waddington}
Mark~L Siegal and Aviv Bergman.
\newblock Waddington's canalization revisited: developmental stability and
  evolution.
\newblock {\em Proceedings of the National Academy of Sciences},
  99(16):10528--10532, 2002.

\bibitem{kauffman2004genetic}
Stuart Kauffman, Carsten Peterson, Bj{\"o}rn Samuelsson, and Carl Troein.
\newblock Genetic networks with canalyzing boolean rules are always stable.
\newblock {\em Proceedings of the National Academy of Sciences},
  101(49):17102--17107, 2004.

\bibitem{manicka2017role}
Santosh Venkatiah~Sudharshan Manicka.
\newblock The role of canalization in the spreading of perturbations in boolean
  networks.
\newblock 2017.

\bibitem{Quine1955truth}
W~V Quine.
\newblock {A Way to Simplify Truth Functions}.
\newblock {\em American Mathematical Monthly}, 62:627--631, 1955.

\bibitem{conrad1990geometry}
Michael Conrad.
\newblock The geometry of evolution.
\newblock {\em BioSystems}, 24(1):61--81, 1990.

\bibitem{mcculloch1943logical}
Warren~S McCulloch and Walter Pitts.
\newblock A logical calculus of the ideas immanent in nervous activity.
\newblock {\em The bulletin of mathematical biophysics}, 5(4):115--133, 1943.

\bibitem{klamt2006methodology}
Steffen Klamt, Julio Saez-Rodriguez, Jonathan~A Lindquist, Luca Simeoni, and
  Ernst~D Gilles.
\newblock A methodology for the structural and functional analysis of signaling
  and regulatory networks.
\newblock {\em BMC bioinformatics}, 7(1):1--26, 2006.

\bibitem{wang2011elementary}
Rui-Sheng Wang and R{\'e}ka Albert.
\newblock Elementary signaling modes predict the essentiality of signal
  transduction network components.
\newblock {\em BMC systems biology}, 5(1):1--14, 2011.

\bibitem{yang2018target}
Gang Yang, Jorge G{\'o}mez Tejeda~Za{\~n}udo, and R{\'e}ka Albert.
\newblock Target control in logical models using the domain of influence of
  nodes.
\newblock {\em Frontiers in physiology}, page 454, 2018.

\bibitem{fiedler2013dynamics}
Bernold Fiedler, Atsushi Mochizuki, Gen Kurosawa, and Daisuke Saito.
\newblock Dynamics and control at feedback vertex sets. i: Informative and
  determining nodes in regulatory networks.
\newblock {\em Journal of Dynamics and Differential Equations}, 25(3):563--604,
  2013.

\bibitem{mochizuki2013dynamics}
Atsushi Mochizuki, Bernold Fiedler, Gen Kurosawa, and Daisuke Saito.
\newblock Dynamics and control at feedback vertex sets. ii: A faithful monitor
  to determine the diversity of molecular activities in regulatory networks.
\newblock {\em Journal of theoretical biology}, 335:130--146, 2013.

\bibitem{zanudo2017structure}
Jorge Gomez~Tejeda Za{\~n}udo, Gang Yang, and R{\'e}ka Albert.
\newblock Structure-based control of complex networks with nonlinear dynamics.
\newblock {\em Proceedings of the National Academy of Sciences},
  114(28):7234--7239, 2017.

\bibitem{simon1962architecture}
Herbert~A Simon.
\newblock The architecture of complexity.
\newblock {\em Proceedings of the American Philosophical Society}, 106, 1962.

\bibitem{kolchinsky2015modularity}
Artemy Kolchinsky, Alexander~J Gates, and Luis~M Rocha.
\newblock Modularity and the spread of perturbations in complex dynamical
  systems.
\newblock {\em Physical Review E}, 92(6):060801, 2015.

\bibitem{karp1972reducibility}
Richard~M Karp.
\newblock Reducibility among combinatorial problems.
\newblock In {\em Complexity of computer computations}, pages 85--103.
  Springer, 1972.

\bibitem{korte2012combinatorial}
Bernhard Korte, Jens Vygen, B~Korte, and J~Vygen.
\newblock {\em Combinatorial optimization}, volume~2.
\newblock Springer, 2012.

\bibitem{li2004yeast}
Fangting Li, Tao Long, Ying Lu, Qi~Ouyang, and Chao Tang.
\newblock The yeast cell-cycle network is robustly designed.
\newblock {\em Proceedings of the National Academy of Sciences},
  101(14):4781--4786, 2004.

\bibitem{chaos2006genes}
Alvaro Chaos, Max Aldana, Carlos Espinosa-Soto, Berenice Garc{\'\i}a~Ponce
  de~Le{\'o}n, Adriana~Garay Arroyo, and Elena~R Alvarez-Buylla.
\newblock From genes to flower patterns and evolution: dynamic models of gene
  regulatory networks.
\newblock {\em Journal of Plant Growth Regulation}, 25(4):278--289, 2006.

\bibitem{zhang2008network}
Ranran Zhang, Mithun~Vinod Shah, Jun Yang, Susan~B Nyland, Xin Liu, Jong~K Yun,
  R{\'e}ka Albert, and Thomas~P Loughran.
\newblock Network model of survival signaling in large granular lymphocyte
  leukemia.
\newblock {\em Proceedings of the National Academy of Sciences}, 2008.

\bibitem{willadsen2007robustness}
Kai Willadsen and Janet Wiles.
\newblock Robustness and state-space structure of boolean gene regulatory
  models.
\newblock {\em Journal of theoretical biology}, 249(4):749--765, 2007.

\bibitem{von2000segment}
George Von~Dassow, Eli Meir, Edwin~M Munro, and Garrett~M Odell.
\newblock The segment polarity network is a robust developmental module.
\newblock {\em Nature}, 406(6792):188, 2000.

\bibitem{ingolia2004topology}
Nicholas~T Ingolia.
\newblock Topology and robustness in the drosophila segment polarity network.
\newblock {\em PLoS biology}, 2(6):e123, 2004.

\bibitem{chaves2005robustness}
Madalena Chaves, Reka Albert, and Eduardo~D Sontag.
\newblock Robustness and fragility of boolean models for genetic regulatory
  networks.
\newblock {\em Journal of theoretical biology}, 235(3):431--449, 2005.

\bibitem{chaves2006methods}
Madalena Chaves, Eduardo~D Sontag, and R{\'e}ka Albert.
\newblock Methods of robustness analysis for boolean models of gene control
  networks.
\newblock {\em arXiv preprint q-bio/0605004}, 2006.

\bibitem{chaves2008studying}
Madalena Chaves and R{\'e}ka Albert.
\newblock Studying the effect of cell division on expression patterns of the
  segment polarity genes.
\newblock {\em Journal of The Royal Society Interface}, 5(Suppl 1):S71--S84,
  2008.

\bibitem{Correia2022.03.02.482557}
Rion~Brattig Correia, Joana~M. Almeida, Margot~J. Wyrwoll, Irene Julca, Daniel
  Sobral, Chandra~Shekhar Misra, Leonardo~G. Guilgur, Hans-Christian Schuppe,
  Neide Silva, Pedro Prud{\^e}ncio, Ana N{\'o}voa, Ana~S. Leoc{\'a}dio, Joana
  Bom, Mois{\'e}s Mallo, Sabine Kliesch, Marek Mutwil, Luis~M. Rocha, Frank
  T{\"u}ttelmann, J{\"o}rg~D. Becker, and Paulo Navarro-Costa.
\newblock The conserved transcriptional program of metazoan male germ cells
  uncovers ancient origins of human infertility.
\newblock {\em bioRxiv}, 2022.

\bibitem{parmer2022influence}
Thomas Parmer, Luis~M Rocha, and Filippo Radicchi.
\newblock Influence maximization in boolean networks.
\newblock {\em Nature communications}, 13(1):1--11, 2022.

\bibitem{gao2014target}
Jianxi Gao, Yang-Yu Liu, Raissa~M D'souza, and Albert-L{\'a}szl{\'o}
  Barab{\'a}si.
\newblock Target control of complex networks.
\newblock {\em Nature communications}, 5(1):1--8, 2014.

\bibitem{correia2018cana}
Rion~Brattig Correia, Alexander~J Gates, Xuan Wang, and Luis~M Rocha.
\newblock Cana: A python package for quantifying control and canalization in
  boolean networks.
\newblock {\em arXiv preprint arXiv:1803.04774}, 2018.

\end{thebibliography}


\begin{thebibliography}{10}

\bibitem{albert2003topology}
R{\'e}ka Albert and Hans~G Othmer.
\newblock The topology of the regulatory interactions predicts the expression
  pattern of the segment polarity genes in drosophila melanogaster.
\newblock {\em Journal of theoretical biology}, 223(1):1--18, 2003.

\bibitem{marques2013canalization}
Manuel Marques-Pita and Luis~M Rocha.
\newblock Canalization and control in automata networks: body segmentation in
  drosophila melanogaster.
\newblock {\em PloS one}, 8(3):e55946, 2013.

\bibitem{chaves2006methods}
Madalena Chaves, Eduardo~D Sontag, and R{\'e}ka Albert.
\newblock Methods of robustness analysis for boolean models of gene control
  networks.
\newblock {\em arXiv preprint q-bio/0605004}, 2006.

\bibitem{gershenson2004introduction}
Carlos Gershenson.
\newblock Introduction to random boolean networks.
\newblock {\em arXiv preprint nlin/0408006}, 2004.

\bibitem{correia2018cana}
Rion~Brattig Correia, Alexander~J Gates, Xuan Wang, and Luis~M Rocha.
\newblock Cana: A python package for quantifying control and canalization in
  boolean networks.
\newblock {\em arXiv preprint arXiv:1803.04774}, 2018.

\bibitem{chaos2006genes}
Alvaro Chaos, Max Aldana, Carlos Espinosa-Soto, Berenice Garc{\'\i}a~Ponce
  de~Le{\'o}n, Adriana~Garay Arroyo, and Elena~R Alvarez-Buylla.
\newblock From genes to flower patterns and evolution: dynamic models of gene
  regulatory networks.
\newblock {\em Journal of Plant Growth Regulation}, 25(4):278--289, 2006.

\bibitem{li2004yeast}
Fangting Li, Tao Long, Ying Lu, Qi~Ouyang, and Chao Tang.
\newblock The yeast cell-cycle network is robustly designed.
\newblock {\em Proceedings of the National Academy of Sciences},
  101(14):4781--4786, 2004.

\bibitem{zhang2008network}
Ranran Zhang, Mithun~Vinod Shah, Jun Yang, Susan~B Nyland, Xin Liu, Jong~K Yun,
  R{\'e}ka Albert, and Thomas~P Loughran.
\newblock Network model of survival signaling in large granular lymphocyte
  leukemia.
\newblock {\em Proceedings of the National Academy of Sciences}, 2008.

\bibitem{fiedler2013dynamics}
Bernold Fiedler, Atsushi Mochizuki, Gen Kurosawa, and Daisuke Saito.
\newblock Dynamics and control at feedback vertex sets. i: Informative and
  determining nodes in regulatory networks.
\newblock {\em Journal of Dynamics and Differential Equations}, 25(3):563--604,
  2013.

\bibitem{mochizuki2013dynamics}
Atsushi Mochizuki, Bernold Fiedler, Gen Kurosawa, and Daisuke Saito.
\newblock Dynamics and control at feedback vertex sets. ii: A faithful monitor
  to determine the diversity of molecular activities in regulatory networks.
\newblock {\em Journal of theoretical biology}, 335:130--146, 2013.

\end{thebibliography}


\end{document}


\maketitle

\section{\textit{Drosophila} Model}

The update rules of the single-cell 
segment polarity network
(SPN) are listed here.

\vspace{0.25cm}
\setlength{\parindent}{1cm}

\textit{Inputs:}


{SLP}$_{t+1}$=SLP$_{t}$

{nWG}$_{t+1}$=nWG$_{t}$

{nHH}$_{t+1}$=nHH$_{t}$


\medskip

\textit{Internal nodes:}


\textit{wg}$_{t+1}$=(CIA$_{t}$ $\wedge$ SLP$_{t}$ $\wedge$ $\lnot$CIR$_{t}$) $\lor$ (\textit{wg}$_{t}$ $\wedge$ (CIA$_{t}$ $\lor$ SLP$_{t}$) $\wedge$ $\lnot$CIR$_{t}$)

WG$_{t+1}$=\textit{wg}$_{t}$

\textit{en}$_{t+1}$=nWG$_{t}$ $\wedge$ $\lnot$SLP$_{t}$

EN$_{t+1}$=\textit{en}$_{t}$

\textit{hh}$_{t+1}$=EN$_{t}$ $\wedge$ $\lnot$CIR$_{t}$

HH$_{t+1}$=\textit{hh}$_{t}$

\textit{ptc}$_{t+1}$=CIA$_{t}$ $\wedge$ $\lnot$EN$_{t}$ $\wedge$ $\lnot$CIR$_{t}$

PTC$_{t+1}$=\textit{ptc}$_{t}$ $\lor$ (PTC$_{t}$ $\wedge$ $\lnot$nHH$_{t}$)

PH$_{t+1}$=PTC$_{t}$ $\wedge$ nHH$_{t}$

SMO$_{t+1}$=$\lnot$PTC$_{t}$ $\lor$ nHH$_{t}$

\textit{ci}$_{t+1}$=$\lnot$EN$_{t}$

CI$_{t+1}$=\textit{ci}$_{t}$

CIA$_{t+1}$=CI$_{t}$ $\wedge$ ($\lnot$PTC$_{t}$ $\lor$ nHH$_{t}$)

CIR$_{t+1}$=CI$_{t}$ $\wedge$ PTC$_{t}$ $\wedge$ $\lnot$nHH$_{t}$

\vspace{0.25cm}

\setlength{\parindent}{17.62482pt} 

The parasegment SPN (size $n=60$) extends the single-cell model by including all 14 internal nodes in each of four cells, where WG and HH influence both of the cell's neighbors (periodic boundary conditions assumed), and an additional node, SLP, acts as an input to each cell (see \cite{albert2003topology} and \cite{marques2013canalization} for a full description).

\section{Algorithms}

Pathway modules can be discovered by a breadth-first search (BFS) on the DCM (\textit{algorithm 1},
see Fig. \ref{fig:algorithm1}), given the caveat of accounting for the temporal dynamics of state updates. 
Input to the algorithm is a DCM, a 
seed set $S^0$, and a perturbation type (pinning or pulse).  The algorithm tracks which s-units and t-units fire in each time step.  As with regular BFS units are visited via a first-in first-out (FIFO) queue, 
but a separate queue is created for each time step.
The time counter is initialized at 0 and increases once all of the units in the associated queue have been visited.

T-units are added to the queue 
associated with
the current time step $t$ 
if their threshold is met, while s-units are added to the queue 
associated with
time step $t+1$. 
If the perturbation type is pinning, 
any given unit is only visited once (as they are afterwards assumed to fire every time step).
If the perturbation type is pulse, 
by contrast,
units that previously fired are allowed to be visited again in future time steps.

The algorithm halts when no new s-units or t-units 
can be added to a queue
(logically, no more units can fire) or the set of s-units and t-units that fire 
at time $t=i$ is equivalent to the set of s-units and t-units 
that fired
at a previous iteration $t={j<i}$ (a cycle is reached)~\footnote{This only applies to pulse perturbation where the set of units that fire is reset after each time step; for pinning perturbation, this condition is equivalent to the previous one (i.e., no new units are added to the queue) since all units that fire at time $t$ also fire at time $t+1$.}.

The algorithm returns a hash table keyed by time step, where each value is the set of s-units and t-units that fire in the associated time step.  The total set of s-units that fire, $S$, can then be easily derived by finding the superset of s-units from each time step (and ignoring all t-units). 


This algorithm can be extended to the general case where each seed s-unit $x$ is associated with a firing schedule, 

$f({x},{t})=$ \{1 if $x$ is forced to fire at time $t$,

        \setlength\parindent{2.3cm}
        0 if $x$ is not forced to fire at time $t$\},
\setlength{\parindent}{0cm}

rather than only allowing pinning or pulse perturbation.
In this case, the external perturbations dictated by the firing schedule take precedence over any units already in the queue.
Additionally, the delay associated with units can be generalized; here
all s-units take one time step to fire (i.e., they have delay $dt=1$);
however, s-units can be allowed a higher delay $dt>1$
such that it takes several (relative) time steps for a signal to reach an s-unit (for example, this may represent cellular processes that operate on a longer time scale than transcription or translation and could be useful for non-synchronous updating schemes \cite{chaves2006methods, gershenson2004introduction}).

\setlength{\parindent}{17.62482pt} 

The set of pathway modules 
discovered by \textit{algorithm 1}
can then be reduced to a set of complex modules ( \textit{algorithm 2},
see Fig.~\ref{fig:algorithm2}).  Input to the algorithm is a DCM, a maximum seed set size \textit{max\_s}, a set of s-units \textit{seeds},
and a perturbation type (pinning or pulse). 
Algorithm 2 initially finds all pathway modules with seed set size $s=1$ 
whose seed is in the set \textit{seeds}
using \textit{algorithm 1} and then removes any subsumed modules 
to create an initial set of complex modules $I_1$.  It then iteratively finds the set of complex modules $I_{s>1}$ with increasing seed set size $s=2,3,4...$\textit{max\_s}.  For each 
size $s$, pathway modules are found 
for all possible seeds sets, 
where seeds are obtained from the set \textit{seeds}.
This set of potential modules is then reduced to a set of complex modules by removing all modules that are not maximal or that contain non-synergistic submodules.  The algorithm halts when the maximum seed set size \textit{max\_s} has been reached and returns the set of complex modules
found thus far.

The set \textit{seeds} can be any set of s-units of interest 
(for example, the s-units representing the network inputs).  If \textit{seeds}$= \mathcal{S}$, the set of all s-units in the network, then all pathway modules (and therefore all complex modules) will be found; if, however, \textit{seeds}$= \Lambda_1$, then only complex modules with maximal seeds will be found~\footnote{$\Lambda_1$ can be found by running \textit{algorithm 2 }with \textit{max\_s}$=1$, \textit{seeds}$= \mathcal{S}$.}.  This is the maximal seed heuristic which finds all core complex modules in the network.

All analysis and code for this paper,
including \textit{algorithm 1} and \textit{algorithm 2}, was written in Python 2.7 with the assistance of the CANA package for finding Boolean network DCMs \cite{correia2018cana}.
The code is publicly available at \url{https://github.com/tjparmer/dynamical\_modularity}.

\bigskip

\begin{figure}[h!] 
\begin{adjustwidth}{-1cm}{}
\hrulefill 

\textbf{Algorithm 1:} Pathway module breadth-first search

\hrulefill 

\textbf{function} thresholded\_BFS (\textit{DCM}, 
$S^0$,
\textit{perturbation\_type}):
	
	\vspace{.25cm}
	\setlength{\parindent}{1cm}
	\indent time \textit{t}=0
	
	\textit{time\_steps}=\{t: 
    queue($S^0$)\}
	
	\textit{unfolding}=\{t: \{\}\} 
	
	\textit{active\_step}=
 \{$S^0$\}
	
	\textit{visited}=\{\}
	
	
	\vspace{.25cm}
	
	\textbf{while} \textit{t}$\leq$max(\textit{time\_steps}):
	
	\vspace{.25cm}
	
	\setlength{\parindent}{2cm}
	
	\textit{visited} $\leftarrow$ \textit{active\_step}
	
	\textbf{if} 
	\textit{active\_step} $=$ \textit{unfolding}[$i<t$]: 
	
	    \setlength{\parindent}{3cm}
	    
	    
	    \textbf{return} \textit{unfolding}
	
	\setlength{\parindent}{2cm}
	\indent \textbf{if} \textit{perturbation\_type}=`pulse': 
	
	\setlength{\parindent}{3cm}
	\indent \textit{active\_step}=\{\}
	
	\textbf{reset} t-unit thresholds
	
	\vspace{.25cm}
	
	\setlength{\parindent}{2cm}
	\indent \textbf{while} \textit{time\_steps}[\textit{t}]:
	
	    \setlength{\parindent}{3cm}
	    \indent \textbf{pop} first unit \textit{x} from queue at \textit{time\_steps}[\textit{t}]
	    
		\textit{unfolding}[\textit{t}] $\leftarrow$ \textit{x}
		
		\textbf{for each} 
        \textit{DCM}
        neighbor (s-unit/t-unit) \textit{n} 
		$\not\in$
		\textit{active\_step}: 
		
		    \setlength{\parindent}{4cm}
			
			    \textbf{if} \textit{n} is not a contradiction \textbf{and} the threshold \textit{$\tau$} of \textit{n} is met:
			
			    \setlength{\parindent}{5cm}
				\indent \textit{dt} = the delay of \textit{n}
				
				\textit{time\_steps}[$t+$\textit{dt}] $\leftarrow$ \textit{n}
				
				\textit{active\_step} $\leftarrow$ \textit{n}
	
	\setlength{\parindent}{2cm}			
	\indent \textit{$t = t+1$}

\vspace{.25cm}
\setlength{\parindent}{1cm}	
\indent
\textbf{return} \textit{unfolding}

\setlength{\parindent}{0cm}
\hrulefill 

\caption{\textbf{Pseudo-code for algorithm 1.}
Input to the function is a DCM, a seed set 
of s-units $S^0$,
and a perturbation type (`pinning' or `pulse').  
The algorithm is similar to breadth-first search (BFS) and explores what part of the DCM is reachable from the seed set, factoring in the perturbation type, relative time steps, logical contradiction, and t-unit thresholds.  FIFO queues hashed by relative time step are stored in \textit{time\_steps} while sets of visited units hashed by relative time step are stored in \textit{unfolding}.  Units (s or t) are only added to the queue associated with the respective time step if they meet the requisite conditions.  
S-units are given a default threshold $\tau=1$ and delay $dt=1$, while t-units have delay $dt=0$.  Units that are visited are added to the sets \textit{visited} and \textit{active\_step}.  Each t-unit must track the s-units that contribute towards its threshold; with pulse perturbation, unit thresholds are reset after each time step.  
The algorithm halts when a repeat set of s-units and t-units is found for a given time step or when no new units can be reached based on their logical conditions.
The function returns the pathway module $\textbf{M}_{S^0}$ initiated by the seed set, i.e., the sets of units visited at each relative time step.  In the above pseudo-code, \{key:value\} represents a hash table and \{\} a set.
}
\label{fig:algorithm1}
\end{adjustwidth}
\end{figure}

\begin{figure}[h!]
\begin{adjustwidth}{-2cm}{}
\hrulefill 

\textbf{Algorithm 2:} Complex module discovery

\hrulefill 

\textbf{function} complex\_module\_search (\textit{DCM}, \textit{max\_s}, \textit{seeds}, \textit{perturbation\_type}): 

    \vspace{.25cm}
    \setlength{\parindent}{1cm}
	iteration \textit{s}=1
	
	
	complex modules \textit{I}=\{\}
	
	
	\vspace{.25cm}
	\textbf{while} $\textit{s} \leq \textit{max\_s}$:
	
	    \setlength{\parindent}{2cm}
	    \textbf{if} \textit{s}=1:
	    
	        \setlength{\parindent}{3cm}
	        \textit{new\_seeds}=\textit{seeds}
	        
	   \setlength{\parindent}{2cm}
	   \textbf{else:}
	   
	        \setlength{\parindent}{3cm}
	        \textit{new\_seeds}=\{$S_i^0 \cup \{\textit{seed}\}$ \textbf{for} $\textit{i} \in \textit{P}$, $\textit{seed} \in \textit{seeds}$\}
	    
	    \vspace{.25cm}
	    \setlength{\parindent}{2cm}
	    
	    pathway modules \textit{P}=\{\}
	    
	    
	    \textbf{for} \textit{seed} in \textit{new\_seeds}:
	    
	        \setlength{\parindent}{3cm}
	        \textit{P} $\leftarrow$ \textbf{thresholded\_BFS}(\textit{DCM}, \textit{seed}, \textit{perturbation\_type})
	   
	   \vspace{.25cm}
	   \setlength{\parindent}{2cm}
	   
	   \textbf{for} module \textit{p} $\in$ \textit{P}:
	   
	        \setlength{\parindent}{3cm}
	        \textbf{if} $p$ is subsumed by $m$ for any module $m \in P$:
	            
	            \setlength{\parindent}{4cm}
	                \textbf{continue} 
	                
	       \setlength{\parindent}{3cm}
	       \textbf{else if} $\mu(S_p^0) - \mu(S_{ps}^0) - \mu(S_p^0 - S_{ps}^0) = \emptyset$ for any submodule \textit{ps} of 
	       \textit{p}:
	            
	                \setlength{\parindent}{4cm}
	                \textbf{continue} 
	                
	       \setlength{\parindent}{3cm}
	       \textbf{else:}
	       
	       \setlength{\parindent}{4cm}
	       \textit{I} $\leftarrow$ \textit{p} 
		
		\setlength{\parindent}{2cm}	
		$\textit{s} = \textit{s}+1$
	
	\vspace{.25cm}
	\setlength{\parindent}{1cm}
	\textbf{return} \textit{I}

\setlength{\parindent}{0cm}		
\hrulefill 

\caption{\textbf{Pseudo-code for algorithm 2.}  
Input to the function is a DCM, the maximum seed set size \textit{max\_s}, a set of s-units
\textit{seeds}, and a perturbation type (`pinning' or `pulse').
At each iteration, new seed sets of size $s=k$ are generated by combining seed sets of 
current pathway modules 
of size $s=k-1$ with the members of \textit{seeds} (size $s=1$).  The pathway modules for these 
new seed sets are found using algorithm 1; this set of modules is then condensed to a set of complex modules by removing those that are subsumed by other modules or that fail a synergy check.  For pinning perturbation, the former condition filters any module $p$ where $S_p \subset S_m$ for any other module $m \in P$, whereas for pulse perturbation the temporal order of when units fire must be checked as well.  
The function returns the set of complex modules $I$.
In the above pseudo-code, \{\} represents a set.
}
\label{fig:algorithm2}
\end{adjustwidth}
\end{figure}

\begin{figure}[]
\centering
\includegraphics[width=1.2\columnwidth]{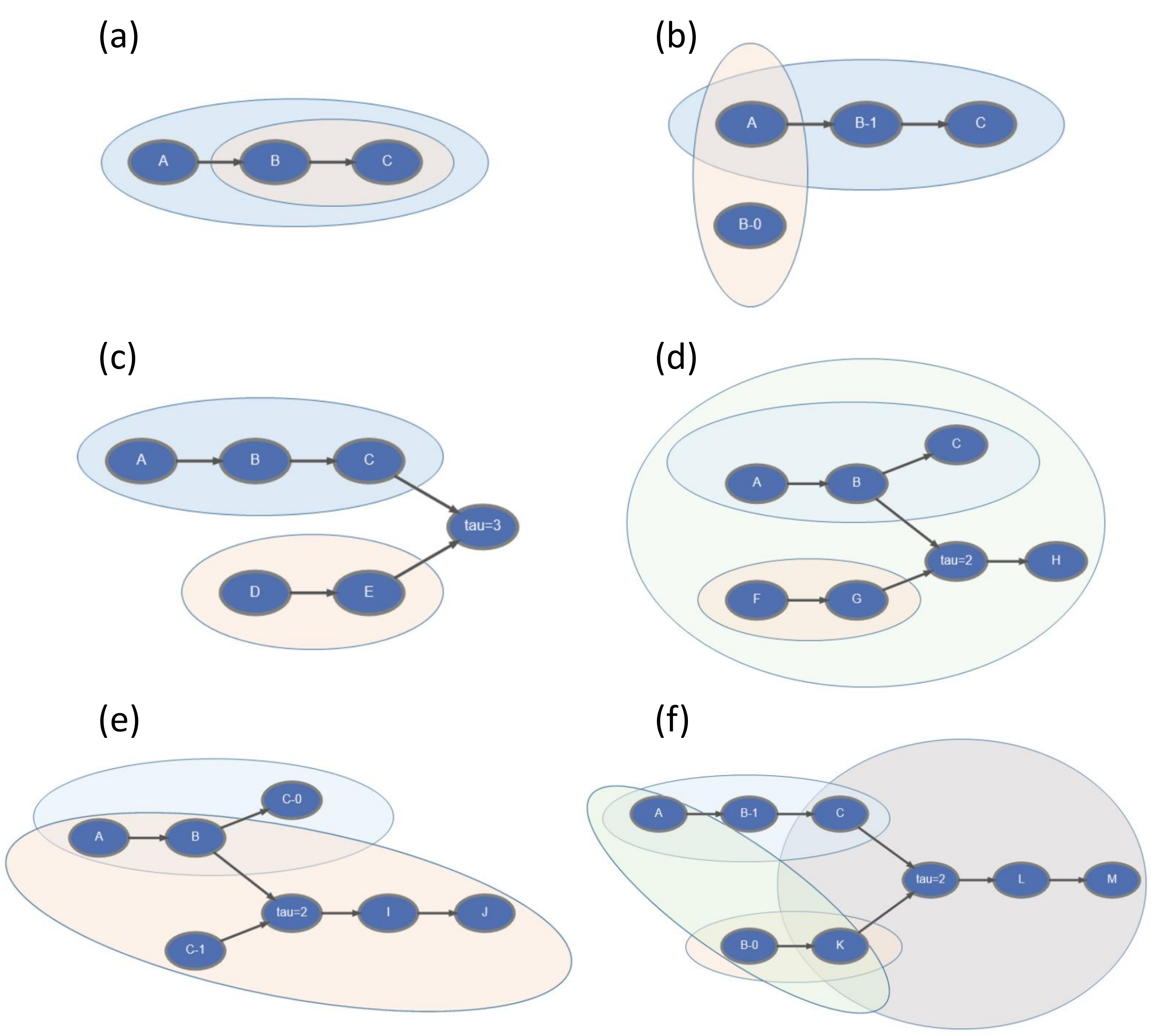}
\caption{\textbf{Example interactions between pathway modules.}  
(a) Module $\textbf{M}_A$ subsumes $\textbf{M}_B$ as the unfolding of $\textbf{M}_B$ is completely contained in $\textbf{M}_A$.  (b) $\textbf{M}_{A, B-0}$ logically obstructs $\textbf{M}_A$ because of the contradiction between B-0 and B-1, which prevents the rest of $\textbf{M}_A$ from unfolding.  (c) $\textbf{M}_A$ and $\textbf{M}_D$ are decoupled
(dynamically independent); even though they share a threshold, $\textbf{M}_{A,D}$ does not fire the t-unit and thus cannot cause any downstream effects.  (d) $\textbf{M}_A$ and $\textbf{M}_F$ are synergistic because $\textbf{M}_{A, F}$ fires a t-unit that neither module can fire by itself, thereby reaching the node H.  (e)  There is both synergy and logical obstruction present between $\textbf{M}_{A}$ and $\textbf{M}_{C-1}$, as a logical contradiction with C-1 blocks the complete unfolding of $\textbf{M}_A$; however, the synergy present also allows $\textbf{M}_{A, C-1}$ to include I and J.  (f)  An example of synergy between non-maximal seeds. $\textbf{M}_C$ is subsumed by $\textbf{M}_A$ and $\textbf{M}_K$ is subsumed by $\textbf{M}_{B-0}$.
Because of the contradiction of B-0 and B-1, $|S_{A, B-0}|=3$ (shown in green).  However, if C and K are used as seeds, then a larger module is seen which includes the extra nodes L and M, $|S_{C, K}|=4$ (shown in brown).
This synergy is found in two complex modules, $|S_{C, B-0}|=5$ and $|S_{A, K}|=6$; however, neither of these are core complex modules 
because $\textbf{M}_C$ and $\textbf{M}_K$ are not maximal.
}
\label{fig:interactions}
\end{figure}

\pagebreak

\section{Complex Modules and Dynamical Unfolding}

\begin{table}[]
\begin{adjustwidth}{-1.5cm}{}
\begin{tabular}{lll} \toprule
$\textbf{M}_i$ 
& dynamical unfolding                      & $|S_i|$ \\ \midrule
\textcolor{ProcessBlue}{$\textbf{M}_{\textbf{en-1}}$  ($\Sigma_\textbf{2}$)}   
& EN-1, \textit{ci-0}, \textit{ptc-0}, CI-0, CIR-0, CIA-0, \textit{hh-1}, HH-1                                                               & 9     \\
\textcolor{VioletRed}{$\textbf{M}_{\textbf{nWG-0}}$  ($\Sigma_\textbf{1}$)}   
& \textit{en-0}, EN-0, \textit{ci-1}, \textit{hh-0}, CI-1, HH-0                                                                                                 & 7      \\
\textcolor{VioletRed}{$\textbf{M}_{\textbf{SLP-1}}$  ($\Sigma_\textbf{1}$)}   
& \textit{en-0}, EN-0, \textit{ci-1}, \textit{hh-0}, CI-1, HH-0                                                                                                 & 7      \\
\textcolor{Emerald}{$\textbf{M}_{\textbf{CIR-1}}$}     
& \textit{hh-0}, \textit{wg-0}, \textit{ptc-0}, HH-0, WG-0                                                                                                      & 6     \\
\textcolor{RedOrange}{$\textbf{M}_{\textbf{PTC-0}}$}     
& SMO-1, PH-0, CIR-0                                                                                                      & 4     \\
\textcolor{ForestGreen}{$\textbf{M}_{\textbf{HH-1}}$}  
& SMO-1, CIR-0                                                                                                                       & 3      \\
\textcolor{BurntOrange}{$\textbf{M}_{\textbf{HH-0}}$}   
& PH-0                                                                                                                               & 2      \\ \textcolor{RedViolet}{$\textbf{M}_{\textbf{wg-1}}$} 
& WG-1   & 2      \\
\textcolor{Maroon}{$\textbf{M}_{\textbf{ptc-1}}$} 
& PTC-1   & 2      \\
\textcolor{gray}{$\textbf{M}_{\textbf{nWG-1}}$} 
&  --  & 1      \\
\textcolor{gray}{$\textbf{M}_{\textbf{SLP-0}}$} 
&  -- & 1      \\
\textcolor{gray}{$\textbf{M}_{\textbf{PH-1}}$}  
&   --  & 1      \\
\textcolor{gray}{$\textbf{M}_{\textbf{SMO-0}}$}  
&  --  & 1      \\
\textcolor{gray}{$\textbf{M}_{\textbf{CIA-1}}$}  
&  --  & 1      \\ 
\midrule 

\textcolor{red}{$\textbf{M}_{\textbf{SLP-1,nHH-1}}$  ($\Sigma_\textbf{3}$)}    
& \begin{tabular}[c]{@{}l@{}}SMO-1, CIR-0, \textit{en-0}, EN-0, \textit{ci-1}, \textit{hh-0}, CI-1, HH-0, CIA-1, \\ \textit{wg-1}, \textit{ptc-1}, WG-1, PTC-1, PH-1\end{tabular} & 16     \\
\textcolor{RedOrange}{$\textbf{M}_{\textbf{SLP-1,PTC-0}}$ ($\Sigma_\textbf{5}$)} 
& \begin{tabular}[c]{@{}l@{}} SMO-1, PH-0, CIR-0, \textit{en-0}, EN-0, \textit{ci-1}, \textit{hh-0}, CI-1, HH-0, \\ CIA-1, \textit{wg-1},  \textit{ptc-1}, WG-1 \end{tabular}                                                   & 15     \\
\textcolor{red}{$\textbf{M}_{\textbf{nWG-0,nHH-1}}$  ($\Sigma_\textbf{3}$)}    
& \begin{tabular}[c]{@{}l@{}}SMO-1, CIR-0, \textit{en-0}, EN-0, \textit{ci-1}, \textit{hh-0}, CI-1, HH-0, CIA-1, \\ \textit{ptc-1}, PTC-1, PH-1\end{tabular} & 14     \\
\textcolor{RedOrange}{$\textbf{M}_{\textbf{nWG-0,PTC-0}}$  ($\Sigma_\textbf{5}$)} 
& \begin{tabular}[c]{@{}l@{}}  SMO-1, PH-0, CIR-0, \textit{en-0}, EN-0, \textit{ci-1}, \textit{hh-0}, CI-1, HH-0, \\ CIA-1, \textit{ptc-1}  \end{tabular}                                                & 13     \\
\textcolor{BlueViolet}{$\textbf{M}_{\textbf{nHH-1,en-1}}$  ($\Sigma_\textbf{4}$)} 
& \begin{tabular}[c]{@{}l@{}} SMO-1, CIR-0, EN-1, \textit{hh-1}, \textit{ci-0}, \textit{ptc-0}, HH-1, CI-0, PTC-0, \\ PH-0,  CIA-0 \end{tabular} & 13     \\
\textcolor{ProcessBlue}{$\textbf{M}_{\textbf{SLP-0,nWG-1}}$  ($\Sigma_\textbf{2}$)}   
& \begin{tabular}[c]{@{}l@{}} \textit{en-1}, EN-1, \textit{ci-0}, \textit{ptc-0}, CI-0, CIR-0, CIA-0, \textit{hh-1}, \textit{wg-0}, \\ HH-1, WG-0  \end{tabular}                                                              & 13     \\
\vspace{5.0pt}
\textcolor{Emerald}{$\textbf{M}_{\textbf{nHH-1,CIR-1}}$}    
& SMO-1, \textit{hh-0}, \textit{wg-0}, \textit{ptc-0}, HH-0, WG-0, PTC-0, PH-0   & 10     \\
\textcolor{RawSienna}{$\textbf{M}_{\textbf{nHH-0,ptc-1}}$}  
& PH-0, PTC-1, SMO-0, CIA-0   & 6      \\ \midrule

\textcolor{BlueViolet}{$\textbf{M}_{\textbf{SLP-0,nWG-1,nHH-1}}$  ($\Sigma_\textbf{4}$)} 
& \begin{tabular}[c]{@{}l@{}}SMO-1, CIR-0, \textit{en-1}, EN-1, \textit{hh-1}, \textit{ci-0}, \textit{ptc-0}, HH-1, CI-0, \\ PTC-0, PH-0, CIA-0, \textit{wg-0}, WG-0\end{tabular} & 17     \\
\textcolor{red}{$\textbf{M}_{\textbf{SLP-1,nHH-1,CIR-1}}$}    
& \begin{tabular}[c]{@{}l@{}}SMO-1, \textit{hh-0}, \textit{wg-0}, \textit{ptc-0}, \textit{en-0}, HH-0, WG-0, PTC-0, EN-0, \\ \textit{ci-1}, PH-0, CI-1, CIA-1 \end{tabular} & 16     \\
\textcolor{red}{$\textbf{M}_{\textbf{nWG-0,nHH-1,CIR-1}}$}    
& \begin{tabular}[c]{@{}l@{}} SMO-1, \textit{hh-0}, \textit{wg-0}, \textit{ptc-0}, \textit{en-0}, HH-0, WG-0, PTC-0, EN-0, \\ \textit{ci-1}, PH-0, CI-1, CIA-1 \end{tabular} & 16    \\
\textcolor{Bittersweet}{$\textbf{M}_{\textbf{SLP-1,nHH-0,ptc-1}}$}  
& \begin{tabular}[c]{@{}l@{}} PH-0, PTC-1, \textit{en-0}, SMO-0, CIA-0, EN-0, \textit{ci-1}, \textit{hh-0}, CI-1, \\ HH-0, CIR-1, \textit{wg-0}, WG-0  \end{tabular}    & 16      \\
\textcolor{Bittersweet}{$\textbf{M}_{\textbf{nWG-0,nHH-0,ptc-1}}$}  
& \begin{tabular}[c]{@{}l@{}} PH-0, PTC-1, \textit{en-0}, SMO-0, CIA-0, EN-0, \textit{ci-1}, \textit{hh-0}, CI-1, \\ HH-0, CIR-1, \textit{wg-0}, WG-0  \end{tabular}    & 16    \\
\textcolor{gray}{$\textbf{M}_{\textbf{SLP-0,nHH-0,ptc-1}}$} 
&  PH-0, PTC-1, SMO-0, CIA-0, \textit{wg-0}, WG-0   & 9      \\
\bottomrule
\end{tabular}
\caption{\textbf{List of the core complex modules in the \textit{drosophila} single-cell SPN with pinning perturbation.}  Modules are grouped by seed set size and ordered by 
resulting size
$|S_i|$.  The module labels are colored to correspond to Figure 5 in the main text, as well as supplementary figures \ref{fig:drosophila_modules1}, \ref{fig:drosophila_modules2}, and \ref{fig:drosophila_modules3}; membership in a set of related modules $\Sigma$ is also indicated in parentheses. 
The s-units are listed in the temporal order that they are visited within the dynamical unfolding process.
}
\label{tab:drosophila_modules}
\end{adjustwidth}
\end{table}

\begin{figure}
\begin{adjustwidth}{-2cm}{}
\centering
\includegraphics[width=1.4\columnwidth]{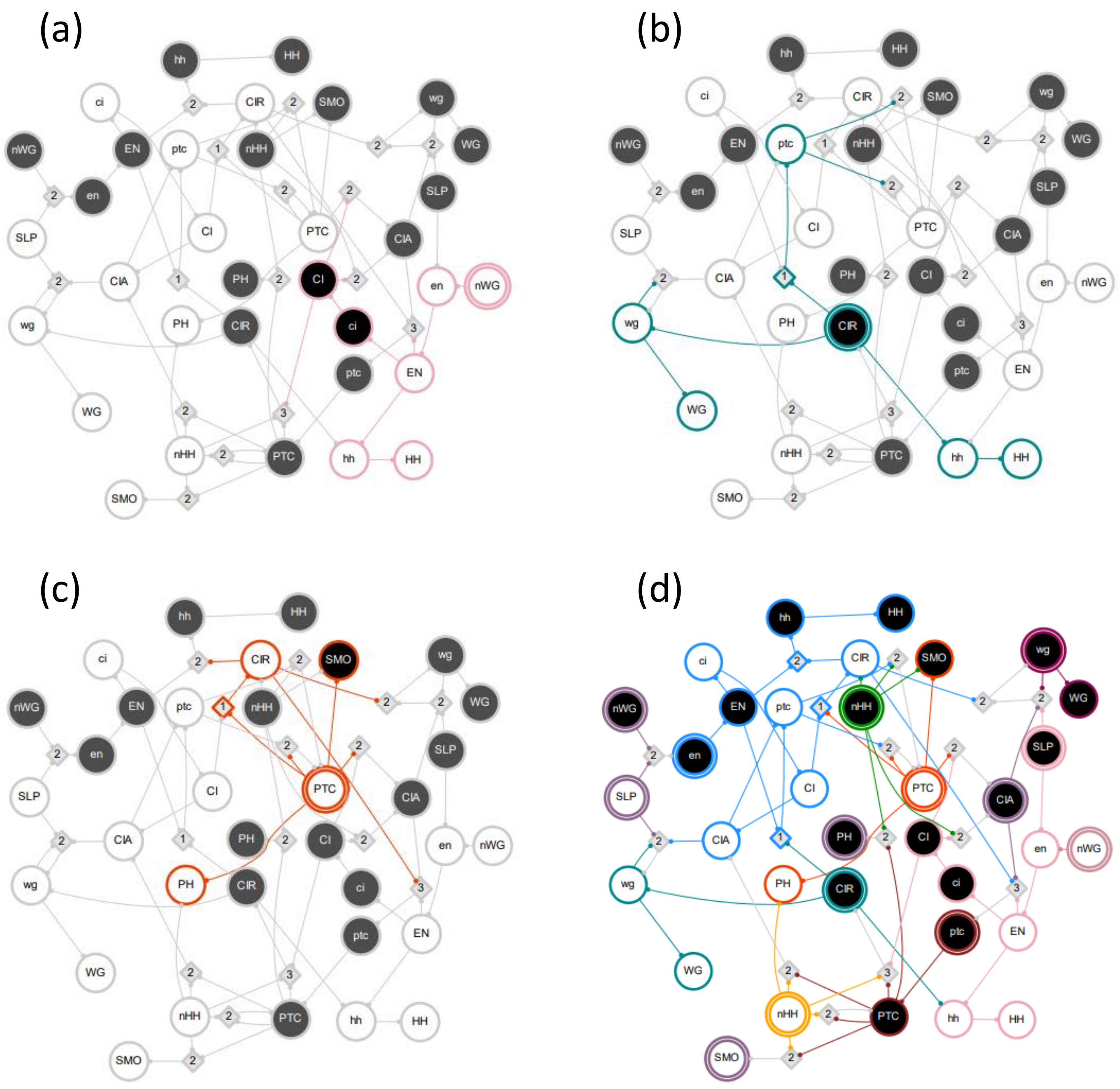}
\caption{\textbf{Complex modules in the \textit{drosophila} single-cell SPN, seed set size $s=1$.}  Select complex modules (indicated by color) are shown: module $\textbf{M}_{nWG-0}$ (a), module $\textbf{M}_{CIR-1}$ (b), module $\textbf{M}_{PTC-0}$ (c), and all 14 complex modules with seed set size $s=1$ (d).  
These modules dynamically unfold from their seed set of size $s=1$ (highlighted with a double edge) to include the s and t-units that are guaranteed to fire given pinning perturbation.  
In panel d, the 14 modules indicated compose $\Lambda_1$ and cover the DCM.  S-units and t-units are colored by the largest module that they are a part of; modules that only contain the seed ($|S_i|=1$) are in purple.
Note that several t-units have been removed or modified in the figure for clarity of the transition function.
}
\label{fig:drosophila_modules1}
\end{adjustwidth}
\end{figure}

\begin{figure}
\begin{adjustwidth}{-2cm}{}
\centering
\includegraphics[width=1.4\columnwidth]{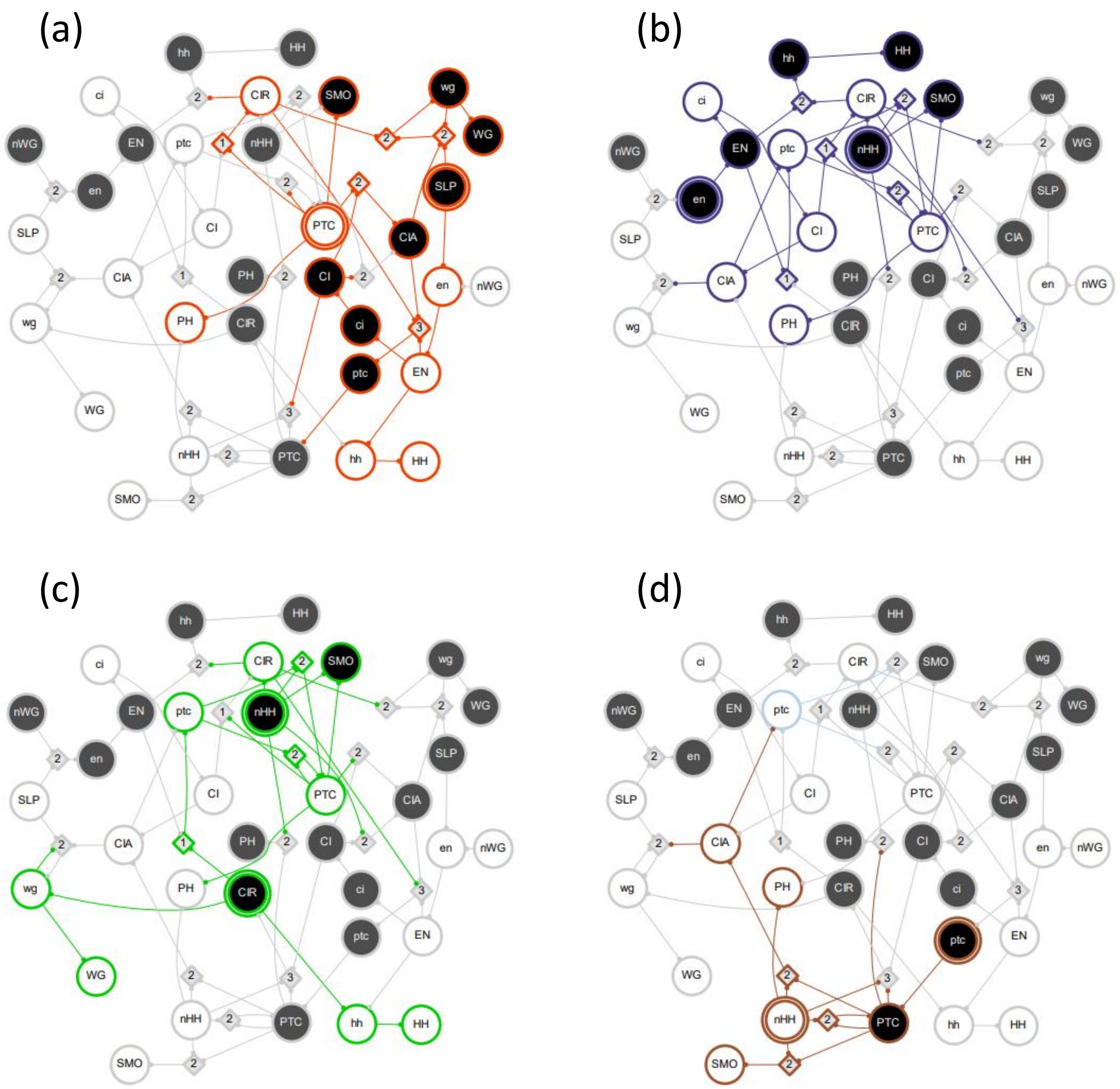}
\caption{\textbf{Complex modules in the \textit{drosophila} single-cell SPN, seed set size $s=2$.}  Select complex modules (indicated by color) are shown: module $\textbf{M}_{SLP-1,PTC-0}$ (a), module $\textbf{M}_{nHH-1,en-1}$ (b), module $\textbf{M}_{nHH-1,CIR-1}$ (c), and module $\textbf{M}_{nHH-0,ptc-1}$ (d).  
These modules dynamically unfold from their seed set of size $s=2$ (highlighted with a double edge) to include the s and t-units that are guaranteed to fire given pinning perturbation.
In panel d, note that s-unit \textit{ptc-0} does not fire with pinning perturbation due to contradiction with \textit{ptc-1}.
Note that several t-units have been removed or modified in the figure for clarity of the transition function.
}
\label{fig:drosophila_modules2}
\end{adjustwidth}
\end{figure}

\begin{figure}
\begin{adjustwidth}{-2cm}{}
\centering
\includegraphics[width=1.4\columnwidth]{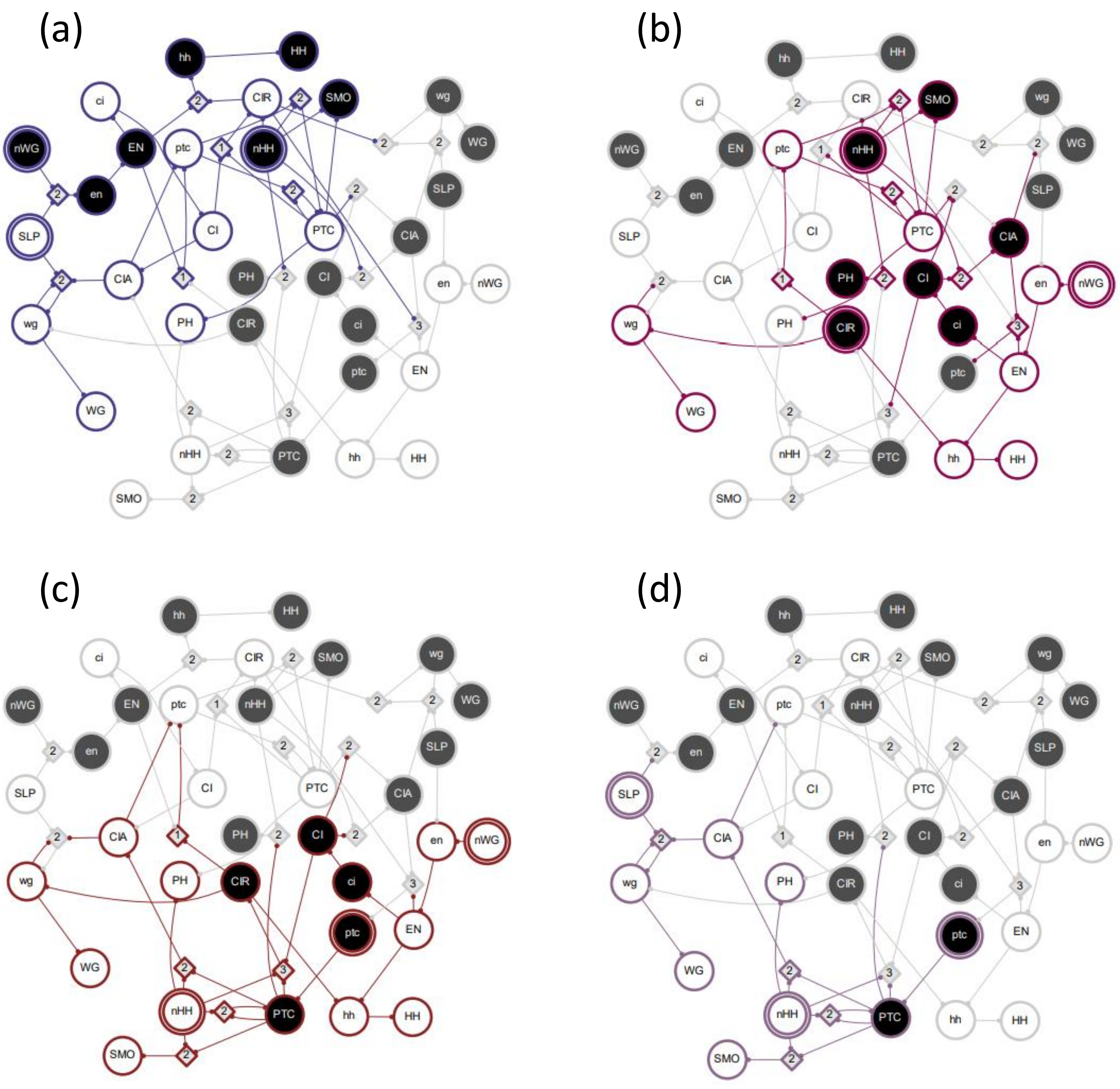}
\caption{\textbf{Complex modules in the \textit{drosophila} single-cell SPN, seed set size $s=3$.}  Select complex modules (indicated by color) are shown: module $\textbf{M}_{SLP-0,nWG-1,nHH-1}$ (a), module $\textbf{M}_{nWG-0,nHH-1,CIR-1}$ (b), module $\textbf{M}_{nWG-0,nHH-0,ptc-1}$ (c), and module $\textbf{M}_{SLP-0,nHH-0,ptc-1}$ (d).  
These modules dynamically unfold from their seed set of size $s=3$ (highlighted with a double edge) to include the s and t-units that are guaranteed to fire given pinning perturbation.
In panel a, the indicated module resolves all 17 variable states and results in one of the network's 10 attractors.
In panel b, the indicated module is identical to $\textbf{M}_{SLP-1,nHH-1,CIR-1}$ but differs only in the seed.
In panel c, the indicated module is identical to $\textbf{M}_{SLP-1,nHH-0,ptc-1}$ but differs only in the seed.
Note that several t-units have been removed or modified in the figure for clarity of the transition function.
}
\label{fig:drosophila_modules3}
\end{adjustwidth}
\end{figure}

\begin{figure}
\begin{adjustwidth}{-2cm}{}
\centering
\includegraphics[width=1.4\columnwidth]{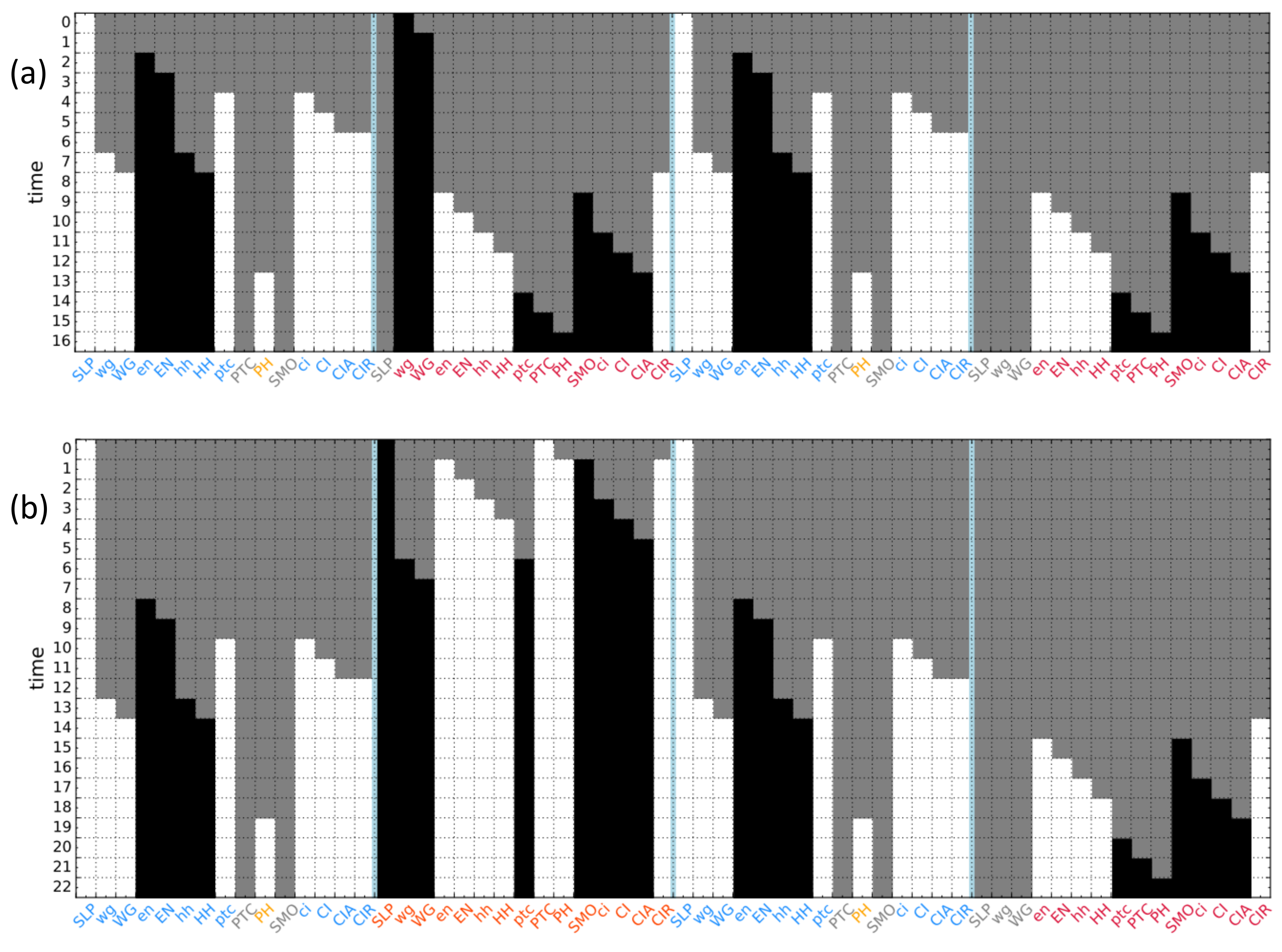}
\caption{\textbf{Dynamical unfolding in the \textit{drosophila} parasegment SPN.}  Cell boundaries are separated by a blue line; white indicates that a node is OFF in that time step, black indicates that a node is ON, and grey indicates that the node's state is unknown.  The color of the node label indicates the associated 
complex module.  
(a) The dynamical unfolding of SLP$_1=$ SLP$_3=0$, \textit{wg}$_2=1$. The presence of WG-1 in cell 2 allows $\textbf{M}_{SLP-0,nWG-1}$ (labeled as blue) to unfold in cells 1 and 3 by acting as an additional input to SLP-0.  These modules turn off WG and turn on HH in cells 1 and 3, which initiates $\textbf{M}_{nWG-0,nHH-1}$ in cells 2 and 4 (labeled as red).  Finally $\textbf{M}_{nWG-0,nHH-1}$ 
in cells 2 and 4
turns off HH which initiates the (small) $\textbf{M}_{nHH-0}$ module in cells 1 and 3 (labeled as yellow).  This resolves almost the entire network (52 variables) with only three inputs.  
(b) The dynamical unfolding of SLP$_1=$ SLP$_3=$ PTC$_2=0$, SLP$_2=1$ is almost identical, except instead of \textit{wg}-1, SLP-1 and PTC-0 are pinned in cell 2.  $\textbf{M}_{SLP-1,PTC-0}$ (labeled as orange) similarly turns on WG, initiating the $\textbf{M}_{SLP-0,nWG-1}$ module in cells 1 and 3.  However, the time differences are more pronounced.  The unfolding can be seen in three distinct epochs where first $\textbf{M}_{SLP-1,PTC-0}$ must unfold in cell 2, then the $\textbf{M}_{SLP-0,nWG-1}$ modules are initiated in cells 1 and 3, and then $\textbf{M}_{nWG-0,nHH-1}$ is initiated in cell 4.  This demonstrates how important timing is in the dynamic interplay 
between cells.}
\label{fig:unfoldings}
\end{adjustwidth}
\end{figure}

\begin{table}
\begin{tabular}{llll} \toprule
$|S_i^0|$ & $S_i^0$    & $|S_i|$ & \begin{tabular}[c]{@{}l@{}}associated \\ modules\end{tabular} \\ \midrule
$s=1$ & \textit{en}-1$_i$                                           & 13          & \textcolor{ProcessBlue}{$\textbf{M}_\textbf{2}$}/\textcolor{ForestGreen}{$\textbf{M}_\textbf{3}$}                                                         \\ \midrule
$s=2$ & \begin{tabular}[c]{@{}l@{}}(\textit{en}-1$_i$, \textit{en}-1$_{i+1}$),\\ (CIR-1$_i$, CIR-1$_{i+2}$)\end{tabular}       & 28          & \begin{tabular}[c]{@{}l@{}}\textcolor{BlueViolet}{$\textbf{M}_\textbf{8}$}/\textcolor{ForestGreen}{$\textbf{M}_\textbf{3}$} ,\\ \textcolor{Emerald}{$\textbf{M}_\textbf{5}$}/\textcolor{VioletRed}{$\textbf{M}_\textbf{1}$}/\textcolor{BurntOrange}{$\textbf{M}_\textbf{4}$} \end{tabular} \\ \midrule
$s=3$ & \{ (SLP-0$_i$, SLP-0$_{i+2}$, \textit{wg}-1$_{i+1}$) \} = X                                          & 52          & \textcolor{ProcessBlue}{$\textbf{M}_\textbf{7}$}/\textcolor{Red}{$\textbf{M}_\textbf{6}$}/\textcolor{BurntOrange}{$\textbf{M}_\textbf{4}$}                                                     \\ \midrule
$s=4$ & \begin{tabular}[c]{@{}l@{}}\{ (X + \textit{en}-1$_{i+1}$), \\ (X + \textit{en}-1$_{i+3}$)  \} = Y\end{tabular}           & 56          & \begin{tabular}[c]{@{}l@{}}\textcolor{BlueViolet}{$\textbf{M}_\textbf{8}$/$\textbf{M}_{\textbf{10}}$}/\textcolor{Red}{$\textbf{M}_\textbf{6}$}, \\ \textcolor{BlueViolet}{$\textbf{M}_\textbf{8}$/$\textbf{M}_{\textbf{10}}$}/\textcolor{Red}{$\textbf{M}_\textbf{6}$} \end{tabular}    \\ \midrule
$s=5$ & \begin{tabular}[c]{@{}l@{}}\{ (SLP-1$_i$, SLP-1$_{i+1}$, SLP-0$_{i+2}$, \textit{wg}-1$_{i+3}$, \textit{en}-1$_{i+3}$),\\ (SLP-1$_i$, SLP-1$_{i+1}$, SLP-0$_{i+3}$, \textit{wg}-1$_{i+2}$, \textit{en}-1$_{i+2}$),\\ (Y + SLP\_1$_{i+3}$),\\ (Y + SLP\_0$_{i+3}$),\\ (SLP-0$_i$, SLP-0$_{i+1}$, SLP-1$_{i+2}$, \textit{wg}-1$_{i+3}$, \textit{en}-1$_{i+3}$),\\ (SLP-0$_i$, SLP-0$_{i+1}$, SLP-1$_{i+3}$, \textit{wg}-1$_{i+2}$, \textit{en}-1$_{i+2}$) \} = Z\end{tabular} & 59          & \begin{tabular}[c]{@{}l@{}}\textcolor{BlueViolet}{$\textbf{M}_\textbf{8}$/$\textbf{M}_{\textbf{10}}$}/\textcolor{Red}{$\textbf{M}_\textbf{6}$} \\ ... \end{tabular}            \\ \midrule
$s=6$ & \begin{tabular}[c]{@{}l@{}}(Z + SLP-0$_j$),\\ (Z + SLP-1$_j$) \\ (\textit{en}-1$_i$, PTC-0$_{i+2}$, SLP-0$_i$, SLP-0$_{i+1}$, SLP-1$_{i+2}$, SLP-1$_{i+3}$) \\
(\textit{en}-1$_i$, PTC-0$_{i+2}$, SLP-0$_i$, SLP-1$_{i+1}$, SLP-1$_{i+2}$, SLP-0$_{i+3}$) \\
(\textit{en}-1$_i$, PTC-0$_{i+2}$, SLP-0$_i$, SLP-0$_{i+1}$, SLP-1$_{i+2}$, SLP-0$_{i+3}$)    \end{tabular}                   

& 60          & \begin{tabular}[c]{@{}l@{}} \textcolor{BlueViolet}{$\textbf{M}_\textbf{8}$/$\textbf{M}_{\textbf{10}}$}/\textcolor{Red}{$\textbf{M}_\textbf{6}$} \\ ... \\ \textcolor{BlueViolet}{$\textbf{M}_{\textbf{10}}$}/\textcolor{Red}{$\textbf{M}_\textbf{6}$}/\textcolor{RedOrange}{$\textbf{M}_\textbf{9}$} \end{tabular}  \\ \bottomrule                                
\end{tabular}
\caption{\textbf{Maximal modules in the \textit{drosophila} parasegment SPN.}  The seed set $S_i^0$ and size $|S_i|$ of the maximal 
pathway module per seed set size $s=|S_i^0|$ are shown, along with the associated intracellular 
complex modules 
found within the maximal module.  Here, $\textbf{M}_1$ indicates $\textbf{M}_{nWG-0}$; $\textbf{M}_2$ indicates $\textbf{M}_{en-1}$; $\textbf{M}_3$ indicates $\textbf{M}_{nHH-1}$; $\textbf{M}_4$ indicates $\textbf{M}_{nHH-0}$; $\textbf{M}_5$ indicates $\textbf{M}_{CIR-1}$; $\textbf{M}_6$ indicates $\textbf{M}_{SLP-1,nHH-1}$; $\textbf{M}_7$ indicates $\textbf{M}_{SLP-0,nWG-1}$; $\textbf{M}_8$ indicates $\textbf{M}_{nHH-1,en-1}$; $\textbf{M}_9$ indicates $\textbf{M}_{PTC-0}$; finally, $\textbf{M}_{10}$ indicates $\textbf{M}_{SLP-0,nWG-1,nHH-1}$.
For $s=5$, all maximal 
pathway modules have the same associated 
intracellular modules; for $s=6$, the first two groupings of seed sets have associated modules $\textbf{M}_8$, $\textbf{M}_{10}$, and $\textbf{M}_6$, while the latter three groupings have associated modules $\textbf{M}_{10}$, $\textbf{M}_{6}$, and $\textbf{M}_9$.
The subscript 
on the s-unit labels
denotes the cell of the parasegment where $i$ and $j$ are arbitrary cells chosen such that there is no logical contradiction between variable states and cellular addition uses modular arithmetic.  As maximal modules with larger seed sets tend to build on the same 
set of 
modules with smaller seed sets, particular seed sets are condensed to variables for readability.  Given the periodic boundaries of the parasegment and the similarity of complex modules, a significant degree of 
redundancy in regards to cellular indices can be seen in the maximal module sets.
A seed set of size $s \geq 6$ is needed to resolve every variable in the network.}
\label{tab:parasegment_maximal_modules}
\end{table}

\begin{figure}
\begin{adjustwidth}{-2cm}{}
\centering
\includegraphics[width=1.4\columnwidth]{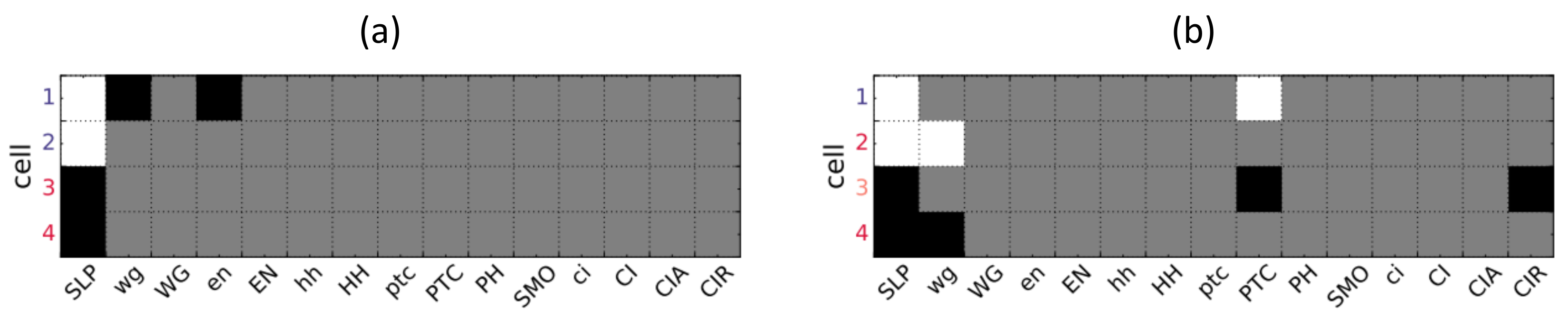}
\caption{\textbf{Minimal schemata 
for attractor control in the \textit{drosophila} parasegment SPN.}
(a) One of the minimal schemata found to reach an attractor, representing the minimal seeds sufficient to 
drive the dynamics to that attractor with pinning control.  
Each row represents a cell in the \textit{drosophila} parasegment and each column the state of the respective node: black represents ON, white represents OFF, and grey represents a wildcard 
value (i.e., the node
state is redundant for convergence 
to the attractor
under pinning perturbation of the 
seeds).
The color of the row label represents the main module group active in that cell during unfolding (pink represents $\Sigma_1$, red represents $\Sigma_3$, and violet represents $\Sigma_4$,
see Table \ref{tab:drosophila_modules}). 
(b) Same as panel a, but for the minimal schema sufficient to reach the wildtype attractor.  The minimal configurations found are much smaller than the previous estimates without pinning perturbation found in \cite{marques2013canalization}.}
\label{fig:minimal_schemata}
\end{adjustwidth}
\end{figure}

\begin{figure}
\begin{adjustwidth}{-2cm}{}
\centering
\includegraphics[width=1.4\columnwidth]{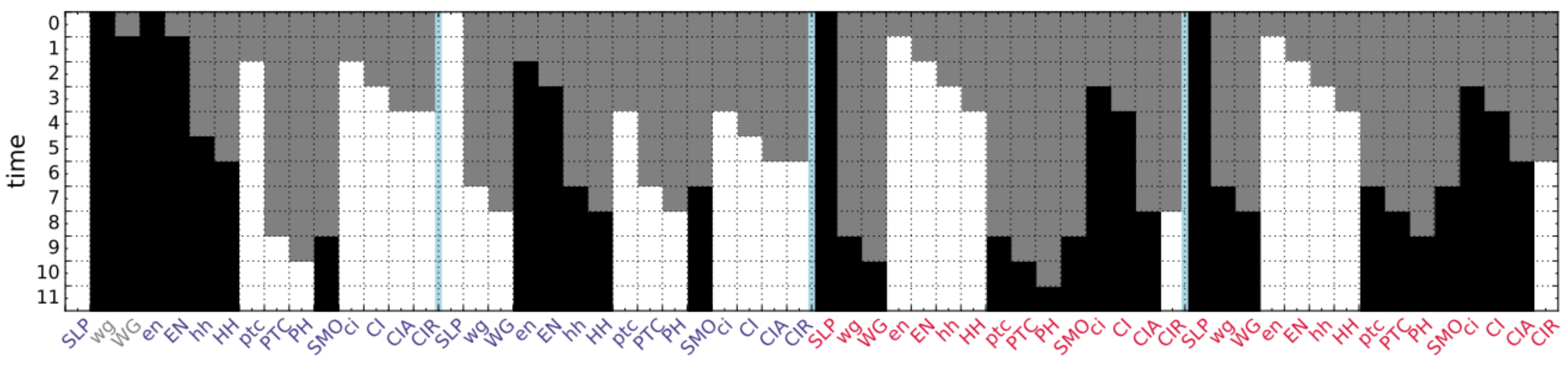}
\caption{\textbf{Dynamical unfolding of a minimal seed set to reach an attractor
in the \textit{drosophila} parasegment SPN.}  
Cell boundaries are separated by a blue line; white indicates that a node is OFF in that time step, black indicates that a node is ON, and grey indicates that the node's state is unknown.  The color of the node label indicates the associated complex module.  All four SLP inputs are pinned; additionally, \textit{wg} and \textit{en} are ON in the first cell.  $\textbf{M}_{en-1}$ unfolds in cell 1, and \textit{wg}-1 initiates $\textbf{M}_{SLP-0,nWG-1}$ in cell 2 by acting as an input along with SLP-0.  This turns on HH in both cells which extends the unfolding in each cell via additional synergy (modules $\textbf{M}_{en-1,nHH-1}$ and $\textbf{M}_{SLP-0,nWG-1,nHH-1}$, labeled as violet).  It also extends the unfolding of $\textbf{M}_{SLP-1}$ in cells 3 and 4 (module $\textbf{M}_{SLP-1,nHH-1}$, labeled as red).
This final configuration is in the basin of attraction of the broad stripes attractor \cite{albert2003topology}, and the network will reach this attractor once the seeds \textit{wg} and \textit{en} in cell 1 are no longer pinned.
}
\label{fig:minimal_unfolding}
\end{adjustwidth}
\end{figure}

\begin{figure}
\centering
\includegraphics[width=1.1\columnwidth]{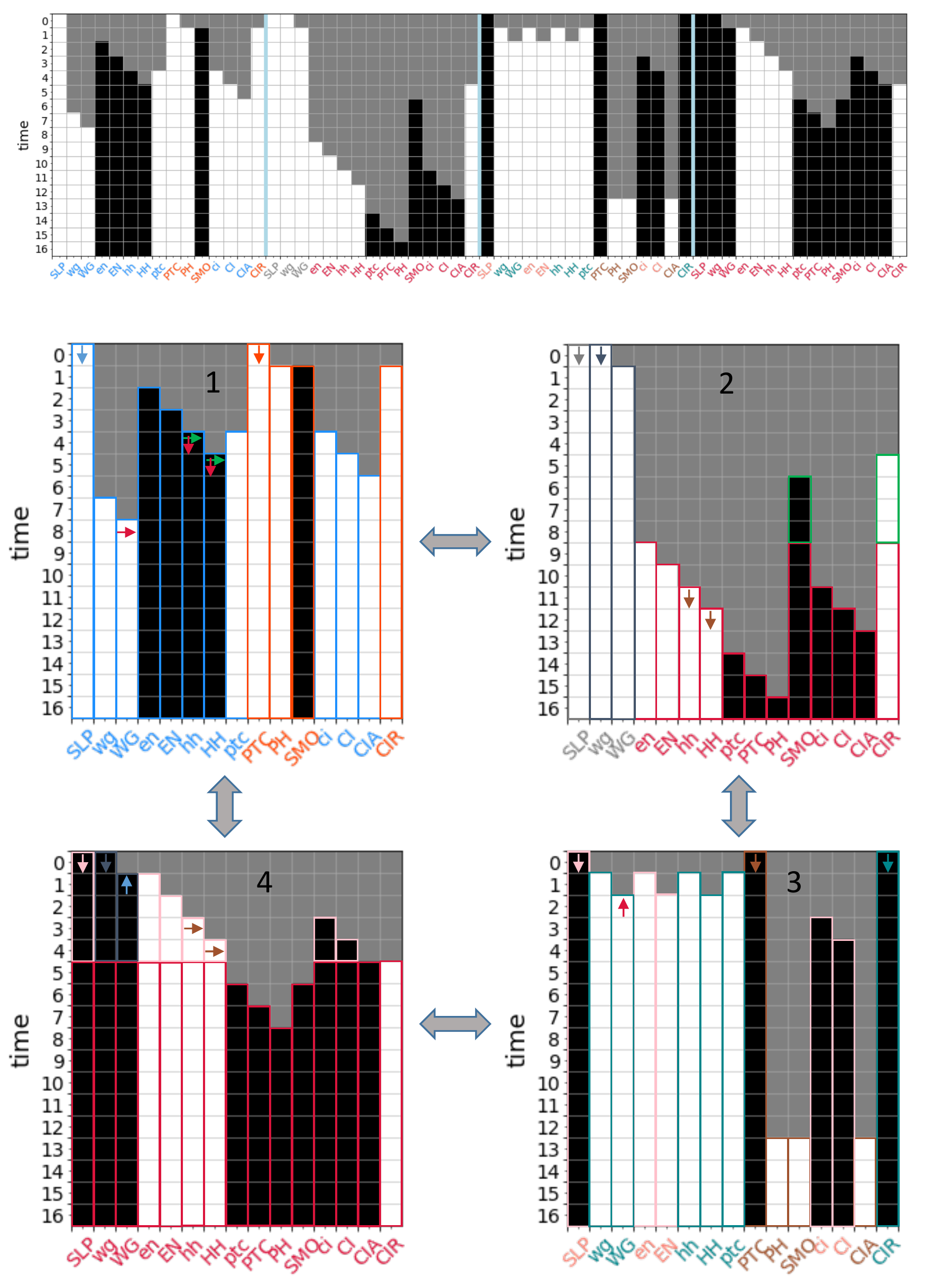}
\caption{\textbf{Dynamical unfolding of the wildtype attractor in the \textit{drosophila} parasegment SPN.} 
This figure is the same as Fig.~6 in the main text except that the full unfolding of the minimal seed set to the wildtype attractor is shown in terms of intracellular complex modules.
$\textbf{M}_{SLP-0}$ is present in cell 1 based on initial conditions (blue arrow).  The influence of WG-1 in cell 4 (blue arrow) initiates $\textbf{M}_{SLP-0, nWG-1}$ in cell 1 (shown by blue borders). This module turns on \textit{hh} and HH which initiates $\textbf{M}_{nHH-1}$ in cell 2 (shown in green) and then turns off WG in cell 1 which, together with WG-0 in cell 3 (indicated by red arrows), initiates $\textbf{M}_{nWG-0}$ in cell 2; the synergy between $\textbf{M}_{nHH-1}$ and $\textbf{M}_{nWG-0}$ results in $\textbf{M}_{nHH-1, nWG-0}$ in cell 2 
(which subsumes both 
modules and is shown by red borders).
Based on initial conditions, $\textbf{M}_{PTC-0}$ is also present in cell 1, $\textbf{M}_{SLP-0}$ and $\textbf{M}_{wg-0}$ are present in cell 2, $\textbf{M}_{SLP-1}$ 
and $\textbf{M}_{CIR-1}$ 
are present in cell 3, and $\textbf{M}_{SLP-1}$ and $\textbf{M}_{wg-1}$ are present in cell 4.  $\textbf{M}_{SLP-1}$ in cell 4 (shown by pink borders) turns off \textit{hh} and HH which, together with \textit{hh}-0 and HH-0 in cell 2 (indicated by brown arrows), initiates $\textbf{M}_{nHH-0}$ in cell 3, which is subsumed by $\textbf{M}_{nHH-0,PTC-1}$ (shown by brown borders).  Finally, the presence of \textit{hh}-1 and HH-1 in cell 1 initiates $\textbf{M}_{nHH-1}$  in cell 4 
, which is subsumed by $\textbf{M}_{SLP-1,nHH-1}$ (shown in red). 
}
\label{fig:wildtype_unfolding_full}
\end{figure}

\pagebreak

\section{\textit{Drosophila} Observability and Control}

The single-cell SPN is small enough that the attractors of the network can be fully enumerated.  Analysis of the dynamical activity within each of the 10 attractors corresponds well with the analysis of complex modules, as each attractor can be associated with one or more modules (Table \ref{tab:control_sets}).
The $\Sigma_4$ module $\textbf{M}_{SLP-0,nWG-1,nHH-1}$ is seen in only one attractor as it requires an AND condition between all three inputs, resulting in the driver set (SLP=0, nWG=1, nHH=1).
The $\Sigma_3$ modules, $\textbf{M}_{SLP-1,nHH-1}$ and $\textbf{M}_{nWG-0,nHH-1}$, are seen in four attractors: the former module resolves every variable except for the input nWG and therefore leads to two attractors depending on the state of this input with driver sets (SLP=1, nWG=1, nHH=1) and (SLP=1, nWG=0, nHH=1); the latter module resolves every variable except for SLP, \textit{wg}, and WG and can lead to three attractors, with driver sets (SLP=1, nWG=0, nHH=1) (seen above), (SLP=0, nWG=0, nHH=1, \textit{wg}=0), and (SLP=0, nWG=0, nHH=1, \textit{wg}=1).
The modules $\textbf{M}_{SLP-1,nHH-0,PTC-1}$ and $\textbf{M}_{nWG-0,nHH-0,PTC-1}$ are identical except for their seeds and lead to three attractors, as both modules resolve every variable state except for the missing input 
(which my be in either state)
\footnote{These are complex modules but not core complex modules as PTC is not a  maximal seed (because $\textbf{M}_{ptc-1}$ subsumes $\textbf{M}_{PTC-1}$).}.  The driver sets for the three attractors are therefore (SLP=1, nWG=0, nHH=0, PTC=1), (SLP=1, nWG=1, nHH=0, PTC=1), and (SLP=0, nWG=0, nHH=0, PTC=1).
The final two attractors require $\textbf{M}_{SLP-0,nWG-1}$ in combination with other modules; one requires $\textbf{M}_{nHH-0}$ and $\textbf{M}_{PTC-0}$, while the other requires $\textbf{M}_{nHH-0,PTC-1}$ (note that there is synergy between nHH-0 and PTC-1 and that this module is also a submodule of $\textbf{M}_{SLP-1,nHH-0,PTC-1}$ and $\textbf{M}_{nWG-0,nHH-0,PTC-1}$, meaning that it appears in four attractors).
The driver sets for these two attractors are therefore (SLP=0, nWG=1, nHH=0, PTC=0) and (SLP=0, nWG=1, nHH=0, PTC=1).
Interestingly, the superset of the nodes associated with the minimal driver sets found via this methodology are the same driver nodes predicted by feedback vertex set theory (inputs SLP, nWG, nHH and internal nodes PTC and \textit{wg}).

The characterization of dynamical modules here highlights the difference between the driver node set with pinning and pulse perturbation control.  If 
a driver node's state must be constant, as with pinning perturbations, PTC and \textit{wg} are needed to fully control the network.  However, if one is allowed to flip the states of the input variables, then no internal nodes need to be pinned.  
For example, to reach the attractor that requires $\textbf{M}_{SLP-0,nWG-1}$, $\textbf{M}_{nHH-0}$, and $\textbf{M}_{PTC-0}$, the module $\textbf{M}_{SLP-0,nWG-1,nHH-1}$ can be used which turns off PTC (and maintains PTC in the OFF state), and nHH can then be switched OFF to reach the attractor.
Furthermore, modules $\textbf{M}_{SLP-1,nHH-1}$ and $\textbf{M}_{nWG-0,nHH-1}$ turn off \textit{ptc}; if the state of nHH is then flipped, the module $\textbf{M}_{nHH-0,PTC-1}$ will unfold, as is seen in four attractors.
In general, perturbing nHH along with the other input
nodes replaces the need to control PTC as nHH is an input to PTC.  Perturbing the internal node \textit{wg}, meanwhile, allows to turn on or off \textit{wg} and WG, when SLP=0, nWG=0, and nHH=1.
However, SLP can also be turned ON 
after $\textbf{M}_{nWG-0,nHH-1}$ unfolds to turn on \textit{wg} and WG and then turned OFF.  Alternatively, SLP can be turned OFF while nWG and nHH are ON to initiate $\textbf{M}_{SLP-0,nWG-1,nHH-1}$, which turns off \textit{wg} and WG; then nWG can be flipped to initiate $\textbf{M}_{nWG-0,nHH-1}$.  Therefore, perturbing SLP and nWG replaces the need to control \textit{wg}.  In a sense, pinning the internal nodes then is just a shortcut to replace the need for input 
node perturbations.  This highlights the usefulness of 
complex modules
in developing a qualitative understanding of control in a system.

Perhaps unsurprisingly, the same intracellular modules that lead to attractors in the single-cell SPN lead to attractors in the parasegment SPN, when considering six biologically-relevant attractors from \cite{albert2003topology}.
In particular, $\textbf{M}_{SLP-0,nWG-1}$, $\textbf{M}_{nWG-0,nHH-1}$, $\textbf{M}_{SLP-1,nHH-1}$, and $\textbf{M}_{SLP-1,nHH-0,PTC-1}$ each play a significant role, appearing in at least four of the attractors
(see Table \ref{tab:control_sets_parasegment}).

The logical inferences made during the dynamical unfolding process are also useful in the problem of observability.  If a network has reached a fixed point, then each variable state is unchanging (i.e., pinned); thus, observing a set of sensor nodes with corresponding s-units $S_i^0$ guarantees that the module $\textbf{M}_{S_i^0}$ has fully unfolded and all the variable states contained in $S_i$ are present in the current state of the  network.  Observing only a small subset of such states (2-3 in the case of the \textit{drosophila} single-cell SPN, equivalent to less than $18\%$ of the network size) is sufficient to figure out which attractor the network is in.  The size of this minimal observability set is naturally upper-bounded by the size of the minimal driver set.
For the \textit{drosophila} parasegment SPN, the minimal observability sets require only two s-units ($3\%$ of the network size) when considering only the six biologically-relevant attractors in Table \ref{tab:control_sets_parasegment}.

\begin{table}[]
\begin{adjustwidth}{-3cm}{}
\centering
\resizebox{1.5\columnwidth}{!}{%
\begin{tabular}{llcc} \toprule
Attractor         & 
Minimal observability sets & 
Minimal control set & Associated modules \\ \midrule
11100001111111010 & \begin{tabular}[c]{@{}l@{}}\{ (nHH-1, SLP-1, nWG-0),\\ (SLP-1, \textit{ptc-1}, nWG-0),\\ (\textit{wg-1}, SLP-1, nWG-0),\\ (SLP-1, nWG-0, WG-1),\\ (CIR-0, SLP-1, nWG-0),\\ (SLP-1, nWG-0, CIA-1),\\ (SLP-1, nWG-0, PH-1),\\ (SLP-1, SMO-1, nWG-0) \}\end{tabular}                           & (nHH-1, SLP-1, nWG-0)                                         & \textcolor{Red}{$\textbf{M}_{\textbf{6}}$,$\textbf{M}_{\textbf{11}}$}                \\ \midrule
01100001111111010 & 
\{(\textit{wg}-1, SLP-0), (SLP-0, WG-1)\}
& (nHH-1, nWG-0, \textit{wg-1}, SLP-0)                                   & \textcolor{Red}{$\textbf{M}_{\textbf{11}}$},\textcolor{RedViolet}{$\textbf{M}_{\textbf{wg-1}}$},\textcolor{gray}{$\textbf{M}_{\textbf{SLP-0}}$}                \\ \midrule
11100001111111011 & \begin{tabular}[c]{@{}l@{}}\{ (nWG-1, \textit{wg-1}), (\textit{ptc-1}, nWG-1),\\ (nWG-1, PH-1), (nWG-1, CIA-1),\\ (nWG-1, WG-1) \}\end{tabular}                      & (nHH-1, nWG-1, SLP-1)                                         & \textcolor{Red}{$\textbf{M}_{\textbf{6}}$},\textcolor{gray}{$\textbf{M}_{\textbf{nWG-1}}$}                \\ \midrule
00011110001000011 & \begin{tabular}[c]{@{}l@{}}\{ (nHH-1, PTC-0), (nHH-1, \textit{en-1}),\\ (EN-1, nHH-1), (nHH-1, \textit{ci-0}),\\ (CI-0, nHH-1), (nHH-1, CIA-0),\\ (nHH-1, \textit{ptc-0}), (nHH-1, \textit{hh-1}),\\ (HH-1, nHH-1), (nHH-1, PH-0) \}\end{tabular}                                                   & (nHH-1, nWG-1, SLP-0)                                         & \textcolor{BlueViolet}{$\textbf{M}_{\textbf{10}}$}             \\ \midrule
10000000100110101 & \{(CIR-1, nWG-1)\}                                      & (SLP-1, nHH-0, PTC-1, nWG-1)                                  & \textcolor{Bittersweet}{$\textbf{M}_{\textbf{13}}$},\textcolor{gray}{$\textbf{M}_{\textbf{nWG-1}}$} 
\\ \midrule
00011110001000001 & 
\{(nHH-0, PTC-0), (SMO-1, nHH-0)\}
& (nWG-1, PTC-0, nHH-0, SLP-0)                                  & \textcolor{ProcessBlue}{$\textbf{M}_\textbf{7}$}, \textcolor{RedOrange}{$\textbf{M}_{\textbf{9}}$}, \textcolor{BurntOrange}{$\textbf{M}_\textbf{4}$}  
\\ \midrule
00000001111111010 & \begin{tabular}[c]{@{}l@{}}\{ (\textit{wg-0}, \textit{ptc-1}), (\textit{ptc-1}, WG-0),\\ (\textit{wg-0}, PH-1), (\textit{wg-0}, CIA-1),\\ (WG-0, PH-1), (WG-0, CIA-1) \}\end{tabular}                                                & (nWG-0, nHH-1, SLP-0, \textit{wg-0})                                   & \textcolor{Red}{$\textbf{M}_{\textbf{11}}$},\textcolor{RedViolet}{$\textbf{M}_{\textbf{wg-0}}$},\textcolor{gray}{$\textbf{M}_{\textbf{SLP-0}}$}                \\ \midrule
00011110100000001 & \begin{tabular}[c]{@{}l@{}}\{ (PTC-1, \textit{en-1}),  (SMO-0, \textit{en-1}),\\ (SMO-0, EN-1), (PTC-1, EN-1),\\ (PTC-1, \textit{ci-0}), (SMO-0, \textit{ci-0}),\\ (CI-0, SMO-0), (PTC-1, CI-0),\\ (PTC-1, \textit{hh-1}), (SMO-0, \textit{hh-1}),\\ (PTC-1, HH-1), (CIR-0, SMO-0),\\ (HH-1, SMO-0) \}\end{tabular} & (nWG-1, PTC-1, SLP-0, nHH-0)                                  & \textcolor{ProcessBlue}{$\textbf{M}_\textbf{7}$}, \textcolor{RawSienna}{$\textbf{M}_{\textbf{12}}$}        \\ \midrule
00000000100110100 & \{(CIR-1, SLP-0)\}                                           & (nWG-0, nHH-0, PTC-1, SLP-0)              &  \textcolor{Bittersweet}{$\textbf{M}_{\textbf{14}}$},\textcolor{gray}{$\textbf{M}_{\textbf{SLP-0}}$}   
\\ \midrule
10000000100110100 & \begin{tabular}[c]{@{}l@{}}\{ (nWG-0, SLP-1, \textit{wg-0}),\\ (nWG-0, nHH-0, SLP-1),\\ (CIR-1, SLP-1, nWG-0),\\ (SLP-1, nWG-0, CIA-0),\\ (SLP-1, SMO-0, nWG-0),\\ (SLP-1, nWG-0, WG-0),\\ (SLP-1, nWG-0, PH-0),\\ (SLP-1, \textit{ptc-0}, nWG-0) \}\end{tabular}                           & (SLP-1, nWG-0, nHH-0, PTC-1)      &  \textcolor{Bittersweet}{$\textbf{M}_{\textbf{13}}$,$\textbf{M}_{\textbf{14}}$} 
\\ \bottomrule     
\end{tabular} }
\caption{\textbf{Minimal observability and control sets for the \textit{drosophila} single-cell SPN.}  
The attractors label the node states in order of SLP, \textit{wg}, WG, \textit{en}, EN, \textit{hh}, HH, \textit{ptc}, PTC, PH, SMO, \textit{ci}, CI, CIA, CIR, nHH, nWG.
The associated complex modules are colored to correspond to Table \ref{tab:drosophila_modules}.
Here, 
$\textbf{M}_4$ indicates $\textbf{M}_{nHH-0}$; 
$\textbf{M}_6$ indicates $\textbf{M}_{SLP-1,nHH-1}$; $\textbf{M}_7$ indicates $\textbf{M}_{SLP-0,nWG-1}$; 
$\textbf{M}_9$ indicates $\textbf{M}_{PTC-0}$, $\textbf{M}_{10}$ indicates $\textbf{M}_{SLP-0,nWG-1,nHH-1}$,
$\textbf{M}_{11}$ indicates $\textbf{M}_{nWG-0,nHH-1}$, $\textbf{M}_{12}$ indicates $\textbf{M}_{nHH-0,PTC-1}$, $\textbf{M}_{13}$ indicates $\textbf{M}_{SLP-1,nHH-0,PTC-1}$; finally, $\textbf{M}_{14}$ indicates $\textbf{M}_{nWG-0,nHH-0,PTC-1}$.
}
\label{tab:control_sets}
\end{adjustwidth}
\end{table}

\begin{table}[]
\begin{adjustwidth}{-3cm}{}
\centering
\resizebox{1.5\textwidth}{!}{%
\begin{tabular}{llll} \toprule
Attractor         & Expressed nodes                                                  & Minimal observability sets                                                    & Associated modules                                                                                 \\ \midrule
Wild-type         & 
\begin{tabular}[c]{@{}l@{}}(\textit{wg}$_4$; WG$_4$; \textit{en}$_1$; EN$_1$; \textit{hh}$_1$; \\ HH$_1$; \textit{ptc}$_{2,4}$;  PTC$_{2,3,4}$; \textit{ci}$_{2,3,4}$; \\ CI$_{2,3,4}$; CIA$_{2,4}$; CIR$_3$)\end{tabular} & \begin{tabular}[c]{@{}l@{}}\{ (CIR$_3$-1, PTC$_1$-0),\\ (PTC$_1$-0, \textit{ptc}$_2$-1),\\ (CIA$_2$-1, PTC$_1$-0),\\ (PH$_2$-1, PTC$_1$-0),\\ (PTC$_1$-0, SMO$_3$-0) \}\end{tabular}      
& \begin{tabular}[c]{@{}l@{}}\textcolor{ProcessBlue}{$\textbf{M}_\textbf{7}$}/\textcolor{RedOrange}{$\textbf{M}_{\textbf{9}}$},\\   \textcolor{Red}{$\textbf{M}_{\textbf{11}}$},\\  \textcolor{Bittersweet}{$\textbf{M}_{\textbf{13}}$} 
\\   \textcolor{Red}{$\textbf{M}_{\textbf{6}}$}\end{tabular}       \\ \midrule

Broad stripes     & 
\begin{tabular}[c]{@{}l@{}}(\textit{wg}$_{3,4}$; WG$_{3,4}$; \textit{en}$_{1,2}$; EN$_{1,2}$; \\ \textit{hh}$_{1,2}$; HH$_{1,2}$; \textit{ptc}$_{3,4}$; PTC$_{3,4}$; \\ \textit{ci}$_{3,4}$; CI$_{3,4}$; CIA$_{3,4}$)\end{tabular}    & \begin{tabular}[c]{@{}l@{}}\{ (\textit{en}$_1$-1, \textit{en}$_2$-1),\\ (\textit{en}$_2$-1, \textit{ptc}$_4$-1),\\ (\textit{en}$_1$-1, \textit{ptc}$_3$-1),\\ (\textit{en}$_1$-1, \textit{wg}$_3$-1),\\ (\textit{wg}$_4$-1, \textit{en}$_2$-1) \}\end{tabular}

& \begin{tabular}[c]{@{}l@{}}\textcolor{BlueViolet}{$\textbf{M}_{\textbf{10}}$},\\   \textcolor{BlueViolet}{$\textbf{M}_{\textbf{10}}$},\\   \textcolor{Red}{$\textbf{M}_{\textbf{6}}$},\\   \textcolor{Red}{$\textbf{M}_{\textbf{6}}$}\end{tabular}     \\ \midrule

No segmentation   & 
\begin{tabular}[c]{@{}l@{}}(\textit{ci}$_{1,2,3,4}$; CI$_{1,2,3,4}$; \\ PTC$_{1,2,3,4}$; CIR$_{1,2,3,4}$)\end{tabular}                                                       & \begin{tabular}[c]{@{}l@{}}\{ (CIR$_2$-1, CIR$_4$-1),\\ (CIR$_3$-1, CIR$_1$-1),\\ (SLP$_3$-1, CIR$_1$-1),\\ (SLP$_4$-1, CIR$_2$-1),\\ (SLP$_3$-1, CIR$_2$-1),\\ (SLP$_4$-1, CIR$_1$-1) \}\end{tabular}

& \begin{tabular}[c]{@{}l@{}} \textcolor{Bittersweet}{$\textbf{M}_{\textbf{14}}$} 
\\ \textcolor{Bittersweet}{$\textbf{M}_{\textbf{14}}$} 
\\ \textcolor{Bittersweet}{$\textbf{M}_{\textbf{13}}$}
\\ \textcolor{Bittersweet}{$\textbf{M}_{\textbf{13}}$}
\end{tabular} \\ \midrule 

Wild-type variant & 
\begin{tabular}[c]{@{}l@{}}(\textit{wg}$_4$; WG$_4$; \textit{en}$_1$; EN$_1$; \textit{hh}$_1$; \\ HH$_1$; \textit{ptc}$_{2,4}$; PTC$_{1,2,3,4}$; \\ \textit{ci}$_{2,3,4}$; CI$_{2,3,4}$; CIA$_{2,4}$; CIR$_3$)\end{tabular}  & \begin{tabular}[c]{@{}l@{}}\{ (\textit{en}$_1$-1, SMO$_1$-0),\\ (SMO$_1$-0, \textit{ptc}$_4$-1),\\ (SMO$_1$-0, \textit{ptc}$_2$-1),\\ (\textit{wg}$_4$-1, SMO$_1$-0) \}\end{tabular} 

& \begin{tabular}[c]{@{}l@{}}\textcolor{ProcessBlue}{$\textbf{M}_\textbf{7}$}/\textcolor{RawSienna}{$\textbf{M}_{\textbf{12}}$},\\   \textcolor{Red}{$\textbf{M}_{\textbf{11}}$},\\  \textcolor{Bittersweet}{$\textbf{M}_{\textbf{13}}$} 
\\   \textcolor{Red}{$\textbf{M}_{\textbf{6}}$}\end{tabular}               \\ \midrule

Ectopic           & 
\begin{tabular}[c]{@{}l@{}}(\textit{wg}$_3$; WG$_3$; \textit{en}$_2$; EN$_2$; \textit{hh}$_2$; \\ HH$_2$; \textit{ptc}$_{1,3}$; PTC$_{1,3,4}$; \textit{ci}$_{1,3,4}$; \\ CI$_{1,3,4}$; CIA$_{1,3}$; CIR$_{4}$)\end{tabular}    & \begin{tabular}[c]{@{}l@{}}\{ (CIR$_4$-1, PTC$_2$-0),\\ (\textit{ptc}$_1$-1, PTC$_2$-0),\\ (CIA$_1$-1, PTC$_2$-0),\\ (PH$_1$-1, PTC$_2$-0),\\ (SMO$_4$-0, PTC$_2$-0) \}\end{tabular}

& \begin{tabular}[c]{@{}l@{}}\textcolor{Red}{$\textbf{M}_{\textbf{11}}$},\\   \textcolor{ProcessBlue}{$\textbf{M}_\textbf{7}$}/\textcolor{RedOrange}{$\textbf{M}_{\textbf{9}}$},\\   \textcolor{Red}{$\textbf{M}_{\textbf{6}}$},\\  \textcolor{Bittersweet}{$\textbf{M}_{\textbf{13}}$}
\end{tabular}                   \\ \midrule

Ectopic variant   & 
\begin{tabular}[c]{@{}l@{}}(\textit{wg}$_3$; WG$_3$; \textit{en}$_2$; EN$_2$; \textit{hh}$_2$; \\ HH$_2$; \textit{ptc}$_{1,3}$; PTC$_{1,2,3,4}$; \textit{ci}$_{1,3,4}$; \\ CI$_{1,3,4}$; CIA$_{1,3}$; CIR$_{4}$)\end{tabular}  & \begin{tabular}[c]{@{}l@{}}\{ (SMO$_2$-0, \textit{en}$_2$-1),\\ (SMO$_2$-0, \textit{ptc}$_3$-1),\\ (SMO$_2$-0, \textit{wg}$_3$-1),\\ (SMO$_2$-0, \textit{ptc}$_1$-1) \}\end{tabular} 

& \begin{tabular}[c]{@{}l@{}}\textcolor{Red}{$\textbf{M}_{\textbf{11}}$},\\   \textcolor{ProcessBlue}{$\textbf{M}_\textbf{7}$}/\textcolor{RawSienna}{$\textbf{M}_{\textbf{12}}$},\\   \textcolor{Red}{$\textbf{M}_{\textbf{6}}$},\\ \textcolor{Bittersweet}{$\textbf{M}_{\textbf{13}}$}
\end{tabular}    \\ \bottomrule          
\end{tabular} 
}
\caption{\textbf{Attractors and minimal observability sets for the \textit{drosophila} parasegment SPN.} 
A node
or s-unit's index indicates to which cell in the parasegment it belongs.  In addition to the expressed nodes listed, every attractor has input nodes SLP$_{1,2}=$OFF, SLP$_{3,4}=$ON.  The associated intracellular complex modules are listed for cells 1-4 (top to bottom) in the parasegment.  Only the main complex modules for each cell are listed; they are colored to correspond to Table \ref{tab:drosophila_modules}.  
Here, 
$\textbf{M}_6$ indicates $\textbf{M}_{SLP-1,nHH-1}$; $\textbf{M}_7$ indicates $\textbf{M}_{SLP-0,nWG-1}$; 
$\textbf{M}_9$ indicates $\textbf{M}_{PTC-0}$, $\textbf{M}_{10}$ indicates $\textbf{M}_{SLP-0,nWG-1,nHH-1}$,
$\textbf{M}_{11}$ indicates $\textbf{M}_{nWG-0,nHH-1}$, $\textbf{M}_{12}$ indicates $\textbf{M}_{nHH-0,PTC-1}$, $\textbf{M}_{13}$ indicates $\textbf{M}_{SLP-1,nHH-0,PTC-1}$; finally, $\textbf{M}_{14}$ indicates $\textbf{M}_{nWG-0,nHH-0,PTC-1}$.  
Note that observability sets only require two sensors to reveal the given attractor.}
\label{tab:control_sets_parasegment}
\end{adjustwidth}
\end{table}


\begin{table}[]
\begin{adjustwidth}{-2cm}{}
\centering
\begin{tabular}{lccccccc} \toprule
                     &  
                     Nodes     
                     &  S-units & Attractors & 
   $\overline{D}(\Pi^*)$ &               
         $|\Pi^*|$ 
         & avg. $|S_i| \in \Pi^*$ & max $|S_i| \in \Pi^*$ \\ \midrule
Example GRN              & 6                                                              & 12             & 5            & 0.83                                                             & 4                                                             & 3.25  & 5  (83\%)      \\ 
\begin{tabular}[c]{@{}l@{}}\textit{Drosophila} \\ single-cell\end{tabular} & 17                                                             & 34             & 10            & 0.81                                                             & 8                                                              & 5.25      & 16  (94\%)     \\ 
\textit{Thaliana}                                                         & 15                                                             & 30             & 10            & 0.98                                                             & 11                                                              & 2.82     & 11 (73\%)      \\ 
Yeast                                                             & 12                                                             & 24             & 11            & 1.0                                                             & 7                                                              & 3.43  & 12 (100\%)      \\ \midrule

\begin{tabular}[c]{@{}l@{}}\textit{Drosophila} \\ parasegment\end{tabular} & 60                                                             & 120            & ?             & 0.87                                                             & 32                                                             & 4.44  & 33 (55\%)      \\ 
Leukemia                                                          & 60                                                             & 120            & ?             & 0.64                                                             & 41                                                             & 4.95  & 42 (70\%)     \\ \bottomrule
\end{tabular} 
\caption{\textbf{Comparison of dynamical modularity between genetic regulatory networks.}  The optimal cover $\Pi^*$ is estimated for each network by maximizing the mean 
dynamical modularity
among covers across seed set sizes, assuming pinning perturbation.  
Six networks are considered: the example GRN from Fig. 2 in the main text, the \textit{drosophila} single-cell and parasegment SPNs \cite{albert2003topology}, the \textit{Thaliana arabadopsis} cell-fate specification network \cite{chaos2006genes}, the yeast cell-cycle network \cite{li2004yeast}, and the T-LGL leukemia network \cite{zhang2008network}.
For the first grouping of networks, $\Pi^*$ was calculated considering all possible covers of $q \leq 6$ core complex modules with seed set size $s \leq 5$, with higher $q$ values used when computationally feasible.
By contrast, $\Pi^*$ for the \textit{drosophila} parasegment and leukemia networks was estimated using the greedy heuristic mentioned in Section 2.3.
For each network the mean dynamical modularity
of the optimal cover,
$\overline{D}(\Pi^*)$, the size of the estimated optimal cover, $|\Pi^*|$, the average module size within the optimal cover, avg. $|S_i| \in \Pi^*$, and the maximum module size within the optimal cover, max $|S_i| \in \Pi^*$, is shown.
Percentages are calculated based on the number of 
nodes 
in the network.}
\label{tab:comparison_modularity}
\end{adjustwidth}
\end{table}

\begin{table}[]
\begin{adjustwidth}{-2cm}{}
\centering
\begin{tabular}{lccccc} \toprule
 &  \begin{tabular}[c]{@{}c@{}}Average driver \\ set size \end{tabular} & \begin{tabular}[c]{@{}c@{}}Min driver \\ set size\end{tabular} & \begin{tabular}[c]{@{}c@{}}Max driver \\ set size\end{tabular} & \begin{tabular}[c]{@{}c@{}}Driver superset \\ size\end{tabular} & \begin{tabular}[c]{@{}c@{}}FVS \\ size\end{tabular} \\ \midrule
Example GRN   
& 2 & 2 & 2 & 2 & 3 
\\
\begin{tabular}[c]{@{}l@{}}\textit{Drosophila} \\ single-cell\end{tabular} 
& 3.7                                                                        & 3                                                        & 4                                                        & 5                                                             & 5                                                   \\
\textit{Thaliana}                  
& 6.1                                                                        & 5                                                        & 7                                                        & 9                                                             & 9                                                   \\
Yeast         
& 6.45                                                                & 5                                                        & 7                                                        & 7                                                             & 8         \\ \bottomrule                                         
\end{tabular} 
\caption{\textbf{Comparison of attractor controllability between genetic regulatory networks.}  Statistics for the driver sets sufficient to fully resolve each 
fixed point
attractor in the network, assuming pinning perturbation, are shown.  
Driver sets are equivalent to the seed sets of the minimal pathway modules whose unfolding 
includes all node states found in the respective attractor.
The superset of 
driver nodes 
needed (based on the set of s-units sufficient to control to each attractor) is compared to that predicted by feedback vertex set theory \cite{fiedler2013dynamics, mochizuki2013dynamics}.
See the caption of Fig.~\ref{tab:comparison_modularity} for a description of the networks used here.
}
\label{tab:comparison_controllability}
\end{adjustwidth}
\end{table}

\begin{table}[]
\begin{adjustwidth}{-2cm}{}
\begin{tabular}{llll|ll} \toprule
 & Yeast               & \textit{Thaliana}    & \begin{tabular}[c]{@{}l@{}}\textit{Drosophila}\\ single-cell\end{tabular}   & \begin{tabular}[c]{@{}l@{}}\textit{Drosophila}\\ parasegment\end{tabular}                                             & Leukemia                                                   \\ \midrule
Network size 
& 12        & 15            & 17                   & 60                & 60                                                         \\ \midrule
\begin{tabular}[c]{@{}l@{}}Number of\\ attractors\end{tabular}           & 11         & 10          & 10                    & ?    & ?                                                          \\ \midrule

\begin{tabular}[c]{@{}l@{}}$s=1$\\  number of modules,\\  maximum module size\\  (\% of network)\end{tabular} &  \begin{tabular}[c]{@{}l@{}}17,\\ 4 (33\%)\end{tabular}   & \begin{tabular}[c]{@{}l@{}}16,\\ 8 (53\%)\end{tabular}   &
\begin{tabular}[c]{@{}l@{}}14,\\ 9 (53\%)\end{tabular}           & \begin{tabular}[c]{@{}l@{}}40,\\ 13 (22\%)\end{tabular}          & \begin{tabular}[c]{@{}l@{}}48,\\ 24 (40\%)\end{tabular}    \\ \midrule

\begin{tabular}[c]{@{}l@{}}$s=2$\\  number of modules,\\  maximum module size\\  (\% of network)\end{tabular}          & \begin{tabular}[c]{@{}l@{}}12,\\ 10 (83\%)\end{tabular}  & \begin{tabular}[c]{@{}l@{}}7,\\ 12 (80\%)\end{tabular}  &
\begin{tabular}[c]{@{}l@{}}8,\\ 16 (94\%)\end{tabular}          & \begin{tabular}[c]{@{}l@{}}40,\\ 28 (47\%)\end{tabular}  &
\begin{tabular}[c]{@{}l@{}}64,\\ 41 (68\%)\end{tabular}   \\ \midrule

\begin{tabular}[c]{@{}l@{}}$s=3$\\  number of modules,\\  maximum module size\\  (\% of network)\end{tabular}    & \begin{tabular}[c]{@{}l@{}}12,\\ 12 (100\%)\end{tabular} & \begin{tabular}[c]{@{}l@{}}3,\\ 14 (93\%)\end{tabular}  & \begin{tabular}[c]{@{}l@{}}6,\\ 17 (100\%)\end{tabular}         & \begin{tabular}[c]{@{}l@{}}152,\\ 52 (87\%)\end{tabular}      &
\begin{tabular}[c]{@{}l@{}}223,\\ 53 (88\%)\end{tabular}  \\ \midrule

\begin{tabular}[c]{@{}l@{}}$s=4$\\  number of modules,\\  maximum module size\\  (\% of network)\end{tabular} &  \begin{tabular}[c]{@{}l@{}}4,\\ 12 (100\%)\end{tabular} & \begin{tabular}[c]{@{}l@{}}0,\\ \end{tabular} &
\begin{tabular}[c]{@{}l@{}}0,\\ \end{tabular}        & \begin{tabular}[c]{@{}l@{}}472,\\ 56 (93\%)\end{tabular}  & \begin{tabular}[c]{@{}l@{}}646,\\ 54 (90\%)\end{tabular}  \\ \midrule

\begin{tabular}[c]{@{}l@{}}$s=5$\\  number of modules,\\  maximum module size\\  (\% of network)\end{tabular} &  \begin{tabular}[c]{@{}l@{}}2,\\ 12 (100\%)\end{tabular} & \begin{tabular}[c]{@{}l@{}}0,\\ \end{tabular} &
\begin{tabular}[c]{@{}l@{}}0,\\ \end{tabular}        & \begin{tabular}[c]{@{}l@{}}1073,\\ 59 (98\%)\end{tabular}       & \begin{tabular}[c]{@{}l@{}}1446,\\ 56 (93\%)\end{tabular} \\ \midrule

\begin{tabular}[c]{@{}l@{}}$s=6$\\  number of modules,\\  maximum module size\\  (\% of network)\end{tabular} &  \begin{tabular}[c]{@{}l@{}}0,\\ \end{tabular} & \begin{tabular}[c]{@{}l@{}}0,\\ \end{tabular} &
\begin{tabular}[c]{@{}l@{}}0,\\ \end{tabular}        & \begin{tabular}[c]{@{}l@{}}1475,\\ 60 (100\%)\end{tabular}       & \begin{tabular}[c]{@{}l@{}}2353,\\ 57 (95\%)\end{tabular} \\ \midrule

\begin{tabular}[c]{@{}l@{}}Min fixed point seed set size\\ (\% of network)\end{tabular}    & 3 (25\%)     & 4 (27\%)        & 3 (18\%)      & 6 (10\%)        & 8 (13\%)  \\ \bottomrule                
\end{tabular}
\caption{\textbf{Comparison of core complex modules between 
genetic regulatory networks.}
The number of core complex modules, along with the maximum module size, is found per network per seed set size $s$. 
Additionally, the minimal seed set size found to fully resolve every variable state (min fixed point seed set size) is shown; note that this fixed point is not necessarily an attractor of the original network, due to the pinning of the seeds involved.
It is apparent that variable states in 
each network are quickly resolved as $s$ is increased when considering the largest complex module.  
In comparing the larger networks, the leukemia network has more core complex modules than the \textit{drosophila} parasegment per $s$ value, suggesting more complicated dynamics.
Percentages are calculated based on the number of variables 
(nodes) in the network.
See the caption of Fig.~\ref{tab:comparison_modularity} for a description of the networks used here.
}
\label{tab:comparison}
\end{adjustwidth}
\end{table}

\pagebreak

\bibliographystyle{unsrt}

\bibliography{dynamics.bib} 


